\def\kasten{\hfil\vrule height6pt width5pt depth-1pt\par }
\def\BS{{\sf S}}
\def\R{{I\!\! R}}
\def\N{{I\!\! N}}
\def\Z{{Z\!\!\! Z}}
\def\p{\partial}

\def\half{{1 \over 2}}
\def\lra{\longrightarrow}
\def\1{{1\!\!1}}
\def\a{\alpha}
\def\b{\beta}
\def\d{\delta}
\def\e{\epsilon}

\def\k{\kappa}
\def\l{\lambda}
\def\om{\omega}
\def\s{\sigma}

\def\Th{\Theta}
\def\vp{\varphi}
\def\ve{\varepsilon}

\def\D{\Delta}
\def\G{\Gamma}
\def\L{\Lambda}
\def\Om{\Omega}

\def\BC{{\cal B}}

\def\DC{{\cal D}}
\def\EC{{\cal E}}
\def\FC{{\cal F}}
\def\GC{{\cal G}}
\def\HC{{\cal H}}

\def\LC{{\cal L}}

\def\PC{{\cal P}}

\def\SC{{\cal S}}

\def\W{{\sf W}}
\def\lv{\left\vert}
\def\rv{\right\vert}

\def\C{\hbox{\vrule width 0.6pt height 6pt depth 0pt \hskip -3.5pt}C}

\def\ssrd{ {\SC ' (\R^d)} }
\def\srd{ {\SC (\R^d)} }
\def\BS{{\sf S}}

\documentclass[11pt]{article}

\newcommand{\bdm}{\begin{displaymath}}
\newcommand{\edm}{\end{displaymath}}
\newcommand{\be}{\begin{equation}}
\newcommand{\ee}{\end{equation}}
\newcommand{\bea}{\begin{eqnarray}}
\newcommand{\eea}{\end{eqnarray}}
\newcommand{\beas}{\begin{eqnarray*}}
\newcommand{\eeas}{\end{eqnarray*}}

\newtheorem{Theorem}{Theorem}[section]
\newtheorem{Definition}[Theorem]{Definition}
\newtheorem{Proposition}[Theorem]{Proposition}
\newtheorem{Lemma}[Theorem]{Lemma}
\newtheorem{Corollary}[Theorem]{Corollary}
\newtheorem{Remark}[Theorem]{Remark}
\newtheorem{Example}[Theorem]{Example}
\newtheorem{Condition}[Theorem]{Condition}

\begin{document}

\author{Sergio Albeverio \\
        Fakult\"at f\"ur Mathematik, Ruhr--Universit\"at Bochum,
        Germany; \\
        SFB 237 Essen--Bochum--D\"usseldorf; \\
        BiBoS Research Centre, Bielefeld; \\
        CERFIM, Locarno (Switzerland)
  \and
        Hanno Gottschalk \\
        Fakult\"at f\"ur Mathematik, Ruhr--Universit\"at Bochum,
        Germany \\           
  \and
        Jiang-Lun Wu \\
        Fakult\"at f\"ur Mathematik, Ruhr--Universit\"at Bochum,
        Germany; \\
        Probability Laboratory, Institute of Applied Mathematics, \\ 
        Academia Sinica, Beijing 100080, PR China
        }

\title{Convoluted Generalized White Noise, Schwinger Functions
and Their Analytic Continuation to Wightman Functions}

\pagenumbering{arabic}
\maketitle
\begin{abstract}
{\footnotesize We construct Euclidean random fields $X$ over $\R^d$, 
by convoluting
generalized white noise $F$ with some integral kernels
$G$, as $X=G* F$. We study properties of Schwinger (or moment)
functions of $X$. In particular, we give a general equivalent 
formulation of the cluster property in terms of truncated Schwinger 
functions which we then
apply to the above fields. We present a partial negative result on 
the reflection positivity of convoluted generalized white noise.

Furthermore, by representing the kernels $G_\a$ of
the pseudo--differential operators $(-\D + m^2_0 )^{-\a}$ for
$\a \in (0,1)$ and $m_0>0$ as Laplace transforms 
we perform the analytic continuation of
the (truncated) Schwinger functions of $X=G_\a * F$, obtaining the
corresponding (truncated) Wightman distributions on Minkowski space
which satisfy the relativistic postulates on invariance, spectral
property, locality and cluster property. 

Finally we give some remarks on scattering theory for these models.}
\end{abstract}
\tableofcontents

\setcounter{section}{-1}
\section{Introduction}
\label{null}
Since the work of Nelson in the early 70's the problem of the mathematical
construction of models of interacting local relativistic quantum fields has 
been related to the one of the construction of Markovian Euclidean 
generalized random fields. 

In models for scalar fields in space-time dimension two Markovian can be 
understood in the strict "global Markov sense", as proven in \cite{AHk},
\cite{Gie}, \cite{Ze}, \cite{AHkZe}(see also \cite{AZe} and references 
therein). This is also true for a class of vector models in space-time
dimensions $d=2,4$(and $8$), see \cite{AHkI}, \cite{AIK}, \cite{AIK1}, 
\cite{AIK2}, \cite{Os} \cite{Sc2}. The latter models are of 
the gauge-type and the
construction of an associated Hilbert space, in the cases $d=4$ and $8$,
presents difficulties(these do not exist for $d=2$, see \cite{AIK2}; for
a partial result for $d=4$ see \cite{AIK1}). For results on a model of 
a scalar field for $d=3$(with the Markovian property replaced by 
reflection positivity in the sense of \cite{GlJa}) see references in 
\cite{GlJa} and \cite{AZh}, for partial(weak and rather negative)
results for scalar fields for $d=4$ see \cite{AFHkL}, \cite{GlJa},
\cite{F},\cite{FFS}, \cite{Re}, \cite{Reh}. For partial results on 
conformal fields on other types of $d=4$ space-times see \cite{Se}.

A program of constructing Euclidean random fields of Markovian type by 
solving pseudo-stochastic partial differential equations of the form
$LX=F$ with $F$ a Euclidean noise and $L$ a suitable invariant 
pseudodifferential operator was started in \cite{Su1}, \cite{AHk1}, 
\cite{AHk2}, \cite{AHk3}, see 
also \cite{AHkI}, \cite{AIK1}, \cite{AIK2}, \cite{BeGL} in the vector
case, and in \cite{AW} for the scalar case (in a recent note 
J. Klauder, see e.g.
\cite{K} and \cite{K1}, and references therein, also 
advocated the use of non ~Gussian
noises in stochastic PDEs resp. functional integrals to circumvent
triviality for scalar fields, his suggestion is however different from
ours). In the present case we 
continue the work initiated in \cite{AW}, extending the study of random 
fields of the form $X=G* F$(of which the above case $G=L^{-1}$ is a 
special one), to more general $G$ than in \cite{AW}, in particular covering
$G_{\a}=(-\D + m^2_0)^{-\a}$ for $\a \in (0,1)$ and $m_0 \ge 0$(only 
the case $m_0=0, \a = 1$ was treated in \cite{AW}). For $\a=\half$, $F$
Gaussian white noise, $X$ is Nelson's Euclidean free field over
$\R^d$\cite{N2}; for $\a={1\over4}$, $X$ is the time zero free field (over
a space--time of dimension $d+1$). The idea of extending $F$ to be a
general, not necessary Gaussian Euclidean noise ("generalized white noise")
(i.e., a generalized random field "independent at every point", in the 
terminology of \cite{GV}, or a "completed scattered random measure", which is
homogeneous with respect to the Euclidean group over $\R^d$) can be motivated
from different points of view. Let us mention three of them (see also e.g.
\cite{AHk1},\cite{AHk2},\cite{AHk3}, and \cite{AHkI}):

\begin{enumerate}
\item From a general Euclidean noise $F$ one can recover a Gaussian
Euclidean (Gaussian white) noise $F^g$ as weak limit, see Remark \ref{1.4rem}
at the end of Section 1. Thus one can look at the fields $X=G\ast F$
constructed from $F$ as perturbations of the free fields $X=G\ast F^g$
constructed from $F^g$. This perturbation is of another type than the
usual perturbations of (Euclidean) quantum field theory (given by additive
Feynman--Kac type functionals, see e.g. \cite{AFHkL},\cite{AZe},\cite{Si},
\cite{GlJa}).

\item As explained in details in Remark \ref{fuenf.4rem} below, the Schwinger
functions $S_n$ of our model, suitably scaled by a factor $\l^{-{n\over
2}}$, $\l>0$, can be written in terms of the free field Schwinger
functions plus a polynomial of finite order in the "coupling constant"
$\l^{-1}$ without a constant term, the coefficients being products of 
truncated Schwinger functions of order $2\leq l \leq n$. In this sense 
we have a parameter $\l^{-1}$ which measures "the amount of Poisson 
component" which can also be seen as "the amount of interaction" present in
the given Schwinger functions.

\item One can give a discrete approximation or "lattice approximation" of the 
models, by replacing $\R^d$ by a lattice $\d\Z^d$, and correspondingly $L$
and $F$ by discrete analogues, see e.g. for special cases, \cite{AIS}, 
\cite{AW}, and \cite{AW1}. One can then interpret 
the probabilistic law of $X$ in a 
bounded region as given by a Euclidean action with a non quadratic kinetic 
energy part (depending essentially on the L\'evy measure characterizing the 
distribution of $F$). 
\end{enumerate}
Let us explain this a little further, starting from be the 
lattice approximation $F^\L_\d $ of a
generalized white noise $F$, where the superscript $\L $ indicates a
cutoff outside a bounded region $\L\subset \R^d$. We assume that $F$ is
determined by a L\'evy characteristic $\psi$ (cf. Section 1) such that
the convolution semigroup $(\mu_t)_{t> 0}$ generated by $\psi$
\cite{BeFo} is absolutely continuous w.r.t. Lebesgue measure on $\R$. Let 
$\varrho_t$, $t> 0$, be the corresponding densities. Then the probability 
distribution of $F_\d^\L$ in every lattice point $\d n \in \L_\d := \d \Z^d\cap
\L$ is given by $\mu_{\d^d}=\varrho_{\d ^d}dx$. We denote by $L_\d^\L$
the lattice discretization of a partial (pseudo-) differential operator
$L$ over the lattice $\L_\d$. Furthermore, we assume that $L_\d^\L$ as
a $|\L_\d|\times |\L_\d|$-matrix is invertible. Then the solvable
discrete stochastic equation $L_\d^\L X_\d^\L=F_\d^\L$ is the lattice
analogue of $LX=F$.We set
$$W_\d(x):=-\d^{-d}\log \varrho_{\d^d}(\d^dx), \ x\in \R .$$
The lattice measure $P_{X_\d^\L}$ is defined as the measure with
respect to which $X_\d^\L$ is the coordinate process. Then we have
(see \cite{AW} and \cite{AW1})
\begin{eqnarray*}
&& P_{X_\d^\L}\{X_\d^\L\in A\}\\
&& = Z^{-1}\int_A e^{-\sum _{\d n\in \L_\d}
\d^dW_\d ((L^\L_\d X_\d^\L ) (\d n)) } \prod_{\d n\in\L_\d}dX_\d^\L(\d n), 
\end{eqnarray*}
for Borel measurable subsets $A\subset \R^{\L_\d}$. Here $\prod_{\d
n\in\L_\d}dX_\d^\L(\d n)$ denotes the flat lattice measure and $Z$ is a
normalization constant which depends on $\d$, $\L$ and $L$. For
$(\mu_t)_{t> 0}$ the Gaussian semigroup (of mean zero and variance $t$) and 
$L_\d^\L$ the discretization
of $(-\D + m_0^2)^\half$ we get the usual lattice approximation of
the Nelson's free field with mass $m_0>0$. If $(\mu _t)_{t> 0}$ is not
the Gaussian semigroup, the action $W_\d$ is no longer quadratic and
therefore contains terms which can be identified with some kind of
interaction.

In this sense the models can be looked upon as 
quantized versions of nonlinear field models (with some analogy with models 
like the Einstein--Infeld field model, see \cite{AHk1},\cite{AHk2},
\cite{AHk3}, and \cite{AHkI}).

In this paper we construct $X$, as given in general by $G\ast F$, study its 
regularity properties and the properties of the associated moment 
functions(Schwinger functions), proving invariance and cluster property. 
We also perform their analytic continuation to relativistic Wightman    
functions, which are shown to satisfy all Wightman axioms (possibly except
for positivity),including the cluster property (a point not discussed in
\cite{AW} for the special case considered there). We also provide a 
counterexample to the reflection positivity condition, for a particular 
choice of noise $F$(with a "sufficiently strong" Poisson component). We also 
indicate that despite the possible absence of reflection positivity in non
Gaussian cases, one can associate scattering states which partly express a 
"particle structure" of the models. We also make several comments concerning the 
interplay of properties of $G$ with the Markov property respectively the
reflection positivity of $X$.

One main method used in our analysis is the study of truncated Schwinger 
functions. This, as most of the results of the present work, is based on
\cite{Go}. Some of our results have been announced in \cite{AGW}.

Here are some details on the single sections in this paper. In section 1
we introduce the basic (white) noises $F$ used in this work. In
section 2 we describe the kernels $G$ and the random fields given by 
$X=G* F$. In section 3 we discuss the basic invariance properties of the
moment functions $S_n$(Schwinger functions) of $X$. In section 4 we
discuss the cluster property of the $S_n$. Section 5 is devoted to a
discussion of the reflection positivity property and to an example
showing that it does not hold for general $F$. In sections 6 and 7 the 
analytic continuation of the Schwinger functions to Wightman functions
is discussed, first(section 6) for the two-point function and then for
the general $n$-point Schwinger functions(section 7). Section 7
also contains the discussion of "positivity properties" in a "scattering
region".

\section{The Generalized White Noise}
\label{eins}
In this section, we shall present some basic concepts as background 
for our discussions in later sections.

As is well--known in probability theory, an infinite divisible probability
distribution $P$ is a probability distribution having the property
that for each $n \in \N$ there exists a probability distribution $P_n$
such that the $n$--fold convolution of $P_n$ with itself is $P$, i.e., 
$P = P_n* \ldots* P_n$ (n times). By L\'evy--Khinchine theorem 
(see e.g. Lukacs \cite{Lu}) we know that the Fourier transform (or
characteristic function) of $P$, denoted by $C_P$, satisfies 
the following formula
\be 
\label{cpt}
  C_P (t) := \int_\R e^{ist} dP (s) = e^{\psi (t)} \ , 
  \ \ t \in \R \ ,
\ee 
where $\psi :\R\to\C$ is a continuous function, called the L\'evy
characteristic of $P$, which is uniquely represented as follows 
\begin{equation}
\label{psit.eq}
  \psi (t) = iat - {\s^2 t^2 \over 2} + \int_{\R \setminus \{ 0 \} }
  \left( e^{ist} - 1 - {ist \over 1 + s^2} \right) dM(s) \ , \ \ 
  t \in \R \ ,
\end{equation}
where $a \in \R$, $\s \ge 0$ and the function $M$ satisfies the
following condition
\be 
\label{intr0}
  \int_{\R \setminus \{ 0 \} } \min (1,s^2) dM(s) < \infty \ .
\ee 
On the other hand, given a triple $(a,\s,M)$ with $a\in\R,\s\ge0$ 
and a measure $M$ on $\R\setminus\{0\}$ which ~fulfils (\ref{intr0}), there
exists a unique infinitely divisible probability distribution $P$
such that the L\'evy characteristic of $P$ is given by (\ref{psit.eq}).

Let $d \in \N$ be a fixed space time dimension. Let $\SC (\R^d )$
(resp. $\SC_{\C} (\R^d )$) be the Schwartz space of all rapidly
decreasing real -- (resp. complex--) valued $C^\infty$--functions on
$\R^d$ with the Schwartz topology. Let $\SC ' (\R^d)$ (resp. 
$\SC '_{\C} (\R^d)$) be the topological dual of $\SC (\R^d)$ (resp.
$\SC_{\C} (\R^d)$). We denote by $< \cdot , \cdot >$ the dual pairing
between $\SC (\R^d)$ (resp. $\SC_{\C} (\R^d)$) and $\SC ' (\R^d)$ (resp.
$\SC ' _{\C} (\R^d)$). Let $\BC$ be the $\s$--algebra generated by
cylinder sets of $\SC ' (\R^d)$. Then $(\SC ' (\R^d) , \BC)$ is a
measurable space.

By a characteristic functional on $\SC (\R^d)$, we mean a functional
$C:\SC (\R^d) \to {\C}$ with the following properties
\begin{enumerate}
\item $C$ is continuous on $\SC (\R^d)$;
\item $C$ is positive--definite;
\item $C(0) = 1$.
\end{enumerate}

By the well-known Bochner-Minlos theorem(see e.g. \cite{GV}) there exists 
a one to one correspondence between characteristic functionals $C$ 
and probability measures $P$ on $(\SC ' (\R^d),\BC)$ given by the
following relation
\be 
C(f)=\int_{\SC'(\R^d)} e^{i<f,\xi>}dP(\xi), \ \ f \in \SC(\R^d) \ .
\ee 
We have the following result
\begin{Theorem}
\label{eins.1theo}
Let $\psi$ be a L\'evy characteristic defined by (1). Then
there exists a unique probability measure $P_\psi$ on 
$(\SC'(\R^d), \BC)$ such that the Fourier transform of $P_\psi$
satisfies 

\be 
\label{intscr.eq}
  \int_{\SC ' (\R^d)} e^{i<f,\xi>} dP_\psi (\xi) = \exp
  \left\{ \int_{\R^d} \psi (f(x) ) dx \right\} \ ,
  \ \ f \in \SC (\R^d) \ .
\ee 
\end{Theorem}

{\bf Proof.}
It suffices to show that the right hand side of (\ref{intscr.eq})
is a characteristic functional on $\SC (\R^d)$. This is 
true, e.g., by Theorem 6 on p. 283 of \cite{GV}. 
\kasten

\begin{Definition}
\label{eins.1def}
We call $P_\psi$ in Theorem \ref{eins.1theo} a generalized white noise
measure with L\'evy characteristic $\psi$ and 
$(\SC ' (\R^d) , \BC , P_\psi )$ the generalized white noise space
associated with $\psi$. The associated coordinate process 
$$F:\SC (\R^d)\times(\SC ' (\R^d) , \BC , P_\psi ) \to \R$$
 defined by
\be 
\label{ffxi.eq}
  F(f,\xi) = <f , \xi > \ , \ \ f \in \SC (\R^d) \ , 
  \ \ \xi \in \SC ' (\R^d)
\ee 
is called {\bf generalized white noise}.
\end{Definition}

\begin{Remark}
\label{1.3rem}
In the terminology of \cite{GV}, $F$ is called a generalized random 
process with independent values at every point, i.e., the random 
variables $<f_1, \cdot>$ and $<f_2 , \cdot>$ are independent whenever
$f_1 (x) f_2 (x) = 0$ for $f_1, f_2 \in \SC (\R^d)$. 
\end{Remark}

Combining (\ref{psit.eq}) and (\ref{intscr.eq}), we get, for $f \in 
\SC(\R^d )$, that
\begin{eqnarray}
\label{intscr2.eq}
  &   &  \int_{\SC ' (\R^d)} e^{i<f,\xi >} dP_\psi (\xi) \nonumber\\
  & = & \exp \left\{ ia \int_{\R^d} f(x) dx - {\s^2 \over 2} 
        \int_{\R^d} [f(x)]^2 dx \right. \nonumber\\
  &   & \left.+ \int_{\R^d} \int_{\R \setminus \{ 0 \} } \left( 
        e^{isf(x)} - 1 - {isf(x) \over 1 + s^2} \right)
        dM (s) dx \right\} \ .   
\end{eqnarray}
From (\ref{intscr2.eq}), we see that a generalized white noise $F$ is
composed by three independent parts, namely, we can give an equivalent
(in law) realization of $F$ as the following direct sum
\be 
\label{xi.eq}
 F(f,\cdot) = F_a(f,\cdot)\oplus F_{\s}(f,\cdot)\oplus F_M(f,\cdot)
\ee 
for $f \in \SC (\R^d)$ with $F_a, F_{\s}$ and $F_M$ the coordinate
processes on the probability spaces $(\SC'(\R^d), \BC, P_a),
(\SC'(\R^d), \BC, P_{\s})$ and $(\SC'(\R^d), \BC, P_M)$, respectively,
where $P_a, P_{\s}$ and $P_M$ are defined, by Theorem 1.1, by the 
following relations  
\beas
  \int_{\SC'(\R^d)} e^{i<f,\xi>} dP_a(\xi) &=& \exp
  \left\{ ia \int_{\R^d} f(x) dx \right\}\\
 \int_{\SC'(R^d)} e^{i<f,\xi>} dP_{\s} (\xi)
   & =& \exp \left\{ - {\s^2 \over 2} \int_{\R^d} [f(x)]^2dx \right\}
\eeas
\begin{eqnarray*}
& &  \int_{\SC'(\R^d)} e^{i<f,\xi>} dP_M(\xi) \\
& = & \exp \left\{ \int_{\R^d} 
   \int_{\R \setminus \{ 0 \} } \left( e^{isf(x)}
   - 1 - {isf(x) \over 1 + s^2} \right) dM(s) dx \right\}
\end{eqnarray*}
for all $f \in \SC (\R^d)$.
We call $F_a$, $F_{\s}$ and $F_M$ in order as degenerate (or
constant), Gaussian and Poisson (with jumps given by $M$) noises,
respectively. The first two terms $F_a$ and $F_{\s}$ can be
clearly understood. Let us discuss further the Poisson noise. 
Its characteristic functional is given by the following formula
\begin{eqnarray}
\label{1.8eqa}
& &  \int_{\SC ' (\R^d)} e^{i<f,\xi>} dP_M(\xi) \nonumber\\
& = & \exp \left\{ \int_{\R^d} \int_{\R \setminus \{ 0 \} }
      \left( e^{isf(x)} - 1 - {isf(x) \over 1 + s^2} \right)
      dM(s) dx \right\}
\end{eqnarray}
for $f \in \SC (\R^d)$. The existence and uniqueness of the Poisson 
noise measure $P_M$ is assured by Theorem \ref{eins.1theo}.

In what follows, we will give a representation of Poisson noise
in terms of a corresponding Poisson distribution. To do this, we 
first introduce some
notions. Let $\DC (\R^d)$ denote the Schwartz space of all (real--
valued) $C^\infty$--functions on $\R^d$ with compact support 
and $\DC'(\R^d)$ its topological dual space. Clearly, 
$\DC (\R^d) \subset \SC (\R^d)$. As was pointed 
out e.g. in \cite{GV} and \cite{I}, 
the Bochner Minlos theorem (and therefore our Theorem 1.1) also holds on 
$\DC (\R^d)$. Especially, (\ref{1.8eqa}) holds for $P_M$ on 
$\DC' (\R^d)$ and $f \in \DC (\R^d)$. Namely, there exists a unique
$P_M$ such that
\begin{eqnarray}
\label{1.9eqa}
  &   & \int_{\DC' (\R^d)} e^{i<f,\xi>} dP_M(\xi) \nonumber\\
  & = & \exp \left\{ \int_{\L (f)} \int_{\R \setminus \{ 0 \} }
        \left( e^{isf(x)} - 1 - {isf(x) \over 1 + s^2} \right)
        dM(s) dx \right\}  
\end{eqnarray}
for $f \in \DC (\R^d)$, where $\L (f) :=$ supp$f (\subset \R^d)$ is
the support of $f$. We assume henceforth that the 
first moment of $M$ exists. In
this case we can drop the third term in the exponential of the right
hand side of (\ref{1.9eqa}) and Theorem \ref{eins.1theo} assures that
there exists a unique measure $\tilde P_M$ such that for 
$f \in \DC (\R^d)$
\be 
\label{intdcr2.eq}
  \int_{\DC ' (\R^d)} e^{i<f,\xi>} d\tilde P_M(\xi) = \exp
  \left\{ \int_{\L (f)} \int_{\R \setminus \{ 0 \}}
  \left( e^{isf(x)} - 1\right) dM(s) dx \right\} \ .
\ee 

Set 
$\k_f = \int_{\L (f)} \int_{\R \setminus \{ 0 \}} dM(s) dx$, which is
a finite and strictly positive number. Then by
Taylor series expansion of the exponential and 
dominated convergence, we have from (\ref{intdcr2.eq}) that
\begin{eqnarray}
\label{intdcr3.eq}
  &   & \int_{\DC ' (\R^d)} e^{i<f,\xi>}d\tilde P_M(\xi) \nonumber\\
  & = & e^{-\k_f} \sum_{n \ge 0} {1 \over n!} \left[ \int_{\L (f)}
        \int_{\R \setminus \{ 0 \}} e^{isf(x)} dM(s) dx \right]^n
        \nonumber\\
  & = & e^{-\k_f} \sum_{n \ge 0} {1 \over n!} \int_{\L (f)}
        \int_{\R \setminus \{ 0 \}} \cdots \nonumber\\
  &   & \cdots  \int_{\L (f)} \int_{\R \setminus \{ 0 \}} 
        e^{i \sum^n_{j=1} s_j f(x_j)} \prod^n_{j=1} dM(s_j) dx_j \ .
\end{eqnarray}

Formula (12) might be interpreted as a representation 
of Poisson chaos and from it we get the following equivalent (in law) 
representation of Poisson noise
\be 
\label{fxi.eq}
  <f,\xi> = <f, \sum^{N_f}_{j=1} \l_j \d_{X_j} > \ , 
  \ \ f \in \DC (\R^d) \ , 
\ee 
where $\delta_x$ is the Dirac-distribution concentrated in $x \in \R$ and 
$N_f$ is a compound Poisson distribution (with intensity
$\k_f$) given as follows
\be 
\label{pr.eq}
  \Pr \{ N_f = n \} = {e^{-\k_f} (\k_f)^n \over n!} \ , 
  \ \ n = 0, 1, 2, \ldots
\ee 
where $\{ (X_j , \l_j )\}_{1 \le j \le N_f}$ is a family of 
~independent, identically distributed random variables
distributed according to the probability measure 
$\k^{-1}_f dx \times dM(s)$ on 
$\L (f) \times (\R \setminus \{ 0 \})$.

Concerning with Poisson noise $F_M$ on $\SC ' (\R^d)$ determined by
formula (9), we note that $\DC(\R^d)$ is dense in $\SC(\R^d)$
with respect to the topology of $\SC(\R^d)$. Therefore, by the 
continuity of the right hand side of (9), the chaos decomposition (12)
determines the law of the coordinate process $F_M$. 

\begin{Remark}
\label{1.4rem}
It is interesting to note, in the relation with the short discussion
given in the introduction, that one can recover Gaussian fields from 
Poisson ones in a limit. In fact, let $\{M_n\}_{n\in\N}$ be a certain
sequence of functions satisfying (\ref{intr0}) and $\{P^{n}_M\}_{n\in\N}$
be the sequence of Poisson noise measures determined by 
(\ref{1.9eqa}), then 
$$\int_{\DC'(\R^d)}e^{i<f,\xi>} dP^{n}_M(\xi)\lra
  e^{ia}\int_{\R^d}f(x)dx - {\s^2\over2}\int_{\R^d}[f(x)]^2dx$$
as $n\to\infty$,
which is the characteristic functional of a Gaussian law on $\DC'(\R^d)$
with mean
$${\bf E}[<f,\cdot>] = a \int_{\R^d}f(x)dx,\ f\in \DC (\R^d)$$
and covariance
$${\bf E}[<f,\cdot><g,\cdot>] = \s^2\int_{\R^d}f(x)g(x)dx, \ f,g\in \DC (\R^d)$$
iff
\begin{eqnarray*}
&& \int_{\R^d}\int_{\R\setminus\{0\}}\left( e^{isf(x)}-1-{isf(x) \over 1+s^2}
 \right) dM_n(s) dx \\
&& \lra ia\int_{\R^d}f(x)dx-{\s^2\over2}
  \int_{\R^d}[f(x)]^2dx
\end{eqnarray*}
as $n\to\infty$. An example is given by Poisson laws where 
the left hand side is 
$$\int_{\R^d}\int_{\R\setminus\{0\}}(e^{isf(x)}-1)dM_n(s)dx $$
which converges to ${\s^2\over2}\int_{\R^d}[f(x)]^2dx$(i.e. to the right 
hand side for $a=0$) if, e.g., $dM_n(s)=n^2\s^2\d_{1\over n}(s)ds$.  
\end{Remark}

\section{Euclidean Random Fields as Convoluted Generalized White 
Noise}
Let us first give the notion of random fields.
\begin{Definition}
\label{zwei.1def}
Let $(\Om , \EC , P)$ be a probability space. By a (generalized) random 
field $X$ on $(\Om , \EC , P)$ with parameter space $\SC (\R^d)$, we
mean a system ${\{ X(f,\om ), \om \in \Om \}}_{f \in \SC (\R^d)}$, of
random variables on $(\Om , \EC , P)$ having the following properties.

\begin{enumerate}
\item $P\{ \om \in \Om : X(c_1f_1 + c_2f_2 , \om) = c_1 X(f_1 ,
       \om) + c_2 X (f_2 , \om ) \} = 1$, $c_1 , c_2 \in \R$, 
       $f_1 , f_2 \in \SC (\R^d)$;
\item $f_n \to f$ in $\SC (\R^d)$ implies that 
      $X(f_n , \cdot) \to X(f,\cdot )$ in law.
\end{enumerate}
\end{Definition}

The coordinate process $F$ in Definition \ref{zwei.1def}
is a random field on the generalized white noise space 
$(\SC ' (\R^d) , \BC , P_\psi)$, which follows immediately from the
facts that the above property 1 is fulfilled pointwise and the property 2
is implied by the pointwise convergence $F(f_n,\om) \to F(f,\om)$ as 
$n \to \infty$ for all $\om \in \SC' (\R^d)$ and $f_n \to f$ in 
$\SC (\R^d)$ since the latter is slightly stronger than convergence in law.

Let $\GC :\SC (\R^d) \to \SC (\R^d)$ be a linear and continuous mapping. 
Then by the known Schwartz theorem, there exists a distribution 
$K \in \SC' (\R^{2d})$, hereafter called the kernel of $\GC$, such that  
\begin{equation}
\label{gcf.eqa}
(\GC f) (x) = \int_{\R^d} K (x,y) f(y) dy \ , \ \ f\in \SC (\R^d).
\end{equation}
It is clear that the conjugate operator 
$\tilde \GC : \SC ' (\R^d) \to \SC ' (\R^d)$ is a measurable
transformation from $(\SC ' (\R^d) , \BC)$ into itself. 

\begin{Example}
\label{zwei.1exa}
Let $\D$ be the Laplace operator on $\R^d$. Let $G_\a$ be the Green
function (i.e., fun\-da\-men\-tal solution) of the 
pseudo--dif\-fe\-ren\-ti\-al 
operator $(-\D + m^2_0)^\a$ for some arbitrary (but fixed) $m_0 >0$ and 
$0 < \a$. Take $K(x,y) = G_\a (x-y)$, $x,y \in \R^d$. Then
$\GC = (-\D + m^2_0)^{-\a}$ is a linear continuous mapping from 
$\SC (\R^d)$ to $\SC (\R^d)$. 

To see this, let $\FC$ and $\FC^{-1}$
denote the Fourier and inverse Fourier transforms, respectively.
Namely,
$$(\FC f) (y) = (2 \pi)^{-{d \over 2}} \int_{\R^d} e^{-ixy} f(x) dx
  $$
$$(\FC^{-1} f) (x) = (2 \pi)^{-{d \over 2}} \int_{\R^d} e^{ixy} f(y)
  dy$$
for $ f \in \SC (\R^d)$. Then we have
$$(\GC f) (x) = \int_{\R^d} G_\a (x-y) f(y) dy \ , \ \ f \in \SC
  (\R^d)$$
and
\begin{equation}
\label{zwei.2eqa}
(\FC G_\a ) (k) = {1  \over {(2 \pi )^{d \over 2}(\vert k \vert^2
+ m^2_0)^\a}} \ , \ \ k \in \R^d \ .
\end{equation}
Thus
\begin{eqnarray*}
     (\FC (\GC f)) (k) 
& = & (2\pi)^{-{d \over 2}} \int_{\R^d} e^{-ixk} \int_{\R^d}
      G_\a (x-y) f(y) dy dx \\
& = & (2\pi)^{-{d \over 2}} \int_{\R^d} e^{-iyk} G_\a (y) dy 
      \int_{\R^d} e^{-ixk} f(x)   dx \\ 
& = & (2\pi)^{{d \over 2}} (\FC G_\a ) (k) \cdot (\FC f) (k)  \\
& = & {1 \over {(\vert k \vert^2 + m^2_0)^\a}} \cdot
     (\FC f) (k) \ . \\
\end{eqnarray*}
Therefore
$$\GC f = (\FC^{-1} \left({1 \over (\vert k \vert^2 + m^2_0)^\a} 
  \right) \cdot \FC f) (k) \ , \ \ f \in \SC (\R^d) \ .$$
We notice that $(-\D +m^2_0)^{-\a}$ maps real test functions to real
test functions. Furthermore, by 
Theorem IX.4 of \cite{RS}, $\FC$ and $\FC^{-1}$ are linear continuous
from $\SC_{\C} (\R^d)$ to $\SC_{\C} (\R^d)$. Hence it suffices to verify 
that the multiplicative operator defined by
$f(\cdot) \to {1 \over (\vert k \vert^2 + m^2_0)^{\a}} \cdot f(\cdot)$ is 
linear and continuous from 
$\SC_{\C} (\R^d)$ to $\SC_{\C} (\R^d)$. The linearity is
obvious. The continuity is derived from the fact that 
$M_h f := hf$ defines a continuous multiplicative operator from
$\SC_{\C}(\R^d)$ to $\SC_{\C} (\R^d)$ if $h$ is $C^\infty$--differentiable and
$h$ itself with all its derivatives are of at most polynomial increasing.
This is true because in our case
$$h(k) = {1 \over (\vert k \vert^2 + m^2_0)^\a} \ .
  $$
\end{Example}

In section 1, we had already defined the generalized white noise
measure $P_\psi$ on $(\SC ' (\R^d) , \BC)$ associated with a L\' evy
characteristic $\psi$. Now let $P_K$ denote the image (probability)
measure of $P_\psi$ under $\tilde \GC$, i.e., $P_K$ is a measure on
$(\SC ' (\R^d) , \BC)$ defined by

\begin{equation}
\label{zwei.3eqa}
P_{K} (A) := P_\psi (\tilde \GC^{-1} A) \ , \ \ A \in \BC \ .
\end{equation}
Then we have the following result

\begin{Proposition}
\label{zwei.1prop}
The Fourier transform of $P_K$ is given by 
\begin{eqnarray}
\label{zwei.4eqa}
& &   \int_{\SC ' (\R^d)} e^{i<f,\xi>} dP_K (\xi) \nonumber\\
& = & \exp \left\{ \int_{\R^d}
   \psi \left( \int_{\R^d} K(x,y)f(y)dy\right) dx \right\} \ , 
   \ \ f \in \SC (\R^d) \ .
\end{eqnarray}
Conversely, given a linear and continuous mapping $\GC :\SC (\R^d)
\to \SC (\R^d)$ and a L\'evy characteristic $\psi$, there exists a 
unique probability measure $P_K$ such that (\ref{zwei.4eqa}) is valid. 
\end{Proposition}

{\bf Proof.}
For $f \in \SC (\R^d)$, by (\ref{zwei.3eqa}) and Theorem \ref{eins.1theo}, we
derive that
\beas
  \int_{\ssrd} e^{i<f,\xi >} dP_K (\xi)
& = & \int_{\ssrd} e^{i<f,\tilde \GC  \xi>} dP_\psi ( \xi ) \\
& = & \int_{\ssrd} e^{i< \GC f,\xi>} dP_\psi (\xi ) \\
& = & \exp \left\{ \int_{\R^d} \psi \left( \int_{\R^d} K(x,y)
     f(y) dy \right) dx \right\} \ .  
\eeas

The converse statement is derived ~analogously to the proof of Theorem
\ref{eins.1theo} since the operator $\GC$ is continuous from
$\srd$ to $\srd$ and thus the RHS of (\ref{zwei.4eqa}) defines a 
characteristic functional.
\kasten

Clearly, from Proposition \ref{zwei.1prop}, the associated coordinate
process $X:\srd\times ( \ssrd , \BC , P_K ) \to \R$ given by
$$X(f,\xi) = <f,\xi> \ , \ f \in \srd \ , \ \xi \in \ssrd
  $$
is a random field on $(\ssrd , \BC , P_K )$. In fact, $X$ is nothing
but $\tilde \GC F$ which is defined by
$$(\tilde \GC F) (f, \xi ) := F(\GC f ,\xi ) \ , \ f\in \srd \ , \ \xi \in
   \ssrd  .
  $$

\begin{Remark}
\label{zwei.1rem}
Concerning Example \ref{zwei.1exa}, the above construction does not
always work if $m_0 = 0$, because $(-\D)^{-\a}$ does not map $\SC (\R^d)$ 
onto $\SC (\R^d)$ for $\a >0$. In
this case, if $F$ is a Gaussian white noise, then we can obtain a
random field $X = (-\D)^{-\a} F$ for $\a < {d \over 4}$ by the following
argument. Take $\a < {d\over 4}$, then the scalar product

$$
       {(f_1 , f_2)}_\a 
 :=  \int_{\R^d}
       { \overline{(\FC f_1) (k)} \cdot (\FC f_2)(k) \over 
       \vert k \vert^{4\a}} dk \ , 
\ \ f_1 , f_2 \in \srd
$$
is continuous with respect to the (Schwartz) topology of $\srd$, where
$\FC$ denotes the Fourier transform as introduced previously. Thus 
$f \in \srd \mapsto \exp \{ - {\sigma ^2\over 2} \Vert f \Vert^2_\a \} \in [0 ,
\infty)$ is a characteristic functional on $\srd$. By Boch\-ner-Minlos
theorem, we get a unique measure $P_K$ satisfying (\ref{zwei.4eqa})
and hence the associated coordinate process is precisely  
$X= (-\D )^{-\a} F$. However, if one wants to follow a corresponding
procedure with a generalized white noise, one needs an explicit calculation
to show the continuity of the functional
$$        f \in \srd 
\to  \exp \Bigl\{ \int_{\R^d} \psi((-\D)^{-\a} f) (x))dx \Bigr\} \in \C \
,
$$
where $\psi$ is a L\'evy characteristic. Then, by Bochner--Minlos theorem, 
one can directly construct the measure $P_K$. 

In order to derive a suitable condition for the continuity of the above
functional, we note that the characteristic functional of generalized
white noise (\ref{intscr.eq}) extends continuously from $\srd $ to
$L^2(\R^d,dx)$, provided the corresponding generalized white noise $F$
has mean zero and finite moments of second order (see Prop. 4.3 of \cite{Be}).
Furthermore, by the above considerations for the Gaussian case and the
fact that $\FC $ is unitary on $L_{\C}^2(\R^d,dx)$, we get that $(-\D)^{-
\a}:\srd\to L^2(\R^d,dx)$ is continuous, if $0<\a<{d\over 4}$. Thus, we can
construct "mass zero" random fields $X=(-\D)^{-\a}F$ for 
$0<\a<{d\over 4}$ and $F$ as characterized above.

In what follows, we will always assume that the continuity 
of $\GC :\SC (\R^d) \to \SC (\R ^d)$
holds.  
\end{Remark}

Let us now turn to discuss the invariance of the random field $X$
under Euclidean transformations. We need to introduce some notations
at first. A proper Euclidean transformation of $\R^d$ is, by
definition, an element of the proper Euclidean group $E_0 (\R^d)$ over
$\R^d$. In fact, the proper Euclidean group $E_0 (\R^d)$ is generated
by
\begin{enumerate}
\item all translations $T_a : x\in\R^d \to T_a x := x-a \in \R^d$,
  $a \in \R^d$;
\item all rotations $R : x \in \R^d \to R x \in \R^d$.
\end{enumerate}

The (full) Euclidean group $E(\R^d)$ over $\R^d$ is generated by all
transformations in $E_0 (\R^d)$ and by all reflections. The group
$E(\R^d)$ is an inhomogeneous orthogonal group, that is, 
$E(\R^d)$ is the group of all nonsingular inhomogeneous linear 
transforms which preserve the Euclidean inner product. It is easy to
see that among the reflections it is enough to consider the ''time
reflection'' $\theta$, defined by writing $x \in \R^d$ as
$x:= (x_0 , \vec x)$, $x_0 \in \R$, $\vec x \in \R^{d-1}$, calling
$x_0$ ''time coordinate'' and setting
$$\theta x := (-x^0 , \vec x) \ \ , \ \ x = (x^0 , \vec x) \in 
  \R \times \R^{d-1} \ .
  $$
If $T$ is a transformation in the Euclidean group $E(\R^d)$, the
corresponding transformation on a test function $f \in \srd$ is
defined by 
$$(Tf) (x) := f(T^{-1} x) \ \ , \ x \in \R^d \ ;
  $$
and on $\ssrd$ is defined by duality as follows
$$<f,T\xi > := <T^{-1}f, \xi > \ \ , \ f\in \srd \ , \ \xi \in \ssrd \ .
  $$
The corresponding transformation on the random field $X$ is defined by
$$(TX) (f,\xi ) := X(T^{-1}f , \xi ) \ \ , \ f \in \srd \ , \ \xi \in \ssrd
  \ .
  $$
By the invariance of the dualization under Euclidean transformations
(which follows from the invariance of Lebesgue measure), we have
$$(TX) (Tf,\xi) = X(f,\xi ) \ \ , \ f \in \srd \ , \ \xi \in \ssrd \ .
  $$

Concerning Euclidean invariance of random fields, we have the
following
\begin{Definition}
\label{zwei.2def}
By Euclidean invariance of the random field $X$ we mean that the laws of
$X$ and $TX$ are the same, for each $T \in E(\R^d)$, i.e., the
probability distributions of $\{ X(f,\cdot) : f \in \srd \}$ and
$\{ (TX) (f,\cdot ):f\in \srd \}$ coincide for each $T \in E(\R^d)$. In
particular, if the laws of $X$ and $\theta X$, where $\theta$ is the ''time--
reflection'' defined above, are the same, we say that the random field
$X$ is (time--)reflection invariant.
\end{Definition}

From Bochner--Minlos Theorem, the probability distribution of 
$\{ X(f,\cdot) : f \in \srd\}$ is uniquely determined by the
characteristic functional $C_X (f)$, $f \in \srd$, and vice versa.
Thus, the property of Euclidean invariance of random fields is also
determined by means of characteristic functionals.

We say that $\GC$ is $T$--invariant, for some 
$T \in E(\R^d)$, if $\GC T = T \GC$. $\GC$ is called
Euclidean invariant if $\GC$ is invariant under all $T \in E_0 (\R^d)$.
In case $\GC$ is translation invariant its kernel $K$ has the form 
$K(x,y) = G(x-y)$(cf. p.39 of \cite{StWi}). If the kernel $G$ of $\GC$
is also invariant under orthogonal transformations, then
$\GC$ is invariant under all $T \in E(\R^d)$. In this case we also say 
for simplicity that $G$ is the Euclidean invariant kernel of $\GC$. The
action of $\GC$ on test function in $\srd$ (and by duality on $\ssrd$ as
well as on random fields) in the translation invariant case is by
convolution
$$(\GC f)(x) = \int_{\R^d} K(x,y) f(y) dy = \int_{\R^d} G(x-y) f(y) dy \
  \ x \in \R^d \ .
  $$
We then also write $\GC f$ as $G \ast f$.

\begin{Remark}
\label{zwei.2rem}
The kernel $G_\a$ determined by formula
(\ref{zwei.2eqa}) in Example \ref{zwei.1exa} is given by
$$G_\a (x) 
= {1 \over {(2\pi)^d}}\int_{\R^d} {e^{ikx} \over (\vert k \vert^2 + m^2)^\a}
  dk \ \ , \ x \in \R^d \ ,
  $$
where the integral has to be understood in the sense of a 
Fourier--transform of a tempered distribution. 
It is invariant under all orthogonal transformations. This can be
verified directly by changing integral variables in the above formula
since orthogonal transforms leave $\vert k \vert$ and $dk$
invariant.  
\end{Remark}

Moreover, we have following result

\begin{Proposition}
\label{zwei.2prop}
Assume that the mapping $\GC :\SC (\R^d) \to \SC (\R^d)$ is 
Euclidean--invariant, then the random field $X = \GC F$ is 
Euclidean--invariant.
\end{Proposition}

{\bf Proof.}
By Bochner-Minlos Theorem, it is sufficient to show that
$$C_X (f) = C_{TX} (f) \ \ , \ f \in \srd 
  $$
for every $T\in E(\R ^d)$.

In fact, we have
\beas
C_{TX} (f)
& = & E \left[ e^{iTX(f, \cdot)} \right]\\
& = & E \left[ e^{iX(T^{-1}f, \cdot)} \right]\\
& = & \exp \left\{ \int_{\R^d} \psi (\GC (T^{-1}f) (x)) dx \right\} \ \ ,
     \ f \in \srd \ , 
\eeas
where the last equality follows from by (\ref{zwei.4eqa}). So we need
only to verify that
$$\int_{\R^d} \psi ((\GC(Tf)) (x)) dx = \int_{\R^d} \psi
  ((\GC f)(x)) dx \ .
  $$
This holds by using the invariance of Lebesgue measure under the
transformation $x \to T^{-1}x$ and the fact that
$$(T^{-1}(\GC f)) (x) = (\GC (T^{-1}f)) (x) \ . 
  $$

\kasten

Hereafter, we only deal with Euclidean invariant kernels, the
derived random fields are then also Euclidean invariant. We call such
random fields Euclidean random fields. From Remark \ref{zwei.2rem},
the integral kernel $G_\a$ defined in Example \ref{zwei.1exa} is
Euclidean invariant. Moreover, since the translation invariance
implies that the integral kernels are of convolution type, the
corresponding Euclidean random fields are convoluted generalized white
noise. We simply denote the convoluted generalized white
noise $X$ by $X:=G_\a \ast F$.

\section{The Schwinger Functions of the Model and their Basic
Properties}
In 1973, E. Nelson \cite{N1} showed how to construct 
a relativistic quantum field theory (QFT) from
an Euclidean Markov field.
Inspired by this, in \cite{OS1} and \cite{OS2} Osterwalder and Schrader (see
also \cite{Gl}, \cite{Heg} , \cite{Zi}) gave a set of axioms, where Schwinger
functions $\{ {\sf S}_n \}_{n \in \N_0}$ defined on the Euclidean
space--time
$E_d$ can be analytically continued to Wightman distributions, i.e. to the
vacuum expectation values of a relativistic QFT.(Here and in the following 
we use the 
"sans-serif" ${\sf S}_n$ for Schwinger functions in general, whereas 
the Schwinger functions of our concrete model are denoted by "italic" $S_n$.) 

Apart from existence of
an analytic continuation, these axioms are $(E0)$ Temperedness, $(E1)$
Euclidean invariance, $(E2)$ Reflection positivity, $(E3)$ Symmetry
and $(E4)$ cluster property. In the case of Euclidean Markov fields and also in the
more general case of Euclidean reflection positive fields \cite{F}, Schwinger 
functions fulfilling $(E0)$--$(E4)$ are obtained
as the moments of the Euclidean field.

In this Section we will calculate the ''Schwinger functions'' $S_n$ of
$X$, which are by definition the moment functions of the convoluted generalized white 
noise $X$. We 
will verify $(E0)$,
$(E1)$ and $(E3)$. A proof of $(E4)$ is given in Section 4. In section 5 a
partial negative result on $(E2)$ is derived for the case of convoluted 
generalized white noises with a non--zero Poisson part.

We now fix a L\'evy characteristic $\psi$, such that the L\'evy
measure $M$ has moments of all orders. With $F$ we denote the
generalized white noise determined by $\psi$.

\begin{Lemma}
\label{drei.1lem}
Let $C^T_F$ denote the functional which maps 
$\vp \in \SC (\R^d)$ to \linebreak $\int_{\R^d}$ $\psi (\vp (x)) dx$.
Then partial derivatives of all orders of $C^T_F$ exist
everywhere on $\SC (\R^d)$. 
For $\vp_1 \ldots \vp_n \in \SC (\R^d)$ we get that
\begin{equation}
\label{drei.1eqa}
{1 \over i^n} {\p^n \over \p \vp_1 \ldots \p \vp_n}
  C^T_F \mid_0 = c_n \int_{\R^d} \vp_1 \ldots \vp_n dx \, .
\end{equation}
Here we introduced the constants $c_n$ defined as
\beas
\label{drei.1aeqa}
c_1 & := & a + \int_{\R \setminus \{ 0 \}} {s^3 \over 1 + s^2}
          dM(s)  \nonumber \\
c_2 & := & \s^2 + \int_{\R \setminus \{ 0 \}}  s^2
          dM(s)\nonumber \\
c_n & := & \int_{\R \setminus \{ 0 \}} s^n dM(s) \; , n\geq 3 \; .\\ 
\eeas
\end{Lemma}
As it will be explained shortly, throughout this paper the superscript
"$T$" stands for the operation of "truncation" (cf. \cite{Ha}).

{\bf Proof.}
Since the differentiability of $\vp \to ia \int_{\R^d} \vp dx$ and
$\vp \to \linebreak {\s^2 \over 2} \int_{\R^d}$ $\vp^2 dx$ is immediate, 
we only
have to deal with the Poisson part of $\psi$. We remark that
$\vert e^{iy} - 1\vert \le \vert y \vert$ for all $y \in \R$. Therefore
$${1 \over t_1} \vert e^{is [\vp (x) + t_1 \vp_1 (x)]} - 
  e^{is \vp (x)} \vert \leq \vert s \vp_1 (x) \vert \, ,  
$$
for all $t_1>0, s \in \R$.
This shows that the LHS is uniformly bounded (in $t_1$) by a function in
$L^1 (\R \setminus \{0\} \times \R^d , dM \otimes dx)$--function. 
~Analogously, for all $t_n>0$,
$${1 \over t_n} \vert s^{n-1} \vp_1 (x) \ldots \vp_{n-1} (x)
  (e^{is [\vp (x) + t_n \vp_n (x)]} - e^{is \vp (x)} )\vert
  \leq \vert s^n \vp_1 (x) \ldots \vp_n (x) \vert \ ,
  $$
and again the RHS is an uniform $L^1(\R \setminus \{0\} \times \R^d ,dM \otimes dx)$
--bound. Thus we may interchange
partial derivatives and integration by the dominated convergence theorem:
\beas
 & & {1 \over i^n} {\p^n \over \p \vp_1 \ldots \p \vp_n} \int_{\R^d}
    \int_{\R \setminus \{ 0 \}} \left( e^{is\vp} - 1 -
    {is \vp (x) \over 1 + s^2} \right) dM(s) dx \\
 & = & \int_{\R^d} \int_{\R \setminus \{ 0 \}} e^{is\vp}  
     \vp_1 (x) \ldots \vp_n (x) s^n dM(s) dx  
\eeas
if $n \ge 2$ and
$$\cdots = \int_{\R^d} \int_{\R \setminus \{ 0 \}} e^{is \vp(x)} 
  \vp_1 (x) {s^3 \over 1 + s^2} dM(s) dx
  $$
if $n=1$. Setting $\vp = 0$ and taking into account the linear and
Gaussian part, we get (\ref{drei.1eqa}).
\kasten

We have $C_F = \exp C^T_F$. Consequently $C_F$ has partial
 derivatives of any order, and it follows, that all moments of $F$ --
i.e. the expectation values of $<\vp_1 , F > \ldots <\vp_n ,F >$, exist
and are equal $i^{-n}$ times the $n$--th order partial derivative of $C_F$
w.r.t. $\vp_1 \ldots \vp_n$ at the point $\vp = 0$, cf. \cite{Lu}.

\begin{Definition}
\label{drei.1def}
Let $\vp_1 \ldots \vp_n \in \SC (\R^d)$. We define $M^F_n$, the 
$n$--th moment function of $F$, by

\begin{equation}
\label{drei.2eqa}
M^F_n (\vp_1 \otimes \ldots \otimes \vp_n) = \int_{\SC ' (\R^d)}
<\vp_1 , \om> \ldots < \vp_n , \om> dP_\psi (\om)
\end{equation}
By the above remark this equals

\begin{equation}
\label{drei.3eqa}
= {1 \over i^n} {\p^n \over \p \vp_1 \ldots \p \vp_n} C_F
\mid_0
\end{equation}
\end{Definition}

In order to calculate partial derivatives of $C_F$ of 
any order we need a generalized chain rule:

\begin{Lemma}
\label{drei.2lem}
Let $V$ be a vector space. Let $g : V\to \C$ be infinitely
often partial differentiable and let 
$f : \C \mapsto \C $ be analytic. Then for $v_1 \ldots v_n \in V$

\begin{equation}
\label{drei.4eqa}
{\p^n \over \p v_1 \ldots \p v_n} f \circ g = \sum^n_{k = 1}
f^{(k)} \circ g \sum_{I \in P^{(n)}_k} 
\prod_{\{ j_1 \ldots j_l\} \in I} 
{\p^l \over \p v_{j_1} \ldots \p v_{j_l}} g
\end{equation}
holds on $V$. Here $P^{(n)}_k$ is the collection of partitions of
$\{ 1 \ldots n\}$ into exactly $k$ disjoint subsets, 
$f^{(k)} (x) = \left( {d^k \over dx^k} f \right) (x)$.
\end{Lemma}

{\bf Proof.}
We proceed by induction over $n$. The statement for 
$n = 1$ is Leibniz' chain rule.
Observe that 
\beas
&   &  {\p^{n+1} \over \p v_1 \ldots \p v_n \p v_{n+1} } g \circ f \\
& = &  {\p \over \p v_{n+1} } \left( \sum^n_{k = 1}
       f^{(k)} \circ g \sum_{I \in P^{(n)}_k} 
       \prod_{\{ j_1 \ldots j_l \} \in I} {\p^l g \over
       \p v_1 \ldots \p v_l} \right) \\
& = & \sum^n_{k = 1} \biggl\{ f^{(k + 1)} \circ g 
      \sum_{I \in P^{(n)}_k} \prod_{\{ j_1 \ldots j_l \} \in I} 
      {\p^l g \over \p v_{j_1} \ldots \p v_{j_l}}
      {\p g \over \p v_{n+1}} \biggr. \\
&  &  \biggl. + f^{(k)} \circ g \sum_{I \in P^{(n)}_k} 
      \sum^k_{m=1} \prod^k_{ i=1 , i \ne m}
      {\p^l g \over \p v_{j^i_1} \ldots \p v_{j^i_l} }
      {\p^{l+1} g \over \p v_{j^m_1} \ldots \p v_{j^m_l} \p v_{n+1}}
      \biggr\} 
\eeas
We denote $I \in P^{(n)}_k$ by $I = \{ I_1, \ldots ,I_k\}$ and
set $I_i = \{ j^i_1 , \ldots j^i_l \}$ (where, of course, $l$ depends on
$i$). Using collections of
partitions $\PC ^{(n+1)}_k$, we can "reindex" the sums in the latter
expression and get:
\beas
& = & \sum^n_{k = 1} \biggl\{ f^{(k + 1)} \circ g 
     \sum_{I \in P^{(n+1)}_{k+1} , \{ n+1 \} \in I} 
     \prod_{\{ j_1 \ldots j_l \} \in I} 
     {\p^l g \over \p v_{j_1} \ldots \p v_{j_l}} \biggr. \\
&  & \biggl. + f^{(k)} \circ g \sum_{I \in P^{(n+1)}_{k} 
     ,\{ n+1 \} \not\in I} \prod_{\{ j_1 \ldots j_l \} \in I}
     {\p^l g \over \p v_{j_1} \ldots \p v_{j_l}} \biggr\} \\
& = & \sum^{n+1}_{k = 1} \biggl\{ f^{(k)} \circ g 
     \sum_{I \in P^{(n+1)}_{k} , \{ n+1 \} \in I} 
     \prod_{\{ j_1 \ldots j_l \} \in I} 
     {\p^l g \over \p v_{j_1} \ldots \p v_{j_l}} \biggr. \\
&  & \biggl. + f^{(k)} \circ g \sum_{ I \in P^{(n+1)}_{k} 
     \{ n+1 \} \not\in I} \prod_{\{ j_1 \ldots j_l \} \in I}
     {\p^l g \over \p v_{j_1} \ldots \p v_{j_l}} \biggr\}  
\eeas
On the RHS we have also reindexed $k + 1$ in the first sum in the braces to
$k$ with the sum ranging from 2 to $n+1$. Since the only
~partition in $P^{(n+1)}_1$, $\{ \{ 1,\ldots ,n+1\} \}$, does not contain 
$\{ n + 1\}$, we may sum
from 1 to $n+1$. Similarly in the second sum we may extend the sum
from $k=1 \ldots n$ to $k = 1 \ldots n+1$, since $P^{(n+1)}_{n+1}$
contains only $\{ \{ 1 \}, \ldots ,\{ n+1\} \}$ and thus gives no
contributions to this sum. But the last expression obtained obviously
equals (\ref{drei.4eqa}) with $n$ replaced by $(n+1)$.
\kasten

We remark that for the case of only one variable $v = v_1= \ldots =v_n$
we get Faa di Brunos expansion formula \cite{Jor}.

\begin{Corollary}
\label{drei.1cor}
(Cumulant formula) Let $V,g$ as in Lemma \ref{drei.2lem} and let $f$
be the exponential function. Furthermore assume $g(0) = 0$. Then
$f^{(k)} \circ g (0) = 1$ for all $k\in \N$. Let $P^{(n)}$ stand for
the collection of all partitions $I$ of $\{ 1 \ldots n \}$ into 
disjoint subsets. It follows that for $v_1,\ldots ,v_n \in V$ we get:
\begin{equation}
\label{drei.5eqa}
{\p^n \over \p v_1 \ldots \p v_n} \exp g \mid_0 = 
\sum_{I \in P^{(n)}} \prod_{\{ j_1 \ldots j_l\} \in I} 
{\p^l  \over \p v_{j_1} \ldots \p v_{j_l} } g \mid_0 \; .
\end{equation}
\end{Corollary}

\begin{Corollary}
\label{drei.2cor}
(Wick's Theorem) Let $f$ be the exponential function and $g$ a quadratic
function on a vector space $V$, i.e. $g(tv) = t^2 g(v)$ for all $v
\in V$. Then by Corollary \ref{drei.1cor}
\begin{equation}
\label{drei.7eqa}
{\p^n \over \p v_1 \ldots \p v_n} \exp g \mid_0 =
\cases{{\displaystyle \sum_{ I \in \hbox{pairings} }\prod_{\{ j_1 , j_2 \} \in I}
       {\p^2 g \over \p v_{j_1} \p v_{j_2}} \mid_0 } & for $n$ even \cr
       0 & for $n$ odd \cr}
\end{equation}
Here the pairings are those partitions $I$ of $\{ 1 \ldots n\}$,
where all subsets in $I$ do contain exactly two elements. 
This follows from the fact that all partial derivatives of
 $g$ are zero, except for partial derivatives of
order 2.
\end{Corollary}

\begin{Proposition}
\label{drei.1pro}
Set $\vp_1 \ldots \vp_n \in \SC (\R^d)$. Then
\begin{equation}
\label{drei.8eqa}
M^F_n (\vp_1 \otimes \ldots \otimes \vp_n) = \sum_{I \in P^{(n)}}
\prod_{\{ j_1 \ldots j_l \} \in I} c_l \int_{\R^d} 
\vp_{j_1} \ldots \vp_{j_l} dx
\end{equation}
\end{Proposition}

{\bf Proof.}
(\ref{drei.8eqa}) follows directly from (\ref{drei.3eqa}), 
$C_F = \exp C^T_F$, Lemma \ref{drei.1lem} and Corollary
\ref{drei.1cor}.
\kasten

\begin{Remark}
\label{drei.1rem}
Choosing $a$ in (\ref{drei.1aeqa}) such that $c_1 = 0$ implies that $F$
has mean zero. If furthermore $M$ is a symmetric measure w.r.t.
reflections at zero, then all $c_n$ vanish for $n$ odd. For such $n$ 
also $M^F_n$ is zero , since in this
case at least one $c_l$ with $l$ odd
appears in every summand on the RHS of (\ref{drei.8eqa}). 
\end{Remark}

We now fix, as in section 2, a linear continuous map
$\GC : \SC (\R^d) \to \SC (\R^d)$ and denote its dual by $\tilde
\GC$. Furthermore, we assume that $\GC $ is Euclidean invariant. Then 
there exists a convolution kernel $G$ which is invariant under orthogonal 
transformations, such that $\GC \varphi = G * \varphi$ for all 
$\varphi \in \SC (\R^d)$ (cf. Section 2). 
Define, as explained in section 2, 
$P_G = P_\psi \circ (\tilde \GC)^{-1}$ and let $X$ be the coordinate
process w.r.t. $P_G$.

\begin{Definition}
\label{drei.2def}
(''Schwinger functions of $X$'') Set 
$\vp_1 \ldots \vp_n \in \SC (\R^d)$. We define the $n$-th Schwinger function
$S_n$ as the $n$--th moment of $X$, i.e.
\begin{equation}
\label{drei.9eqa}
S_n (\vp_1 \otimes \ldots \otimes \vp_n) = \int_{\SC' (\R^d)}
<\vp_1 , \om > \ldots < \vp_n , \om > dP_G (\om) \ \
n \in \N_0
\end{equation}
\end{Definition}

\begin{Proposition}
\label{drei.2pro}
The Schwinger functions $S_n$ defined above are symmetric and Euclidean invariant
tempered distributions, i.e. $S_n \in \SC ' (\R^{dn})$ for $n\geq 1$. Furthermore for
$\vp_1 \ldots \vp_n \in \SC (\R^d)$ we have
\begin{equation}
\label{drei.10eqa}
S_n (\vp_1 \otimes \ldots \otimes \vp_n ) = \sum_{I \in P^{(n)}}
\prod_{\{ j_1 \ldots j_l \} \in I} c_l \int_{\R^d} G* \vp_{j_1} \ldots 
G* \vp_{j_l} dx \; .
\end{equation}
\end{Proposition}

{\bf Proof.}
The symmetry follows directly from Definition \ref{drei.2def}. Euclidean
invariance of the moment functions follows from the Euclidean
invariance in law of the random field $X$. 

Now we first prove
(\ref{drei.10eqa}). By the transformation formula, the RHS of 
(\ref{drei.9eqa}) is equal to
\begin{eqnarray*}
  &  & \int_{\SC ' (\R^d)} < \vp_1 , \tilde \GC \om > \ldots
       < \vp_n , \tilde \GC \om > dP_\psi (\om) \\
  & = & \int_{\SC ' (\R^d)} < G* \vp_1 ,  \om > \ldots
        < G* \vp_n ,\om > dP_\psi (\om) 
\end{eqnarray*}
This together with Proposition \ref{drei.1pro} now implies 
(\ref{drei.10eqa}). 

Concerning temperedness we remark that by (\ref{drei.10eqa}) $S_n$ is 
a sum of tensor products of linear functionals, say $S^T_l$, which map 
$\vp_1 \otimes \ldots \otimes \vp_l$ into 
$c_l \int_{\R^d} G* \vp_1 \ldots G* \vp_l dx$. Fix
$\vp_1 \ldots \vp_{j-1} \vp_{j+1} \ldots \vp_l \in \SC (\R^d)$. Then
$\vp_j \mapsto G* \vp_i \mapsto G* \vp_j \prod_{m=1,m
\ne j} G* \vp_m \mapsto c_l \int_{\R^d} G* \vp_1 \ldots G* \vp_l
dx$ is a map composed of $\SC (\R^d)$--continuous mappings and
therefore is continuous in $\vp_j$ alone, provided the other
$\vp$'s are fixed. A use of Schwartz nuclear theorem yields the
temperedness of the $S^T_l$ and a second application of the nuclear
theorem then implies $S_n \in \SC '(\R^d )$.
\kasten

\section{Truncated Schwinger Functions and the Cluster Property}     
\label{vier}
In this section, let $\{ \BS _n \}_{n_\in \N_0}$ be a sequence of distributions, 
where $\BS_0 =1$ and $\BS_n \in \SC ' (\R^{dn})$ for $n\geq 1$. We define truncated 
distributions $\BS^T_n$
$n \ge 1$ in the sense of Haag \cite{Ha}. The ${\{ \BS_n \}}_{n \in \N_0}$
determine the corresponding sequence of truncated distributions uniquely
and vice versa, and we can translate many properties of one sequence
into properties of the other. This is quite obvious e.g. for $(E0)$,
$(E1)$ and $(E3)$.

Making use of arguments in the classical papers on the so--called
asymptotic condition in axiomatic QFT (\cite{Ha},\cite{Ar1}, \cite{Ar2}, 
\cite{HeJo}) we obtain the equivalence of the cluster property $(E4)$ of
translation invariant $\{ \BS _n\}_{n\in \N_0}$ and the cluster property
of the truncated Schwinger functions $\{\BS_n^T\}_{n\in \N}$.

As an immediate consequence of formula (\ref{drei.10eqa}) we have explicit
formulae for the
truncated Schwinger functions of our model and we can easily check
their ''truncated'' cluster property in order to verify $(E4)$.

\begin{Definition}
\label{vier.1def}
Let ${\{ \BS_n \}}_{n \in \N_0}$ be a sequence of distributions 
with $\BS_0 =1$ and $\BS_n \in \SC ' (\R^{dn})$ for $n\geq 1$. Let
$\vp_1 \ldots \vp_n \in \SC (\R^d)$.
By the relation

\begin{equation}
\label{vier.1eqa}
\BS_n (\vp_1 \otimes \ldots \otimes \vp_n) = \sum_{I \in P^{(n)}}
\prod_{\{ j_1 \ldots j_l\} \in I} \BS^T_l (\vp_{j_1} \otimes
\ldots \otimes \vp_{j_l}) \ \ n \ge 1
\end{equation}

we recursively define the $n$--th truncated distribution $\BS^T_n$.
Here, for \linebreak $\{ j_1,\ldots,j_l\} \in I$ we assume 
$j_1<j_2<\cdots <j_l$.
\end{Definition}

\begin{Remark}
\label{vier.1rem}

\begin{enumerate}
\item By the Schwartz nuclear theorem the sequence 
${\{ \BS_n \}}_{n \in \N_0}$ u\-ni\-que\-ly determines the sequence
${\{ \BS_n^T \}}_{n \in \N}$ and vice versa.

\item All $\BS^T_n$ are Euclidean (translation) invariant if and only
if all $\BS_n$ are euclidean (translation) invariant. The same
equivalence holds for temperedness and symmetry (see e.g. \cite{Ar1}
for the symmetry).
\end{enumerate}
\end{Remark}

From now on we assume at least translation invariance for
$\{ \BS_n \}_{n\in \N_0}$, $\{ \BS^T_n \}_{n\in \N}$ respectively. And we will call 
these distributions
(truncated) Schwinger functions, even through at this level they
might have little to do with QFT.

\begin{Definition}
\label{vier.2def}
Let $a \in \R^d$,$a \ne 0$ $\l \in \R$. Let $T_{\l a}$ denote the
representation of the translation by $\l a$ on $\SC (\R^{dn})$, 
$n \in \N$. Take $m,n \ge 1$, $\vp_1 \ldots \vp_{n+m} \in \SC (\R^d)$

\begin{enumerate}
\item {\bf cluster property $(E4)$} A sequence of Schwinger functions 
${\{ \BS_n \}}_{n \in \N_0}$ has the cluster property if for all $n,m
\ge 1$

\begin{eqnarray}
\label{vier.2eqa}
& & \lim_{\l \to \infty} \Biggl\{ \BS_{m+n} (\vp_1 \otimes \ldots \vp_m
    \otimes T_{\l a} ( \vp_{m+1} \otimes \ldots \otimes \vp_{m+n}))
    \Biggr. \nonumber\\
& - & \Biggl. \BS_m (\vp_1 \otimes \ldots \vp_m )\BS_{n}
      ( \vp_{m+1} \otimes \ldots \otimes \vp_{m+n}) \Biggr\} =  0 \; .
\end{eqnarray}

\item {\bf cluster property of truncated Schwinger functions $(E4T)$}
A sequence of truncated Schwinger functions ${\{ \BS_n^T \}}_{n \in \N}$
has the cluster property of truncated Schwinger functions, if for all
$n,m \ge 1$

\begin{equation}
\label{vier.3eqa}
\lim_{\l \to \infty}   \BS^T_{m+n} (\vp_1 \otimes \ldots \otimes\vp_m
\otimes T_{\l a} ( \vp_{m+1} \otimes \ldots \otimes \vp_{m+n}))  = 0
\end{equation}
\end{enumerate}
\end{Definition}

\begin{Remark}
\label{vier.2rem}
We could also replace $\lim_{\l \to \infty}(\cdot)$in (\ref{vier.2eqa}) and 
(\ref{vier.3eqa}) by \linebreak $\lim_{\l \to \infty} \vert \l \vert ^N(\cdot)$  
for $N$ arbitrary. Such conditions should be seen as  cluster--
properties for short--range interactions. Indeed they would exclude 
the mass zero cases. 
Take e.g. the free Markov field of mass zero 
$X=(-\D)^{- \half} F$ in $d$ dimensions, $d \ge 3$, where $F$ is a 
Gaussian white
noise. Then (\ref{vier.2eqa}) and (\ref{vier.3eqa}) tend to zero only as
$\l^{-d + 2}$, as $\l \to \infty$. Nevertheless, by the same ~proofs as in 
Theorem \ref{vier.1theo} 
and Corollary \ref{vier.1cor} below we can also show that the Schwinger 
functions 
of our model have the ''short-range'' cluster
property. This already indicates the existence of a "mass-gap" in the
corresponding relativistic theory, obtained by the analytic
continuation of the (truncated) Schwinger functions in Sections 6 and 7.
\end{Remark}

The following result is the Euclidean analogue to one proven 
by Araki \cite{Ar1}, \cite{Ar2} for the
case of (truncated) Wightman functions.

\begin{Theorem}
\label{vier.1theo}
Let $\{ \BS _n\}_{n\in\N}$, $\{ \BS^T_n \}_{n\in\N}$ 
be as in Definition \ref{vier.1def}
and let the $\{ \BS_n \}_{n\in \N_0}$, $\{ \BS^T_n \}_{n\in\N }$ be 
translation
invariant. Then $\{ \BS_n \}_{n\in \N_0}$ has the cluster property, if 
and only if 
$\{ \BS^T_n \}_{n\in\N} $ has the cluster property of the truncated Schwinger
functions.
\end{Theorem}

{\bf Proof $(E4) \Rightarrow (E4T)$}
Assume there exist $m,n \ge 1$, such that $\vp_1 \ldots \vp_{n+m}\linebreak
 \in \SC (\R^d)$
and a $a \in \R^d \setminus \{ 0 \}$, such that the limit in 
(\ref{vier.1eqa})
is not zero or does not exist. Let furthermore $n+m$ be 
minimal w.r.t. this property.

Define $\vp^\l_k = \vp_k$ for $k = 1 \ldots m$ and 
$=T_{\l a} \vp_k$ for $k = m+1 \ldots m+n$.
By translation invariance the LHS of (\ref{vier.2eqa}) is then equal to

$$\lim_{\l \to \infty} \sum_{I \in P^{(m+n)}_{m,n}} 
  \prod_{\{ j_1 \ldots j_l \} \in I} \BS^T_l (\vp^\l_{j_1} \otimes
  \ldots \otimes \vp^\l_{j_l})
  $$

The symbol $P^{(m+n)}_{m,n}$ stands for all partitions $I$ of 
$\{ 1 \ldots n+m\}$ into disjoint subsets, such that in each partition $I$
there is at least
one subset $\{ j_1,\ldots ,j_l\} \in I$ such that $\{ j_1,\ldots ,j_l\} \cap 
\{ 1 \ldots m\} \ne \emptyset$, 
$\{ j_1,\ldots ,j_l\} \cap \{ m+1 \ldots m+n\} \ne \emptyset$. By the
assumption $(E4)$, the above expression equals zero.

Furthermore, every summand
except for the one indexed by $I = \{ \{ 1 \ldots n+m \} \}$ tends to
zero, since in each such summand at least one truncated Schwinger
function in the
product is evaluated on some $\vp^\l_k$'s $1 \le k \le m$ and
some $\vp^\l_k$'s $m+1 \le k \le m+n$, at the same time. By the minimality
of $n+m$ this factor tends to zero as $\l \to \infty$. The other
factors in the product either tend to zero (by minimality of $n+m$)
as $\l \to \infty$ or are constant, either by the definition of the 
$\vp ^{\l}_k$
for $k=1,\ldots ,m$ or by the translation invariance of the ${\sf S}^T_l$.

~Consequently also the summand indexed by $I=\{ \{1,\ldots ,n+m\} \}$
has to converge to zero as $\l \to \infty$, which is in contradiction 
with the above assumptions on $n,m$.

{\bf $(E4T) \Rightarrow (E4)$}
Fix $\vp_1 \ldots \vp_{n+m}$ and define $\vp^\l_k$ as above. Then

\beas
&   & \lim_{\l \to \infty }\BS_{m+n} (\vp^\l_1 \otimes \ldots \otimes
      \vp^\l_{m+n})\\
& = & \lim_{\l \to \infty} \Biggl\{ \sum_{I \in (P^{(m+n)}_{m,n})^c}
      \prod_{\{ j_1 \ldots j_l \} \in I} \BS^T_l (\vp^\l_{j_1}
      \otimes \ldots \otimes \vp_{j_l}^\l) \Biggr. \\ 
& &  + \Biggl. \sum_{I \in  P^{(m+n)}_{m,n} }
      \prod_{\{ j_1 \ldots j_l \} \in I} \BS^T_l (\vp^\l_{j_1}
      \otimes \ldots \otimes \vp_{j_l}^\l) \Biggr\} \; ,
\eeas
where $(P^{(m+n)}_{m,n})^c := P^{(m+n)} \setminus P^{(m+n)}_{m,n}$. In
the second term all products contain at least one factor that tends to
zero as $\l \to \infty$ by $(E4T)$. The other factors are constant
(either by the definition of the $\vp^\l_k$ or by the translation 
invariance of the ${\sf S}^T_l$) or tend to zero by $(E4T)$, again. Thus, 
the whole second sum vanishes for $\l \to \infty$.

In the first term all
$\BS^T_l$ are evaluated on $\vp^\l_{j_1} \ldots \vp^\l_{j_l}$ such that
either $\{ j_1 \ldots j_l \}  \subset \{ 1 \ldots m\}$ or
$\subset \{ m+1 \ldots m+n\}$. It follows from the definition of the 
$\vp^\l_k$ or the translation 
invariance of the ${\sf S}_l^T$, that all factors do not depend on $\l$.
In the first term, we may omit the $\l$'s therefore. 
The first term by this argument equals

\beas
&  & \left( \sum_{I \in P^{(m)}} \prod_{\{ j_1 \ldots j_l\} \in I}
      \BS^T_l (\vp_{j_1} \otimes \ldots \otimes \vp_{j_l} )\right) \\
&\times  & \left( \sum_{I \in P^{(n)}} \prod_{\{ j_1 \ldots j_l\} \in I}
      \BS^T_l (\vp_{j_1 + m} \otimes \ldots \otimes \vp_{j_l + m}) 
      \right) \\
& = & \BS_m (\vp_1 \otimes \ldots \otimes \vp_m) \BS_n (\vp_{m+1}
      \otimes \ldots \otimes \vp_{m+n}) \; ,
\eeas

from which we get (\ref{vier.2eqa}).
\kasten

\begin{Remark}
\label{vier.3rem}
Let ${\{ S_n \}}_{n \in \N_0}$ be defined according to Definition
\ref{drei.2def}. Let ${\{ S^T_n \}}_{n \in \N}$ denote the sequence 
of truncated Schwinger functions. Then, of course, by comparison
of (\ref{vier.1eqa}) and (\ref{drei.10eqa}) we get 

\begin{equation}
\label{vier.4eqa}
S^T_n (\vp_1 \otimes \ldots \otimes \vp_n) = c_n \int_{\R^d}
G* \vp_1 \ldots G* \vp_n dx
\end{equation}

for $\vp_1 \ldots \vp_n \in \SC (\R^d)$.
\end{Remark}

\begin{Corollary}
\label{vier.1cor}
Let ${\{ S_n \}}_{n \in \N_0}$ be as in Definition \ref{drei.2def}. Then 
${\{ S_n \}}_{n \in \N_0}$ has the cluster property $(E4)$.
\end{Corollary}

{\bf Proof.}
By Theorem \ref{vier.1theo} it suffices to show $(E4T)$ for 
$\{ S^T_n \}_{n \in \N} $. Fix $a \ne 0$ in $\R^d$ and let $\l \in \R^d$, 
$\vp_1 \ldots \vp_{n+m} \in \SC (\R^d)$, $m,n \ge 1$. We have

\beas
&  & \lim_{\l \to \infty} S^T_{m+n} (\vp_1 \otimes \ldots \otimes
      \vp_m T_{\l a} (\vp_{m+1} \otimes \ldots \otimes \vp_{m+n})) \\
& = & \lim_{\l \to \infty} c_{m+n} \int_{\R^d} G* \vp_1 \ldots G* \vp_m
      G* T_{\l a} \vp_{m+1} \ldots G* T_{\l a} \vp_{m+n} dx \\
& = & \lim_{\l \to \infty} c_{m+n} \int_{\R^d} G* \vp_1 \ldots G* \vp_m
       T_{\l a} (G* \vp_{m+1} \ldots G*  \vp_{m+n} )dx  
\eeas

On the RHS we shift away a fast falling function 
$G* \vp_{m+1} \ldots G*  \vp_{m+n}$ from the fixed fast falling
function $G* \vp_{ 1} \ldots G*  \vp_{m }$. The RHS therefore
approaches zero faster than any negative power of $\l$ falls to zero.
\kasten

\section{On Reflection Positivity}     
\label{fuenf}
In Section 4 we saw that many properties of Schwinger functions can be
directly translated into related properties of truncated
Schwinger functions. How about reflection positivity then?

Let us first discuss a simple related case. If $C_\mu$ denotes the
Fourier--transform of a probability measure $\mu$ on the real line,
and $C_\mu$ is analytic in a ~neighborhood of zero, then the derivatives of 
$C_\mu$ fulfill the positivity condition

\begin{equation}
\label{fuenf.1eqa}
\left( {a_n \over i^n} {d^n \over dx^n} + \ldots +
{a_1 \over i} {d  \over dx } +a_0\right)^2 C_\mu \mid_0 \ge 0
\end{equation}
for all $a_0 \ldots a_n \in \R$. This is related to $(E2)$.
Now suppose, that $C_\mu = \exp (C^T_\mu)$ and consider 
$C^\l_\mu = \exp (\l C^T_\mu)$. Sch\"onberg´s theorem (see
\cite{BeFo}) says that $C^\l_\mu$ is positive definite for all 
$\l \in \R _+$ if and only if $C^T_\mu$ is a L\'evy characteristic 
cf. section 1). One can
show, that the derivatives of $C^T_\mu$ at zero (the cumulants or ''truncated
moments of $\mu$'') fulfill the same positivity condition
(\ref{fuenf.1eqa}) if we set $a_0 = 0$, since a L\'evy characteristic
can be approximated by a sequence $b_n - C_n (0) + C_n$, 
where for $n\in \N$ $C_n$ is a positive definite function and $b_n \ge 0$ 
(cf. \cite{BeFo}). On
the other hand, if we can disprove one of the latter positivity
conditions, we will, as a consequence of Sch\"onberg´s theorem, find
some $\l \in \R_+$ such that $C^\l_\mu$ is not positive definite and
therefore (\ref{fuenf.1eqa}) does not hold for some $\l$, $n$, 
$a_l \ \ l = 0, \ldots , n$.

Along these lines we now construct some counter examples of 
convoluted generalized white noises $X=G* F$ with nonzero Poisson part, 
which do not have
the property of reflection positivity. It is interesting that such 
$X$ do exist even in such cases where the corresponding convoluted 
Gaussian white 
noise is reflection positive. Roughly speaking, the Schwinger 
functions $\{S_n\}_{n\in \N_0}$ belonging to $X$ 
do not have the 
property of reflection positivity, if the terms in the $S_n$ emerging
from the "interaction" are large in comparison with the "free" terms. 
It remains an open question, whether reflection positivity holds or 
does not hold in other cases.

Let us start with some definitions borrowed from the theory of
infinitely divisible random distributions.

\begin{Definition}
\label{fuenf.1def}
Let ${\{ \BS_n \}}_{n \in \N_0}$, $\BS_n \in \SC ' (\R^{d_n})$
$n \ge 1$, $\BS_0 = 1$ be a sequence of distributions and 
${\{ \BS^T_n \}}_{n \in \N}$ the corresponding truncated sequence. By $\R^d_+$ we
denote the set of $x \in \R^d x = (x^0 , \vec{x} )\in \R \times \R ^{d-1}$ $x^0 > 0$. 
Let $\SC ((\R^d_+)^n)$ be the Schwartz--functions on $\R^{dn}$ with
support in $(\R^d_+)^n$. $\theta$ is the ''time--reflection'', i.e. 
$\theta (x^0 , \vec{x} ) = (-x^0 , \vec{x} )$.

\begin{enumerate}
\item We call ${\{ \BS_n \}}_{n \in \N }$ reflection
infinitely divisible if for all $\l \in \R_+$ the sequence of Schwinger
functions ${\{ \BS^\l_n \}}_{n \in \N_0}$ determined by the truncated sequence 
$\{ \l \BS^T_n \} _{n\in \N}$ is reflection positive.
\item We say, a sequence of truncated Schwinger functions 
$\{\BS _n^T\}_{n\in \N}$
is conditional reflection
positive, if for test functions 
$\vp_k \in \SC ((\R^d_+)^k )$, $k = 1, \ldots , n$ the inequality 
\begin{equation}
\label{fuenf.2eqa}
\sum^n_{k , l = 1} \BS^T_{k + l} (\theta \vp_k \otimes
\vp_l ) \ge 0
\end{equation}
holds.
\end{enumerate}
\end{Definition}

Observe that the sum in (\ref{fuenf.2eqa}) is over 
$k,l = 1 \ldots n$, which distinguishes conditional reflection
positivity from reflection positivity $(E2)$, where the sum is over
$k , l = 0 \ldots n$ and $\vp_0 \in \R$.

\begin{Remark}
\label{fuenf.1rem}
The positivity--condition introduced here is a little more strict than
the original positivity--condition in
\cite{OS1}. Here we demand ''positivity'' of the Schwinger functions 
${\{ \BS_n \}}_{n \in \N_0}$ for test--functions 
$\vp_k \in \SC ((\R^d_+)^k)$ $k = 1 \ldots n$ $\vp_0 \in \R$ instead
of $\vp_k \in \SC_{+<} ((\R^d)^k)$ $k = 1 \ldots n$, which is the
original condition from \cite{OS1}. Here $\SC_{+<} ((\R^d)^k)$ is the space
of all test functions with support in 
$(\R^d_+)_< = \{ (x_1 \ldots x_k) \in \R^{dk} x_j = (x^0_j , \vec x_j)
\ j = 1 \ldots 2 \ 0 < x^0_1 < \ldots < x^0_k \}$.
Since the latter space is contained in the former, our condition is
more strict.

But in the present case, they are equivalent: The reason is that for a
large class of convolution kernels $G$ the Schwinger functions $S_n$ belonging to
$X=G* F$ (cf. section 3), are more regular than tempered distributions namely they are
locally integrable functions (cf. Lemma \ref{7.1lem}). One can therefore
evaluate such $S_n$ on test functions with ''jumps''. Thus, by application of ~symmetry,
we may calculate for $\vp_k \in \SC ((\R^d_+)^k)$, $k = 1 \ldots n$, 
$\vp_0 \in \R$:

\beas
&   & \sum^n_{k,j=0} S_{k+j} (\theta \vp_k \otimes \vp_k) \\
& = & \sum^n_{k,j=0} \sum_{ \pi \in \hbox{ Perm}(k)}
      \sum_{\pi' \in \hbox{ Perm} (j)} S_{k+j} (\theta
      (1_{\{ x^0_{\pi (1)} < \ldots < x^0_{\pi (k)}\} } \vp_k ) \\
&   & \otimes (1_{\{ x^0_{\pi ' (1)} < \ldots < x^0_{\pi ' (j)}\}} 
       \vp_j ) ) \\
& = & \sum_{k,j=0}^n S_{k+j}(\theta \tilde{\vp}_k \otimes \tilde{\vp}_j) \,
,
\eeas
where $\hbox{ Perm} (k)$ is the group of permutations of $\{1,\ldots
,k\} $ and $\tilde{\vp } _k$ is defined as
$$ \tilde{\vp }_k(x_1,\ldots ,x_k) := 1_{\{ x_1^0<\ldots <x_k^0\}}
(x_1,\ldots ,x_k)\sum _{\pi \in \hbox{ Perm} (k)} \vp (x_{\pi ^{-1}(1)},
\ldots x_{\pi ^{-1}(k)}). $$
Now the functions $\tilde{\vp }_k$ can be approximated by functions
from $\SC_{+<} (\R^{dk})$ and by application of the dominated
convergence theorem we can derive the sharpened reflection positivity 
condition (E4) from the original one of \cite{OS1}.
\end{Remark}

\begin{Lemma}
\label{fuenf.1lem}
Let $\{ \BS_n \}_{n\in \N_0}$ and $\{ \BS^T_n \}_{n\in \N}$ be as in Definition
\ref{fuenf.1def}. If \linebreak $\{ \BS_n \}_{n\in \N_0}$ is reflection 
infinitely divisible, then
$\{ \BS^T_n \}_{n\in \N }$ is conditional reflection positive.

More precisely: If $\{ \BS _n^T\} _{n\in \N} $ is not conditionally
reflection positive, then there exists a $\l _0>0 $, such that for all
$\l $, $ 0<\l <\l_0$, the sequence of Schwinger functions $\{ \BS _n^\l\}
_{n\in \N_0} $ is not reflection positive.
\end{Lemma}

{\bf Proof.}
Suppose, $\{ \BS _n^T \} _{n\in \N} $ is not conditionally reflection
positive. Then there are test functions $\vp_1 \ldots \vp_n$ as in definition
\ref{fuenf.1def} such that the LHS of (\ref{fuenf.2eqa} ) is negative.
Since 

$$ \lim_{ \l \to + 0}
  \left[ {1 \over \l} \sum^n_{k,l = 1} \BS^\l_{k + l}
  (\theta \vp_k \otimes \vp_l) \right] = \sum^n_{k , l =1}
  \BS^T_{k + l} (\theta \vp_k \otimes \vp_l) <0
  $$
there exists a $\l_0 >0$ such that for all $\l$ , $0<\l <\l_0$ , also 
the LHS of the above equation is negative. This implies the statement of
the lemma.
\kasten

\begin{Remark}
\label{fuenf.2rem}
Provided a quite weak growth condition in $n$ is fulfilled by the 
$\{ \BS^T_n \}_{n\in \N}$ and the $\BS_n$, $\BS^T_n$ are symmetric,
 we also have:
If $\{ \BS^T_n \}_{n\in \N}$ is conditional reflection positive then 
$\{ \BS_n \}_{n\in\N_0}$ is reflection infinitely divisible. 
(see \cite{Go}).
\end{Remark}

From now on we want to impose some restrictions on the kernel $G$.
These restrictions are typical for the Green's functions of a large
class of (pseudo) differential operators.

\begin{Condition}
\label{fuenf.1con}
We want to consider kernels $G$ that are continuous, real functions
on $\R^d \setminus \{ 0 \}$, and have a singularity
at the origin, i.e. $\lim_{\vert x \vert \to 0} G(x) \linebreak 
= \pm \infty$ and 
fall to zero as $|x| \to \infty $.
Furthermore assume that the mapping $f \mapsto G* f$ is well
defined and continuous from $\SC (\R^d)$ to $\SC (\R^d)$.
Finally, $G$ is assumed to be invariant under orthogonal transformations.
\end{Condition}

\begin{Remark}
\label{fuenf.3rem}
Of course, the Green's functions of the pseudo differential operators
$(-\D + m^2)^\a$, $\a \in (0,1)$, fulfill Condition \ref{fuenf.1con} 
(see e.g. the representation we give in Section 6).
\end{Remark}

\begin{Definition}
\label{fuenf.2def}
Let $\vp \in \SC (\R^d_+)$. For $x \in \R^d$ we write $(x^0 , \vec x)$.
Define $\vp^s (x) = \half \vp (x)$ for $x^0 > 0$ and $\vp^s(x)= \half \vp
(\theta x)$ for $x^0 < 0$. Let $\vp^a =$ sign$(x^0) \vp^s$. By $q(\vp)$ we 
denote the function
\begin{equation}
\label{fuenf.3eqa}
q(\vp) = G* \theta \varphi \, G* \varphi = (G* \vp ^s )^2 - (G*
\vp ^a )^2 \ .
\end{equation}
\end{Definition}

Clearly $\vp ^s,\vp ^a$ and $q(\vp )$ are fast falling functions. For
$\vp_1, \ldots, \vp _n \in \SC (\R_+^d)$ we get by the $\theta $-invariance of
$G$ and the definition of
$q$:
$$
S^T_{2n} (\theta (\vp_1 \otimes \ldots \otimes \vp_n) \otimes \vp_1
\ldots \otimes \vp_n) = 2c_{2n} \int_{\R^d_+} q(\vp_1) \ldots q(\vp_n) dx \, .
$$
We concentrate on
\begin{equation}
\label{fuenf.4eqa}
S^T_{4n +2} (\theta (\vp_1 ^{\otimes 2n}\otimes \vp_2) 
\otimes \vp_1 ^{\otimes 2n}\otimes \vp_2) = 2c_{4n + 2} \int_{\R^d_+} 
\left( q(\vp_1) \right)^{2n}    q(\vp_2) dx 
\end{equation}
and try to choose $\vp_1, \vp_2 \in \SC (\R^d)$ such that the RHS of
(\ref{fuenf.4eqa}) is smaller than zero, provided $c_{4n + 2} >0$.
To this aim, we at first need to proof some ~technical lemmas. 

Here and in the following $B_{\e}(x), x\in \R^d $ shall denote the open
ball of radius $\e $ around $x$.

\begin{Lemma}
\label{fuenf.2lem}
Fix $x^0 > 0$ and let $x = (x^0,0) \in \R^d$. Then there exists a function
$\vp_2 \in \SC (\R^d_+)$ and an $\e_2 > 0$ such that $q(\vp_2) < 0$ on
$B_{\e_2} (x)$.
\end{Lemma}

{\bf Proof.}
$q(\vp)(y)$ is continuous in $y$ for all $\vp \in \SC (\R^d_+)$. Therefore it
suffices to pick a $\vp_2$ s.t. $q(\vp_2) (x) < 0$. The existence of
an $\e_2$ follows.

Define continuous linear functionals $F^\pm$ on $\SC (\R^d_+)$ by
$$F^\pm : f \mapsto \int_{\R^d_+} G (x \mp y) f(y) dy.$$ 
$F^+ (y)$ on $\R_+^d$ has a singularity in $y=x$, but $F^- (y)$ has none. It
follows, that $F^+$ and $F^-$ are linear independent and so are 
$F^s := F^+ + F^-$ and $F^a := F^+ - F^-$. ~Consequently there is a
function $\vp _2 \in \hbox{kernel}(F^s)$, $\vp_2 \not \in \hbox{kernel}(F^a)$. 
It follows fom equation (\ref{fuenf.3eqa}), that 
$$q (\vp_2 ) (x) = (F^s \vp_2)^2 - (F^a \vp_2 ) ^2 = -(F^a \vp_2)^2 <0$$.
\kasten

\begin{Lemma}
\label{fuenf.3lem}
Let $x \in \R^d_+$, $x = (x^0 , 0)$ such that $G((2x^0, 0)) \ne 0$. For
every $\e_2 > 0$ there exist a function $\vp_1 \in \SC (\R^d_+)$ and an
$\e_1 , \e_2 > \e_1 > 0$, in such a way, that 
$\vert q (\vp_1 )\vert > 2$ on $B_{\e_1} (x)$ and 
$\vert q (\vp_1 )\vert < \half$ on $\R^d_+ \setminus B_{\e_2} (x)$.
\end{Lemma}

{\bf Proof.}
First we investigate what happens, if we take $\d_x$, the Dirac
measure with mass one in $x$, as a ''test function'':
$$q(\d_x) (y) = G (y - \Th x) G(y-x).
  $$
$G$ is continuous on $\R^d \setminus \{ 0 \}$ and 
$G((2x^0,0)) \ne 0$. It follows, that $q (\d_x)$ on $\R^d_+$ has a unique
singularity in
$y=x$, i.e. $\lim_{\vert x-y\vert \to 0} \vert q(\d_x) (y) \vert =
\infty$, and is continuous on $\overline{\R^d_+} \setminus \{ y \}$.
Now let $\d_{\rho , x} \in \SC (\R^d_+)$ be an approximation of $\d_x$,
i.e. $\d_{\rho , x} \to \d_x$ for $\rho \to 0$. Then $\vert
q(\d_{\rho,x})\vert$ takes arbitrarily large values in a ~neighborhood of
$x$ for $\rho \to 0$. At the same time, on 
$\R^d_+ \setminus B_{\e_1} (x)$ $\vert q(\d_\rho , x)\vert$ 
is bounded uniformly in $\rho$ by a
positive constant, say $D \over 2$. Choose $\e_1$, $0 < \e_1 < \e_2$
small enough, so that $\vert q(\d_{\rho , x})\vert > 2D$ on $B_{\e_1} (x)$
for some small $\rho$. Fix such a $\rho$ and let 
$\vp_1 = D^{-\half} \d_{\rho,x}$. 
Then $\vp_1$ and $\e_1$ fulfill the conditions of the lemma.
\kasten

We are now able to construct very sharp maxima for the functions 
$\vert q(\vp_1)\vert^{2n}$ on  arbitrarily small ~neighborhoods
$B_{\e_1} (x)$ of certain points $x$. Moreover, we can also
achieve that $\vert q(\vp_1)\vert^{2n}$ takes very small values
outside a ball $B_{\e_2} (x)$. Of course, now we want to enforce the
''negative spot'' of $q(\vp_2)$, detected in Lemma \ref{fuenf.2lem}
through multiplication by an adequately ~chosen $\vert q(\vp_1)\vert^{2n}$.

\begin{Lemma}
\label{fuenf.4lem}
It is possible to choose $\vp_1 , \vp_2 \in \SC (\R^d_+)$ and 
$n \in \N$, such that the integral on the RHS of (\ref{fuenf.4eqa})
takes a negative value.
\end{Lemma}

{\bf Proof.}
Fix $x=(x^0,\vec 0)$ as in Lemma \ref{fuenf.3lem}. According to Lemma
\ref{fuenf.2lem} there exists a function $\vp_2 \in \SC (\R^d_+)$ and
an $\e_2 > 0$ such that  $q(\vp_2) < 0$ on $B_{\e_2} (x)$.
Furthermore choose $\vp_1 \in \SC (\R^d_+)$ fulfilling the conditions
of Lemma \ref{fuenf.3lem} with the above fixed $\e_2 , x$. Now

\beas
&   & \int_{\R^d_+} q(\vp_1)^{2n} q(\vp_2) dx \\
& = & \left( \int_{\R^d_+ \setminus B_{\e_2} (x)} + 
      \int_{B_{\e_2} (x) \setminus B_{\e_1}(x)} + \int_{B_{\e_1}(x)}
\right) q (\vp_1)^{2n} q(\vp_2) dx \\
&  & < 2^{-2n} \int_{\R^d_+} \vert q(\vp_2)\vert dx   
      - 2^{ 2n} \int_{B_{\e_1} (x)} \vert q(\vp_2)\vert dx  
\eeas

It is clear, that for $n$ large enough, the RHS of this inequality
becomes negative.
\kasten

The derivation of this section's main result is now easy.

\begin{Proposition}
\label{fuenf.1prop}
Let $G$ fulfill Condition \ref{fuenf.1con}, and
$\psi$ be a L\'evy--characteristic with non--zero Poisson part and a 
L\'evy--measure $M$, such that all moments of $M$ exist. Let $F^\l$
denote the noise determined by $\l \psi$, $\l >0$, and let
 $X^\l = G* F^\l$ in the sense of Section 2.
Then there exists a $\l_0 > 0$ s.t. for all $\l$, $0<\l < \l_0$, the
Schwinger functions of $X^\l$ given by Definition \ref{drei.2def}
are not reflection positive.
\end{Proposition}

{\bf Proof.}
Since $c_{4n+2}>0$ for a $\psi $ with $M \ne 0$, equation (\ref{fuenf.4eqa})
together with Lemma \ref{fuenf.4lem} imply that conditional reflection
does not hold for the truncated Schwinger functions of $X=G* F^1$.
Therefore, Proposition \ref{fuenf.1prop} follows from Lemma
\ref{fuenf.1lem} .
\kasten

\begin{Remark}
\label{fuenf.4rem}
We assume --- and in the next section we will present some examples 
--- that at least the 2--point function $S_2 = S^T_2$ of a convoluted
generalized white noise with mean zero is reflection positive. Furthermore
we may choose the L\'evy measure $M$ of $\psi$ symmetric w.r.t 
the reflection at $0$.
In this case all Schwinger functions $S_n$, $n$
odd, vanish (cf. Remark \ref{drei.1rem}). 

 We choose $\l>0$. By scaling the test functions in the reflection positivity
condition $\vp_0\in\R\mapsto \vp_0\in\R,\vp_k\in
\SC((\R^d_+)^k)\mapsto \l^{k/2}\vp_k\in\SC((\R^d_+)^k)$ for $k=1,\ldots ,n$,
we get that the sequence of Schwinger functions $\{S_n^\l\}_{n\in\N_0}$ is
reflection positive (fulfills the Osterwalder--Schrader axioms) if and
only if the sequence of Schwinger functions $\{\tilde
S_n^\l\}_{n\in\N_0}$ defined by $\tilde S_n^\l:=\l^{-n/2}S_n^\l$ is
reflection positive (fulfills the Osterwalder--Schrader axioms)
\cite{OS1}. 

Take 
$\vp_1 \ldots \vp_{2n} \in \SC (\R^d)$ and write

\beas
&  & \tilde S^\l_{2n} \left( \vp_1 \otimes \ldots \otimes \vp_{2n} \right) \\
& = &   \sum_{I \in \hbox{ pairings }}
    \prod_{\{ j_1,j_2 \} \in I} S^T_2 (\vp_{j_1} 
    \otimes \vp_{j_2}) \\
& & +  \sum^{n-1}_{k = 1} \l^{k-n} \sum_{I \in P^{(2n)}_k}
    \prod_{\{ j_1 \ldots j_l\} \in I} S^T_l (\vp_{j_1} \otimes \ldots
    \otimes \vp_{j_l})  \ .
\eeas

For $\l$ large we may interpret $\tilde S^\l_{2n}$ as the $2n$-point
Schwinger function of a ''perturbed'' Gaussian 
reflection--positive random field with covariance
function $ S_2^T$. The ''perturbation'' is a polynomial without a
constant term of degree $n-1$ in the ''coupling constant'' $\l^{-1}$.
In Proposition \ref{fuenf.1prop}, we have shown that the reflection positivity 
breaks down if the
''coupling constant'' $\l^{-1}$ is larger than a certain ~threshold
$\l^{-1}_0$. It remains an open question what happens if $\l^{-1}$ is
small. We will not study this problem here.
\end{Remark}

On the first look, Proposition \ref{fuenf.1prop} may be discouraging.
The lack of reflection positivity in the general case leads to some
difficulties in the physical interpretation of the model. In the
''state space'' of the reconstruction theorem in
\cite{OS1} in general we may find some ''states'' with negative
norm. A straight ~forward probabilistic interpretation is therefore 
difficult in this
case, since "negative probabilities" would occur.

Nevertheless, even for the case of a large ''coupling constant''
$\l^{-1}$, some of the difficulties can be overcome, at least that is
what we hope at the moment: Let us note, that in the situation of 
Proposition \ref{fuenf.1prop} the obstructions to reflection 
positivity come fom the higher order truncated Schwinger 
functions. In Section 6 and Section 7 we analytically
continue the truncated Schwinger functions ''by hand''. The truncated 
Wightman distributions obtained by this procedure fulfill the spectral 
condition of QFTs with a "mass-gap", Poincar\'e invariance and
locality. From Haag--Ruelle theory (see e.g. p.317 of \cite{RS} Vol.
III), we know, that such truncated Wightman distributions of 
order $n\ge 3$ do not
contribute to the norm of a state approaching the asymptotic resp.
scattering region $x^0 \to \pm \infty $, because of the short range of
the forces involved (in the case of a QFT with a "mass-gap").
Therefore, if stable one particle states exist (take e.g. our
model $X=G_\a * F$, where $\a =\half$) the pseudonorm of the states 
approaching the scattering region should get positive. (For a precised
discussion in a special case, see Subsection 7.6).

\section{Analytic Continuation I:Laplace--Representation for the Kernel of 
$(-\D +m^2_0)^{-\a }$, $\a \mbox{$\in $} (0,1)$}
In this section, we will give a representation of the kernel 
$(-\D +m^2_0)^{-\a }$, for $m_0>0$ and $\a \in (0,1)$, in terms of a 
Laplace transform (which is specified later on). In \cite{AW}, an 
analytic continuation of the kernel associated with the pseudo 
differential operator 
$(-\D )^{-1}$ (corresponding to the case that $m_0=0$ and 
$\a =1$) was obtained, the starting point of which was a 
representation of the kernel of $(\D )^{-1}$ as a Laplace transform. 
In \cite{AIK}, based on the same representation of $(\D)^{-1}$, an 
analytic continuation of the (vector) kernel of 
$\partial ^{-1}$ was derived, where $\partial $ is the 
quaternionic Cauchy--Riemann 
operator. This section should be regarded as an introduction to the next 
section, where we extend the methods concerning analytic
continuation of Schwinger functions of random fields in \cite{AIK} and
\cite{AW}. It is also interesting to extend our approach here to the
case of vector kernels including the case of mass $m_0>0$. 
We intend to investigate this problem in forthcoming papers.

We notice that the kernel of $(-\D +m^2_0)^{-\a }$ is given by 

\be
\label{sechs.1eqa}
G_\a (x)=
(2\pi)^{-d}\int _{\R ^d} {e^{ikx} \over (\vert k \vert ^2 +m^2_0)^\a } dk
\ , \ x \in \R ^d
\ee
which is the  Fourier transform of a tempered distribution (see Example 
2.2). The idea we will realize here is that we
represent the integral (\ref{sechs.1eqa}), which is over the conjugate 
variable $k^0$ on the real time axis (i.e.\ , the 
$k^0$--axis in $k=(k^0,\vec k) \in \R \times \R ^{d-1}$), 
by an integral over (a part of) the upper half part of the
imaginary axis $ik^0$ (thus the $k^0$--axis being replaced
by an imaginary axis) so that the above Fourier transform goes over 
to a Laplace transform.

For simplicity of exposition, we assume first that $d=1$. We want to
evaluate the integral

\be
\label{6.2}
\int_{-\infty}^\infty {e^{ikx} \over (k^2 +m^2_0)^\a } dk \ , \ x \in \R
\ee
by some complex integral. This can be done by a contour integral around
a branch point as follows.

We denote by $\log$ the main branch of the complex logarithm which is
~holomorphic on $\C \setminus (-\infty, 0]$. Set
$$
(iz+m_0)^{-\a}=f_1(z)=\exp\{-\a \log (iz+m_0) \}, \, \, \, \, z\in \C \setminus
i[m_0,\infty);$$
$$
(-iz+m_0)^{-\a}=f_2(z)=\exp\{-\a \log (-iz+m_0) \}, \, \, \, \, z\in \C \setminus
i(-\infty, -m_0].$$
Clearly, $f_1$ and $f_2$ are holomorphic functions on the indicated
domains, respectively. Therefore, for arbitrarily fixed $x \in \R$
$$
h(z):=e^{izx}f_1(z)f_2(z)$$
is a holomorphic extension of the function $k\in \R \to {e^{ikx}\over
(k^2+m^2_0)^\a}\in \C$, which is defined on the real line, to the domain 
$\C \setminus \{iy: y\in \R, |y|>m_0 \}$. Take $C \subset
\C \setminus \{iy: y\in \R, |y|>m_0 \}$ as indicated in Fig 1. By the
well-known Cauchy integral theorem, we get
\bdm
\int _{C} h(z) dz =0 \ .
\edm

On the other hand
\bdm
\int _{C} h(z) dz = 
\int _{-t}^t {e^{ikx} \over (k^2 
+m^2_0)^\a } dk + \sum _{j=1}^5 \int _{C_j} h(z) dz \ , 
\edm
\begin{figure}[t]


\begin{picture}(300,180)(-180,-100)
\thicklines 
\put(0,50){\line(0,1){76}}
\put(0,-63){\line(0,1){13}}
\thinlines
\put(-125,0){\vector(1,0){250}}
\put(0,-63){\vector(0,1){193}}
\put(0,0){\circle*{3}}
\put(5,-10){0}
\put(-100,0){\circle*{3}}
\put(-103,-10){$-t$}
\put(100,0){\circle*{3}}
\put(100,-10){$t$}
\put(0,50){\circle*{3}}
\put(0,-50){\circle*{3}}
\put(-30,-50){$-im_0$}
\put(0,50){\line(-1,1){30}}
\put(-25,80){$<\!\! )=$}
\put(-14,71){$\beta_2$}
\put(3,4){$<\!\! ) =\beta_1$}
\put(125,-10){$\Re $}
\put(3,122){$\Im $}
\put(0,50){\vector(-1,-1){10}}
\put(-9,47){$s$}
\put(-41,91){\circle*{3}}
\put(-62,94){${\scriptstyle im_0+ie^{i\beta_2}t_1}$}
\put(-127,-83){\parbox{3.7in}{{\bf Fig.1}:{\footnotesize The closed complex 
contour $C$ stays inside the domain of holomorphy of the function $h$. It is
composed of the parts $C_0,\ldots,C_5$. }}}
\put(10,60){\line(1,1){31}}
\put(-10,60){\line(-1,1){31}}
\put(-100,0){\line(1,0){200}}
\put(-50,0){\vector(1,0){1}}
\put(48,0){\vector(1,0){1}}
\put(70,71){\vector(-1,1){1}}
\put(-70,71){\vector(-1,-1){1}}
\put(25,75){\vector(-1,-1){1}}
\put(-25,75){\vector(-1,1){1}}
\put(-2,36){\vector(-1,0){1}}
\put(70,-10){$C_0$}
\put(83,65){$C_1$}
\put(-93,65){$C_5$}
\put(12,80){$C_2$}
\put(-15,28){$C_3$}
\put(-40,70){$C_4$}
\bezier{300}(100,0)(99,60)(41,91)
\bezier{300}(-100,0)(-99,60)(-41,91)
\bezier{300}(0,0)(20.5,45.5)(41,91)
\put(0,50){\circle{28}}
\end{picture}
\end{figure}

the curves $C_j$ being as in Fig.1.

Thus we have derived

\be
\label{6.4}
\int _{-t}^t {e^{ikx} \over (k^2 +m^2_0)^\a } dk = 
- \sum _{j=1}^5 \int _{C_j} h(z) dz \ .
\ee
Moreover, we have the following result

\begin{Lemma}
For every $\a \in (0,1)$ and $x > 0$,

\be
\label{6.5}
\int _{-\infty}^\infty {e^{ikx} \over (k^2 +m^2_0)^\a } dk = 2 \sin (\pi
\a) \int^\infty_{m_0} {e^{-rx} \over (r^2-m^2_0)^\a } dr \ .
\ee
\end{Lemma}

{\bf Proof.}   We remark at first that for any fixed $x \in \R$

\bdm
\int _{-\infty}^\infty {e^{ikx} \over (k^2 +m^2_0)^\a } dk = \lim_{t \to
\infty} \int _{-t}^t {e^{ikx} \over (k^2 +m^2_0)^\a } dk \ ,
\edm
where by Leibniz criterion the right hand side converges for every 
$\a >0$.
Thus we should analyse each curve integral on the right hand side of
(\ref{6.4}). We shall use polar coordinates for each integral.

\begin{enumerate}
\item
Using the polar coordinate representation $z=re^{i\b }$, $r>0$, 
$-{3 \pi \over 2} < \b < {\pi \over 2}$, 
$C_1=\{ (r,\b )\ : \ r=t , 0 \le \b \le \b _1 \}$, we have the 
following derivation for $x\ge 0$ and $\b _1 \in (0, {\pi \over 2})$

\beas
\lv \int _{C_1} h(z) dz \rv 
& = &   \lv \int_0^{\b _1} t i e^{i\b }e^{ixt e^{i\b }}f_1(te^{i\b})  
        f_2(te^{i\b}) d\b \rv   \\
& \le & \int_0^{\b _1} {t e^{-x t \sin \b} \over t^{2\a } \lv 
        e^{2i\b }+{m^2_0 \over t^2} \rv ^\a } d\b  \\
& = &   t^{1-2\a } \int_0^{\b _1} {e^{-x t \sin \b} \over \left[ 
        1+2 \left( {m_0 \over t} \right)^2 \cos 2\b + \left( {m_0 \over t} 
        \right) ^4 \right] ^{{\a \over 2 }} } d\b  \\
& \le & t^{1-2\a } \left[ 1- \left( {m_0 \over t} \right) ^2 \right]^{-\a } 
        \int_0^{\b _1} e^{-x t \sin \b} d\b  \\
& \le & t^{1-2\a } \left[ 1- \left( {m_0 \over t} \right) ^2 \right]^{-\a } 
        {1 \over \cos \b _1} \int_0^{\sin \b _1} e^{-xtr} dr \\
& = &   {1-e^{-x t \sin \b _1} \over x t^{2 \a } \left[ 1- \left( 
        {m_0 \over t} \right) ^2 \right]^{\a } \cos \b _1 } \\
& \to & 0
\eeas
as $t \to \infty$.

\item
Analogously, using the representation $z=re^{i\b }$, $r>0$, 
$-{3 \pi \over 2} < \b < {\pi \over 2}$, 
$C_5=\{ (r,\b )\ : \ r=t , -\pi -\b _1 \le \b \le -\pi \}$, we have for any
arbitrary fixed $\b _1 \in (0, {\pi \over 2})$

\bdm
\lv \int _{C_5} h(z) dz \rv \to 0
\edm
as $t \to \infty$.

\item
Using the polar coordinates representation $z=im_0+ire^{i\b }$, $r>0$,
$0 \le \b <2\pi $, $C_3=\{ (r,\b )\ : \ r=s , \b _2 \le \b \le 2 \pi -
\b _2 \}$, we have for any fixed $\b _2 \in (0, {\pi \over 2})$ and
$s<m_0$ that

\beas
\lv \int _{C_3} h(z) dz \rv 
& = &   \lv - \int_{\b _2} ^{2x  -\b _2} {e^{i \left( im_0+i s e^{i \b } 
        \right) x} \over \left( \left( im_0+i s e^{i \b} \right) ^2 +m^2_0
        \right) ^\a } isie^{i \b} d \b \rv \\
& \le & \int_{\b _2} ^{2\pi  -\b _2} {s^{1-\a } e^{-(m+s\cos \b )x}
        \over (4 m^2_0+4ms \cos \b +s^2) ^{\a \over 2}} d\b \\
& \le & s^{1-\a } \int_{\b _2} ^{2\pi  -\b _2} {e^{-(m-s)x} \over (2m-s) 
        ^{\a \over 2}} d\b \\
& = &   {s^{1-\a } (2 \pi - 2 \b _2) e^{-(m-s)x} \over 2(m-s)^{\a \over 2}}\\
& \to & 0
\eeas
as $s \to 0$.

\item
Using $z=im_0+ire^{i\b }$, $r>0$, $0 \le \b <2\pi $, $C_4=\{ (r,\b )\ : \ 
s \le r \le t_1, \  \b = \b _2 \}$ with $t_1=(t^2+m^2_0-2tm\sin \b
_1)^\half$,
we have, for $\b _2 \in (0, {\pi \over 2})$ and $s<1<t$, the
following derivation

\begin{eqnarray}
\label{sechs.6eqa}
&   &   \int _{C_4} h(z) dz \nonumber \\
& = &   \int_{s} ^{t_1} e^{i \left( im_0+i r e^{i \b_2 } \right) x} 
       f_1(im_0+ire^{i\b _2}) f_2(im_0+ire^{i\b _2}) ie^{i \b_2} dr \nonumber\\
& = &   \int_{s} ^{t_1} {e^{i \left( im_0+i r e^{i \b_2 } \right) x}ie^{i\b _2}
        \over [i(im_0 + ire^{i\b _2})+m_0]^\a[-i(im_0 + ire^{i\b _2})+m_0]^\a} dr
        \nonumber\\
& = &  ie^{i\b _2} \int^{t_1}_s {e^{-(m_0+re^{i\b _2})x} \over
        {(-re^{i\b _2})^\a(2m_0+re^{i\b _2})^\a}}dr \nonumber \\
& = &  ie^{i\b _2} \int^{t_1}_s {e^{-(m_0+re^{i\b _2})x} \over
        {r^\a (e^{i\b _2-\pi})^\a(2m_0+re^{i\b _2})^\a}}dr \nonumber \\
& = &  ie^{i\pi \a}e^{i\b _2(1-\a)} \int^{t_1}_s {e^{-(m_0+re^{i\b _2})x} \over
        r^\a (2m_0+re^{i\b _2})^\a}dr \nonumber \\
& = &  ie^{i\pi \a}e^{i\b_2(1-\a)} \int_s^{t_1} {e^{-
       \left( m_0+re^{i\b _2} \right) x} \over r^\a \left( 2m_0+re{^i\b_2}
       \right) ^\a } dr \ ,
\end{eqnarray}
where we had used the representation $-1=e^{-i\pi}$ in the fourth
equality. Now we want to consider the limit of (\ref {sechs.6eqa}) as $s \to 0$
and $t_1 \to \infty$. In order to do that, by an application of Lebesgue theorem,
we estimate the integrand in (\ref {sechs.6eqa}) as follows:

\begin{eqnarray}
\label{sechs.7eqa}
&     & \lv \int_s^{t_1} {e^{- \left( m_0+re^{i\b _2} \right) x} \over r^\a 
        \left( 2m_0+re^{i\b_2} \right) ^\a } dr \rv \nonumber\\
& \le & \int_s^{t_1} {e^{- \left( m_0+r \cos \b_2 \right) x} \over r^\a 
        \lv 2m_0+re^{i\b_2} \rv ^\a } dr \nonumber\\
& = &   \int_s^{t_1} {e^{- \left( m_0+r \cos \b_2 \right) x} \over r^\a 
        \left( r^2+4m_0r\cos \b_2 +4m^2_0 \right) ^{{\a \over 2}} } dr \nonumber\\
& = &   \left( \int_s^{1} + \int_1^{t_1} \right)
        {e^{- \left( m_0+r \cos \b_2 \right) x} \over r^\a 
        \left( r^2+4m_0r\cos \b_2 +4m^2_0 \right) ^{{\a \over 2}} } dr \ ,
\end{eqnarray}
and  

\begin{eqnarray}
\label{sechs.8eqa}
         \int_s^1 {e^{- \left( m_0+r \cos \b_2 \right) x} \over r^\a 
         \left( r^2+4m_0r\cos \b_2 +4m^2_0 \right) ^{{\a \over 2}} } dr
& \le &  \int_s^1 {e^{-m_0x} \over r^\a \left( r^2+4m^2_0 \right) 
         ^{{\a \over 2}} } dr \nonumber\\
& \le &  {e^{-m_0x} \over (2m_0)^\a } \int_s^1 {dr \over  r^\a } \nonumber\\
& =   &  {e^{-m_0x} \over (2m_0)^\a } {1-s^{1-\a } \over 1-\a } \nonumber\\
& \to &  {e^{-m_0x} \over (2m_0) ^\a (1-\a )}
\end{eqnarray}
as $s \to 0$. Moreover 

\begin{eqnarray}
\label{sechs.9eqa}
&     &  \int_1^{t_1} {e^{- \left( m_0+r \cos \b_2 \right) x} \over r^\a 
         \left( r^2+4m_0r\cos \b_2 +4m^2_0 \right) ^{{\a \over 2}} } dr
         \nonumber\\
& \le &  {e^{-m_0x} \over \left( 1+4m_0\cos \b_2 +4m^2_0 \right) ^{{\a \over 2}}} 
         \int_1^{t_1} e^{-rx\cos \b_2 } dr \nonumber\\
& =   &  {e^{-m_0x} \over \left( 1+4m_0\cos \b_2 +4m^2_0 \right) ^{{\a \over 2}}} 
         {e^{-x\cos \b_2 } -e^{-xt_1 \cos \b_2 } \over x \cos \b_2 } 
         \nonumber\\
& \to &  {e^{-(m_0+ \cos \b_2 )x} \over \left( 1+4m_0\cos \b_2 +4m^2_0 \right) 
         ^{{\a \over 2}} x \cos \b_2 } 
\end{eqnarray}
as $t_1 \to \infty$ (or, equivalently, $t \to \infty$). Using the above facts 
(\ref{sechs.6eqa})--(\ref{sechs.9eqa}), we see that the following limit exists 
and is given by

\bdm
\lim_{s\to 0, t \to \infty} \int _{C_4} h(z) dz =
 ie^{i\pi \a}e^{i(1-\a )\b_2} \int_0^\infty {e^{- \left( m_0+r\cos \b_2 \right) 
x -irx \sin \b_2 } \over r^\a \left( 2m_0+re^{i\b_2} \right) ^\a } dr \ .
\edm
Now setting $\b_2 \to 0$, by Lebesgue theorem again, we get

\beas
\lim_{\b_2 \to 0} \lim_{s\to 0, t \to \infty} \int _{C_4} h(z) dz 
& = &  ie^{i\pi \a} \int_0^\infty {e^{-(m_0+r)x} \over r^\a (2m_0+r) ^\a }
       dr \\
& = &  ie^{\a \pi i} \int_{m_0}^\infty {e^{-rx} \over (r^2-m^2_0)^\a } dr \ . 
\eeas

\item
Similarly, we get

$$
\lim_{\b_2 \to 0} \lim_{s\to 0, t \to \infty} \int _{C_2} h(z) dz 
 =   -ie^{-\a \pi i} \int_{m_0}^\infty {e^{-rx} \over (r^2-m^2_0) ^\a } dr \ .
$$

Finally, combining the above steps 1 to 5, we derive from (\ref{6.4}) that

\beas
\int_{-\infty}^\infty {e^{ikx} \over (k^2+m^2_0)^\a } dk
& = &  -\left[ ie^{\a \pi i}-ie^{-\a \pi i}\right] \int_{m_0}^\infty {e^{-rx} 
       \over (r^2-m^2_0) ^\a } dr \\
& = &  2 \sin (\pi \a) \int_{m_0}^\infty {e^{-rx} \over (r^2-m^2_0) ^\a } dr \ .
\eeas
Hence we obtain (\ref{6.5}). \kasten
\end{enumerate}

Now we give the notion of Laplace transform in one variable. The
definition the of Laplace transform in the multi variable case will
be given in Section 7.

\begin{Definition}
\label{sechs.1def}
Let $M_d$ denote the $d$--dimensional Minkowski space--time with
Minkowski inner product $< \ , \ >_M$. Choose $e_0 \in M_d$ such that
\linebreak $< e_0 , e_0 >_M = 1$ and let $\{ e_0 , \ldots , e_{d-1} \}$ 
be an orthonormal frame in $M_d$.
Let $V^{+}_0$ be the forward light cone, namely,
$$V^{+}_{0} := \{ k \in M_d : k^2 > 0 , < k , e_0 >_M < 0 \},$$
and $V^{*+}$ its closure. We recall that 
$\R^d_+ := \{ x = (x^0 , \vec x) \in \R \times \R^{d-1} : x^0 > 0 \}$.
Notice that for $x=(x^0,\vec{x})\in \R^d_+$, the function 
$e(x,k):=e^{-x^0 k^0 + i\vec{x}\vec{k}}$ on the 
forward light cone $V^{*+}_0$ behaves as a fast falling function, i.e.,
there exists $h_x\in \SC_{\C}(\R^d)$ such that $h_x(k)=e(k,x)$ for
$k\in V^{*+}_0$. Therefore, for a tempered distribution $f\in \SC'(\R^d)$
with $\hbox{supp} f \subset V^{*+}_0$, we can define 

\begin{equation}  
\label{sechs.10eqa}
(\LC f)(x):= <\overline{e(\cdot,x)}, f>:=<\overline{h_x}, f>,
\end{equation}
which is well defined since by the fact that $\hbox{supp} f \subset V^{*+}_0$
there is no ambiguity arising from the choice of $h_x$. We call 
$\LC f$ the Laplace transform
of $f\in \SC'(\R^d)$ with $\hbox{supp} f \subset V^{*+}_0$.
\end{Definition}

Clearly, the Laplace transform of a test function 
$f \in \SC (\R^d_+)$ is given by the following formula
$$
(\LC f) (x) = (2\pi)^{- {d \over 2}} \int_{\R^d_+} 
e^{-x^0 k^0 + i \vec x \vec k} f(k) dk \ , \qquad x \in \R^d_+ \ .
$$
Sometimes $\LC$ (e.g. in \cite{AIK}) is called the Fourier--Laplace
transform. Our definition here is close to the one given in
\cite{StWi}.

The following proposition is the main result of this section, which
gives a representation of $G_\a$ in terms of a Laplace transform.

\begin{Proposition}
\label{sechs.1prop}
For $\a \in (0,1)$ and $x \in \R^d \setminus \{ 0 \}$, we have the
following formula

\be
\label{sechs.11eqa}
G_\a (x) = 2(2\pi)^{-d} \sin (\pi \a) \int_{\R^d_+} 
e^{-\vert x^0 \vert  k^0 + i \vec x \vec k} 
{1_{\{ k^0 > (\vert \vec k \vert^2 + m^2_0)^\half \}}(k) 
\over ({k^0}^2 - \vert \vec k \vert^2 - m^2_0)^\a}dk \ ,
\ee
where $ 1_{\{ k^0 > (\vert \vec k \vert^2 + m^2_0)^\half \}} (k)$ is the
indicator of the subset $\{ k= (k^0 , \vec k) \in \R \times \R^{d-1}$
$: k^0 > ( \vert \vec k \vert^2 + m^2_0 )^\half \} \subset \R^d_+$.
\end{Proposition}

{\bf Proof.}
By (\ref{sechs.1eqa}) and (\ref{6.5}), for $x \in \R^d_+$, we
obtain (\ref{sechs.11eqa}) by Fubini theorem (cf. \cite{Go} for details
on using Fubini theorem) and the following derivation
\beas
&&       G_\a (x) \\
&=&(2\pi)^{-d}\int_\R {e^{ikx} \over (\vert k \vert^2 + m^2_0)^\a } dk \\
&=&(2\pi)^{-d}\int_{\R^{d-1}} e^{i \vec k \vec x} \left[ 
    \int^\infty_{- \infty}
    {e^{ik^0 x^0} \over ({k^0}^2 + \vert \vec k \vert^2 + m^2_0)^\a }
    dk^0 \right] d\vec k \\
&=&(2\pi)^{-d}\int_{\R^{d-1}} e^{i \vec k \vec x} \left[ 2 \sin (\pi \a) 
    \int^\infty_{( \vert \vec k \vert^2 + m^2_0)^\half}
    {e^{- k^0 x^0} \over ({k^0}^2 - \vert \vec k \vert^2 - m^2_0)^\a }
    dk^0 \right] d \vec k \\
&=&2(2\pi)^{-d} \sin (\pi \a) \int_{\R^d_+} e^{- k^0 x^0 + i \vec k \vec x}
     {1_{\{ k^0 > (\vert \vec k \vert^2 + m^2_0)^\half \}}(k)
     \over ({k^0}^2 - \vert \vec k \vert^2 - m^2_0)^\a }dk \ .  
\eeas
Now formula (\ref{sechs.11eqa}) results from the (full) Euclidean
invariance of $G_\a$ (see Remark \ref{zwei.2rem}).  \kasten

By changing variables in (\ref{sechs.11eqa}) as 
$(k^0,\vec k ) \to ((\vert \vec k \vert^2 + m^2_0 )^\half , \vec k)$, we
get a ''K\"allen--Lehmann representation'' for $G_\a$ (we refer the
reader to Theorem IX.33 of \cite{RS} or Theorem II.4 of \cite{Si} for
the K\"allen--Lehmann representation).

\begin{Corollary}
\label{sechs.1cor}
For $\a \in (0,1)$ and $x \in \R^d \setminus \{ 0 \}$, the kernel
$G_\a$ of $(-\D + m^2_0 )^{-\a}$ has the following representation

\begin{equation}
\label{sechs.12eqa}
G_\a (x) = \int^\infty_0 C_m (x)\rho _\a  (dm^2_0) \ ,
\end{equation}
where

\begin{equation}
\label{sechs.13eqa}
C_m (x) = (2\pi)^{-d}\int_{\R^d} e^{-x^0 k^0 + i \vec k \vec x} 
1_{\{ k^0 > 0 \}} (k) \d ({k^0}^2 - \vert \vec k \vert^2 - m^2 )dk\ ,
\end{equation}

\begin{equation}
\label{sechs.14eqa}
\rho_\a (dm^2) = 
2 \sin (\pi \a) 1_{\{ m^2 > m^2_0 \}} {dm^2 \over (m^2-
m^2_0)^\a} \ ,
\end{equation}
\end{Corollary}

\begin{Remark}
\label{sechs.1rem}
In general, the K\"allen--Lehmann representation characterizes 
the 2--point Schwinger functions of a
Lorentz--invariant field--theory. Comparing our formula
(\ref{sechs.13eqa}) with formula (6.2.6) of \cite{GlJa}, we see that
$C_m$ is a representation of the 2--point Schwinger
function (i.e. the free covariance) of the relativistic free
field of mass $m$.

From formula (7.2.2) of \cite{GlJa}, we have for $d \ge 2$ that
\begin{equation}
\label{sechs.15eqa}
C_m (x) = (2\pi)^{d \over 2} ({m \over \vert x \vert})^{d-2 \over 2}
K_{d-2 \over 2} (m \vert x \vert ) \ , 
\ \ x \in \R^d \setminus \{ 0 \} \ ,
\end{equation}
where $K_{d-2 \over 2} > 0$, for $x \in \R^d \setminus \{ 0 \}$, is
the modified Bessel function. Thus $C_m$ is a positive function on
$\R^d \setminus \{ 0 \}$ with a singularity at the origin and
exponential decay as $m \vert x \vert \to \infty$. This implies that
$G_\a$ is singular at the origin, as was assumed in Section 5.
The consequence that $G_\a$ is of exponential decay for 
$\vert x \vert \to \infty$ can be shown by using the precise estimates
in Proposition 7.2.1 of \cite{GlJa} and the fact $\rho_\a$ has a ''mass
gap'', i.e., supp$\vp_\a \subset [m^2_0 , \infty )$. (For $d=1$, the
singularity of $G_\a$ is only due to $\rho_\a$ being an infinite
measure).  \kasten 
\end{Remark}

Concerning the truncated Schwinger function $S^T_2$ associated 
with $X_\a = G_\a * F$, we
have the following result, which gives a representation of the
integral kernel $s$ of $S^T_2$ as a Laplace transform, while
by formula (2--25) on p. 40 of \cite{StWi} and translation invariance 
of s, we have 
$$S^T_2 (f_1 \otimes f_2 ) = c_2 \int_{\R^d} \int_{\R^d}
  f_1 (x) s (x-y) f_2 (y) dx dy \ , \ \ f_1 , f_2 \in \SC (\R^d) \ .
$$

\begin{Corollary}
\label{sechs.2cor}
For $\a \in (0 , \half)$,
\begin{equation}
\label{sechs.17eqa}
s(x-y) = 2(2\pi)^{-{d \over 2}}\sin(\pi 2\a)\LC \left({1_{\{ k^0 > 
(\vert \vec k \vert^2 + m^2_0)^\half \}} (k) 
\over (k^2 - \vert \vec k \vert^2 - m^2_0)^{2\a}} \right)(x-y) \ .
\end{equation}
\end{Corollary}

{\bf Proof.}
By (\ref{drei.10eqa}), we have

\beas
     S^T_2 (f_1 \otimes f_2)
& = & c_2 <(-\D +m^2_0)^{-\a}f_1, (-\D +m^2_0)^{-\a}f_2>_{L^2} \\
& = & c_2 <f_1, (-\D +m^2_0)^{-2\a}f_2>_{L^2} \ , 
     f_1 , f_2 \in \SC (\R^d) \ ,
\eeas
where $c_2$ is given by (\ref{drei.10eqa}). 
Therefore by (\ref{sechs.1eqa}) we get

\beas
     s(x-y)
& = & c_2 G_{2\a} (x-y) \\
& = & 2 (2\pi )^{-d} \sin (2\pi \a) \times \\
&   & \times \int_{\R^d_+} 
     e^{-\vert x^0 - y^0 \vert k^0 + i (\vec x - \vec y )
     \vec k } 
     {1_{\{ k^0 > {(\vert \vec k \vert^2 + m^2_0)}^\half \}} (k)
     \over ((k^{0})^2 - \vert \vec k \vert^2 - m^2_0)^{2\a}}dk \\
& = & 2(2\pi )^{-{d \over 2}} \sin (2\pi \a) \LC \left( {1_{\{ k^0 > 
     (\vert \vec k \vert^2 + m^2_0)^\half \}} (k) \over ((k^{0})^2 - 
     \vert \vec k \vert^2 - m^2_0)^{2\a}} \right) (x-y) \ . 
\eeas

\kasten

\begin{Remark}
\label{sechs.2rem}
We should point out that in section 7 we shall give a different
derivation of $s$. By the above formula
(\ref{sechs.13eqa}), one can analytically continue $s$ to the 
kernel, $w$ say, of the 2--point (truncated) Wightman distribution
$W^T_2=w(x-y)$ with
$$w(x-y) := 2(2\pi )^{-{d \over 2}} c_2 \sin (2\pi \a) \FC^{-1} \left( 
  {1_{\{ k^0 > (\vert \vec k \vert^2 + m^2_0)^\half \}} (k)
  \over ((k^{0})^2 -  \vert \vec k \vert^2 - m^2_0)^{2\a}} \right)
  (x-y)  \ .
$$
\kasten
\end{Remark}

\begin{Remark}
\label{sechs.3rem}
Concerning the Schwinger function $S_2$ and the 2--point Wightman
distribution for the case $\a = \half$, see the discussion Section
II.5 of \cite{Si}. \kasten
\end{Remark}

\begin{Remark}
\label{sechs.4rem}
Suppose $F$ is a Gaussian white noise, then $X_\a=(-\D + m^2_0)^{-\a} F$
is a generalized free field for each $\a \in (0,\half )$ and $m_0>0$. We
refer the reader to e.g. \cite{Heg} and \cite{Si} for the notion of
generalized free field. Our argument here is as follows. In this case, all
Wightman distributions $W_n$ $n \in \N$, are given symbolically by
$$W_n (x_1, \ldots , x_n) =
  \cases{ 0\ , &$n$ is odd \cr
          \sum W^T_2 (x_{j_1} , x_{l_1}) \ldots W^T_2 (x_{j_{n\over 2}} , 
          x_{l_{n\over 2}}) 
          \ , & $n$ is even \cr}
$$
where the sum is over 
$\{ (j_1 , \ldots , j_{n\over 2} , l_1 , \ldots , l_{n \over 2} )$
$\in \N^n : 1 \le j_1 < j_2 < \ldots < j_{n\over 2} < n$ and
$j_k < l_k$ for $1 \le k <  {n \over 2} \}$. This shows that all
$W_n$, $n \in \N$, are determined by $W^T_2$. The corresponding
Schwinger function $S_n$, $n \in \N$, satisfy reflection positivity
since by Corollary \ref{sechs.1cor}, $W^T_2$ has a K\"allen--Lehmann
representation, therefore the field $X_\a$ is a generalized free Euclidean 
field. Especially $X_\a$ is reflection positive in the sense of
\cite{GlJa}.  \kasten
\end{Remark}

\begin{Remark}
\label{sechs.5rem}
Finally in this section, we should point out that concerning the
kernel $G_1$ of $(-\D + m^2_0)^{-1}$, i.e., $\a = 1$, the above
procedure can not be performed. One can apply the residue theorem instead
of Cauchy integral theorem to get a Laplace transform representation
of $G_1$. We therefore have a
representation of $G_1$ by using the following basic formula (see e.g.
(II.3) of \cite{GRS} or (II.4) of \cite{Si}):
$$\int^\infty_{-\infty} {e^{its} \over t^2 + r^2} dt =
  {2\pi e^{-r \vert s \vert} \over r} \ \ , r >  \ \ ,
  s \in (-\infty , \infty) \ .
  $$
In fact, we have the following derivation

\beas
      G_1 (x)
& = & (2\pi)^{-d} \int_{\R^d} {e^{ikx} \over \vert k \vert^2 + m^2_0} dk \\
& = & (2\pi)^{-d}\int_{\R^{d-1}} e^{i \vec k \vec x} \left( 
     \int^\infty_{-\infty} {e^{ik^0x^0} \over {k^0}^2 + 
     \vert \vec k \vert^2 + m^2_0} dk^0 \right) d\vec k \\
& = & (2\pi)^{-d+1} \int_{\R^{d-1}} 
     {e^{- ( \vert \vec k \vert^2 + m^2_0)^\half
     \vert x^0 \vert + i \vec k \vec x} \over 
     2(\vert \vec k \vert^2 + m^2_0)^\half} dk \ . 
\eeas

The above integral representation of $G_1$ in case $m_0=0$ was used in
\cite{AW} (see the corresponding formula (6.6) in Minkowski space in
\cite{AW}) as a starting point for the analytic continuation of the
Schwinger functions.  \kasten
\end{Remark}

\section{Analytic Continuation II: Continuation of the 
(Truncated) Schwinger Functions}
\subsection{Preliminary remarks}
In this section we present the analytic continuation of the Euclidean
(truncated) Schwinger functions obtained from convoluted generalized
white noise $X= (-\D + m^2_0)^{-\a } F$, $m_0 > 0$, 
$\a \in (0,\half ]$, to relativistic (truncated) Wightman distributions. 
More generally, formal expressions are obtained for the class of random
fields $X$ given by $X = G * F$, where $G$ is the Laplace--transform of 
a ~Lorentz--invariant signed measure on the ~forward lightcone. But as we 
will illustrate for $X = (-\D + m^2_0 )^{-\a} F, \a \in (\half , 1)$, the
question whether such relativistic expressions really represent the analytic 
continuation of the corresponding (truncated) Schwinger functions, deserves 
a separate discussion.

We derive manifestly Poincar\'e--invariant formulas for the Fourier--
transform of the (truncated) Wightman functions. The obtained Wightman
distributions ~fulfil the strong spectral condition of a QFT. with a ''mass-
-gap''. Locality, Hermiticity and the cluster property of the (truncated)
Wightman functions follow from the (truncated) Schwinger functions'
symmetry, $\theta$ -invariance and cluster property respectively, as a
result of the general procedures of axiomatic QFT. We continue the
discussion, started at the end of section 5, for the special case
$X=(-\D + m^2_0)^{-\half} F$, $d=4$.

To the readers convenience (and to keep the length of the paper within
reasonable bounds) not every step is presented with all details 
(for complete proofs see \cite{Go}). The
methods applied here are based on and extend those of \cite{AIK},
\cite{AW}.

\subsection{The mathematical background}
Let us first clarify the notations and the mathematical background. In
this subsection $n$ is a fixed integer with $n\ge 2$.
From Definition \ref{sechs.1def} recall the meaning of $M_d$,  $< \ , \ >_M$ 
and let $\{ e_0 , \ldots , e_{d-1} \}$ be an
orthonormal frame in $M_d$, by which $M_d$ is identified with $\R 
\times \R^{d-1} \cong\R^d$. 
From now on we write for $x\in M_d $ $x=(x^0,\vec x)$ and
$x^2 = < x , x>_M = {x^0}^2 - \vert \vec x \vert ^2 $. The forward mass 
cone of mass $m_0>0$ is defined in analogy with the 
forward light cone $V_0^+$ as

\begin{equation}
\label{7.1eqa}
V^+_{m_0} := \{ k \in M_d : k^2 > m_0^2 , < k , e_0 >_M < 0 \} \ \ 
m_0 \ge 0 \ .
\end{equation}
 
By $V^{* +}_{m_0}$ we denote its closure.  Let $\theta$ again denote the
time--reflection. The backward (closed) mass cone/lightcone is defined by 
$V^{(*) -}_{m_0 \backslash 0} := \theta V^{(*) -}_{m_0 \backslash 0}$.
Since $M_d \cong \R^d \subset \C^d$ there is a complexification of the Minkowski 
inner product $<\ , \ >_M$
such that it is analytic in the coordinates with respect to 
$\{ e_0 , \ldots , e_{d-1} \}$. We denote this complexification by
$< \ , \ >^{\C}_M$. Let $T^n$ be the tubular
domain in ${(\C ^d)}^n$ with base $V^-_0$, i.e.

\begin{equation}
\label{7.2eqa}
T^n := \left\{ \underline{z} = (z_1 , \ldots , z_n) \in 
{(\C^d)}^n : z_j - z_{j+1} \in M_d + i V^-_0 \ ,
\ 1 \le j \le n-1 \right\}
\end{equation}
$T^n$ is called the backward tube. Finally define

\begin{equation}
\label{7.3eqa}
e(\underline{k},\underline{z}) := (2 \pi )^{-{dn\over 2}}
\exp \left\{ i \sum^n_{l=1} < k_l , z_l >^{\C}_M \right\} \ , \
\end{equation}
where $\underline{k} =(k_1,\ldots,k_n) \in (\R^d)^n $ and 
$ \underline{z} =(z_1,\ldots ,z_n) \in {(\C ^d)}^n$.
We now give the definition of spectral conditions that are crucial
for the theory of the Laplace--transforms in $n$ arguments 
as well as for the notion of
causality in a relativistic QFT.

\begin{Definition}
\label{7.1def}
\begin{enumerate}
\item Let $\hat {\sf W} \in \SC '_c (\R^{dn})$. We say that $\hat {\sf W}$
~fulfils the spectral condition, if there is a 
$\hat {\sf w} \in \SC (\R^{d(n-1)})$ such that
\begin{equation}
\label{7.4eqa}
\hat {\sf W} (k_1 \ldots k_n) = \hat {\sf w} (k_1 , k_1 + k_2 , \ldots , 
k_1 + \ldots +k_{n-1} ) \d \left( \sum^n_{l=1} k_l \right)
\end{equation}
and supp $\hat {\sf w} \subset {(V^{* -}_0)}^{n-1}$.

\item Let ${\{ \hat {\sf W}_l^T \}}_{l \in \N}$ be a sequence of truncated 
distributions determined by a sequence of distributions $\{ \hat {\sf W}_l\} _{l\in
\N_0}$,
$\hat {\sf W} _l\in \SC_{\C} ' (\R^{dl}), \ \hat {\sf W}_0=1$. We say that the ${\{ \hat
W_l^T \}}_{l \in \N} $ (the ${\{ \hat {\sf W}_l \}}_{l \in \N_0}$ respectively) 
~fulfil the strong spectral condition
with a mass gap $m_0>0$ if all the distributions $\hat {\sf W}_l^T, l\ge 2$,
~fulfil the spectral condition (\ref{7.4eqa}),
where $V^{*-}_0$ is replaced by $V^{*-}_{m_0}$.
\end{enumerate}
\end{Definition}

The following two theorems, taken from the theory of Laplace transforms,
provide us with the necessary mathematical tools for the analytic
continuation of Schwinger functions.

\begin{Theorem}
\label{7.1theo}
Assume, that $\hat {\sf W} \in \SC_{\C} (\R^{dn})$ ~fulfils the spectral
condition. Then

\begin{enumerate}
\item $\LC (\hat {\sf W}) (\underline{z} )= < \overline{ \hat {\sf W} } , 
e( \cdot , \underline{z})>$ is well--defined and holomorphic in
the variables $z_j - z_{j+1}$ $j=1 \ldots n-1$ on the domain 
$\underline{z} \in T^n$. $\LC (\hat {\sf W})$ is called the Laplace
transform of $\hat {\sf W}$.
\item $\FC^{-1} (\hat {\sf W}) (\Re z)$ is the boundary--value of 
$\LC (\hat {\sf W}) (\underline{z})$ for $\Im (z_j - z_{j+1}) \to 0$ inside
$T^n$, i.e. the relation

$$\lim_{ \Im (z_j - z_{j+1})\in \G\to 0} 
\LC (\hat {\sf W}) (\underline{z}) =
  \FC^{-1} (\hat {\sf W}) (\Re \underline{z})
  $$

holds in the sense of tempered distributions in the argument 
$\Re \underline{z} \in \R ^d$. Here $\G \subset V^-_0$ is 
a subcone of $V^-_0$
such that $\G \cup \{ 0 \}$ is closed in $\R^d$, and the Fourier
transform $\FC $ is taken w.r.t. the Minkowski inner product on
$M_d\cong \R^d$.
\end{enumerate}
\end{Theorem}

Well--definedness in Theorem \ref{7.1theo} holds, since $e(\cdot,
\underline{z})$ on the
support of $\hat {\sf W}$ behaves like a fast decreasing function, if 
$\underline{z} \in T^n$ (see also Definition \ref{sechs.1def}). 
We remark, that 
$< \overline{\hat {\sf W} } , e(\cdot , \underline{z}) >$ does only depend
on the variables $z_j - z_{j+1}$ $j=1 \ldots n-1$ as a consequence of
(\ref{7.4eqa}) and up to the factor $(2 \pi)^{d/2}$ equals
the Laplace--transform of $\hat {\sf w}$, defined in the sense of
\cite{StWi}. (In \cite{StWi} a different convention
on the Fourier--transform leads to the interchange of forward and
backward cones). Theorem \ref{7.1theo} is essentially equal to Theorem 2.6
and Theorem 2.9 of \cite{StWi}.

Let us remark, that most textbooks work with other conventions on the 
Laplace--transform as we do here, since we here mostly use the 
variables $z_j$  
and not the difference variables $\zeta_j:=z_j-z_{j+1}$. Lastly, let us
remark that for
distributions that depend only on one variable $k \in \R^d$ we keep
the conventions of Section 6.

We observe that

\begin{eqnarray}
\label{7.5eqa}
{(E_d)}^n_< &:=&  \Bigl\{ \underline{z} \in {(M^{\C}_d)}^n : \Im (z^0_j
- z^0_{j+1}) < 0 \ , \ j=1 \ldots n-1 \ , \nonumber \\
& &  \Im \vec z_j = 0 \ , \
\Re z^0_j = 0  \ j=1 \ldots n \Bigr\}
\end{eqnarray}
is included in $T^n$ and is a $dn$-dimensional real manifold.
We can calculate the coefficients of a local expansion of a function
holomorphic in $T ^n$ around a point $o\in (E_d)_<^n$ by performing
''real'' differentiations inside $(E_d)_<^n$. Therefore,
functions which are single valued holomorphic on $T^n$ are determined
by their values on $(E_d)_<^n$. This gives rise to the following

\begin{Theorem}
\label{7.2theo}
Let ${\sf S} \in \SC _c' (\R^{dn})$ and $\hat {\sf W} \in \SC_c (\R^{dn})$
~fulfil the spectral condition. Assume

\begin{equation}
\label{7.6eqa}
\BS \left( (\Im z^0_1 , \Re \vec z_1),\ldots , (\Im z^0_n , 
\Re \vec z_n ) \right) = \LC (\hat {\sf W} ) (\underline{z}) \ \ 
\underline{z} \in {(E_d)}^n_< 
\end{equation}
Then $\hat {\sf W}$ is determined uniquely by this requirement. If $\BS$ is
invariant under Euclidean transformations, then ${\sf W} =\FC^{-1} (\hat {\sf W})$
is Poincar\'e--invariant. If $\BS$ is, in addition, symmetric, then $W$ is
local in the sense that

\begin{equation}
\label{7.7eqa}
{\sf W}(x_1, \ldots ,x_j, x_{j+1}, \ldots ,x_n) = {\sf W}(x_1, \ldots , 
x_{j+1}, x_j, \ldots ,x_n) 
\end{equation}
if $ x_j - x_{j+1}$ is space--like,
i.e. $(x_j - x_{j+1})^2 < 0$. Furthermore, if $\BS$ is a real
distribution, $\theta$--invariance of $\BS$ implies Hermiticity of ${\sf W}$,
i.e.

\begin{equation}
\label{7.8eqa}
{\sf W}(x_1 , \ldots , x_n) = \overline{ {\sf W}(x_n , \ldots , x_1)}
\end{equation}
\end{Theorem}

Theorem \ref{7.2theo} is part of the reconstruction theorem in\cite{OS1} and 
\cite{Si}. We remark, that for symmetric and Euclidean invariant
${\sf S}$, $\LC (\hat {\sf W})$ possesses a single valued holomorphic 
extension to
the so--called permuted extended tube $T^n_{p,e}$ (see \cite{StWi} for
definitions and proofs). Our restriction of Theorem \ref{7.2theo} to
tempered distributions ${\sf S}$ is motivated only by our model, where
''time--coincident''Schwinger functions are well defined. In \cite{OS1}
Theorem \ref{7.2theo} is proved for ${\sf S} \in \SC_{\ne ,c}' (\R^{dn})$, which is
a larger class of distributions than $\SC'_c (\R^{dn})$.

\begin{Remark}
\label{7.1rem}
Let $y_l = (y_l^0 , \vec y_l ) = (\Im z^0_l , \Re \vec z_l )$. Then
(\ref{7.6eqa}) reads

\begin{equation}
\label{7.9eqa}
\BS (y_1 , \ldots , y_n) = (2 \pi)^{-{dn\over 2}} \int_{\R^{dn}}
e^{- \sum _{l=1}^nk^0_l y^0_l+ i \vec k_l \vec y_l} \hat {\sf W}
(k_1 , \ldots , k_n) \bigotimes^n_{l=1} dk_l
\end{equation}
where $\underline{y} = (y_1 \ldots y_n) \in {(\R^d)}^n_< :=$
$\{ \underline{x} \in \R^{dn} : x^0_1 < x^0_2 < \ldots < x^0_n \}$.
\end{Remark}

\subsection{Truncated Schwinger functions with "sharp masses"}
Our goal in this subsection is to represent the truncated Schwinger 
functions $S^T_n,n\ge 2$ of a convoluted generalized white noise 
as the restriction to the Euclidean region ${(E_d)}^n_<$ of a Laplace 
transform of a tempered distribution $\hat{W}^T_n$ having the 
spectral property. In \cite{AIK} and \cite{AW} this is done for the
truncated Schwinger functions of $X = \D^{-1} F$. The kernel 
$\D^{-1} (x)$ can be represented as the Laplace transform of the
essentially (up to a multiplication with a positive constant) unique 
Lorentz--invariant measure on the forward mass
"hyperboloid" with mass $m=0$. Given that, the crucial step in the
analytic continuation is a change of variables, depending on the value
of $m$.

Since Section 6 shows that we have to deal with a continuum of
masses rather than with a sharp mass, we proceed in three steps: 
First in this
subsection we give an integral representation of $S^T_n$ in terms 
of "truncated
Schwinger functions with sharp masses" $S^T_{\underline{m},n}$ ,
$\underline{m} = (m_1, \ldots ,m_n)$. Then the subsection 7.4 deals with the
representation of $S^T_{\underline{m},n}$ as
the Laplace--transform of a tempered distribution 
$\hat W^T_{\underline{m},n}$ restricted to ${(E_d)}^n_<$. Finally, we
integrate $\hat W^T_{\underline{m},n}$ over the masses to obtain the
Fourier transform of the truncated $n$--point Wightman function
$W^T_n$ (subsection 7.5). In that subsection we also collect the 
properties of the obtained (truncated) Wightman distributions arising at 
the main theorem of the present section.

From now on, we restrict ourselves to convoluted generalized white noises 
$X=G*F$ with a kernel $G$ which admits a representation of the form

\begin{equation}
\label{7.10eqa}
G(x)=\int_{\R^+} C_m(x) \rho (dm^2) \ , \ x \in \R ^d \setminus \{ 0\}
\,
\end{equation}
where $\rho$ is a (possibly signed) Borel measure on $\R_+$. Furthermore
we restrict ourselves to $\rho$'s which fulfill the
following

\begin{Condition}
\label{7.1con}
\begin{enumerate}
\item There exists a mass $m_0 > 0$, such that 
supp$\rho \subset [m^2_0 , \infty)$.
\item $\int_{\R^+} {1 \over m^2} \vert \rho \vert (dm^2) < \infty$.
\end{enumerate}
\end{Condition}

\begin{Remark}
\label{7.2rem}
\begin{enumerate}
\item Since $C_m$ is the Laplace transform of the Lorentz invariant
distribution $(2\pi)^{-{d\over 2}}\d^+_m (k) :=(2\pi)^{-{d\over 2}}
 1_{\{ k^0 > 0\}}(k) \d (k^2 - m^2)$ in the
sense of Section 6, (\ref{7.10eqa}) and Condition \ref{7.1con} 
mean that $G$ is the Laplace transform of a signed Lorentz invariant
measure on the forward mass--cone $V^{*+}_{m_0}$.
\item Condition \ref{7.1con} is also a sufficient condition for 
$\GC :  f \in \SC (\R^d) \mapsto G* f \in \SC (\R^d)$ to be well--
defined and continuous. The proof of this statement can be
verified using similar ~techniques as in the proof of Lemma \ref{7.1lem} 
\cite{Go}.
\item The $\rho$'s obtained in Section 6 obviously ~fulfil Condition
\ref{7.1con}.
\end{enumerate}
\end{Remark}

\begin{Lemma}
\label{7.1lem}
Let $\rho$ and $G$ be as above and $S^T_n$ as in (\ref{vier.4eqa}). Then for
$\vp \in \SC (\R^{dn})$

\begin{equation}
\label{7.11eqa}
< S^T_n , \vp > = c_n \int_{{(\R^+)}^n} < S^T_{\underline{m},n}, \vp>
\underline{\rho} (d \underline{m}^2)
\end{equation}
where $\underline{m} = (m_1 , \ldots , m_n)$ and
$\rho (d\underline{m}^2) = \rho^{\otimes n} (dm^2_1 \times \ldots
\times dm^2_n)$. $S^T_{\underline{m},n}$ is defined by

\begin{equation}
\label{7.12eqa}
< S^T_{\underline{m},n} , \vp_1 \otimes \ldots \otimes \vp_n >:=
\int_{\R^d} C_{m_1} * \vp_1 \ldots C_{m_n} * \vp_n dx \ , \
\vp_1 \ldots \vp_n \in \SC (\R^d)
\end{equation}
\end{Lemma}

{\bf Proof.}
By the nuclear theorem, (\ref{7.12eqa}) well--defines
$S^T_{\underline{m},n} \in \SC ' (\R^{dn})$. To prove (\ref{7.11eqa})
let again $\vp = \vp_1 \otimes \ldots \otimes \vp_n$, with
$\vp_1 \ldots \vp_n \in \SC (\R^d)$. We remark that
$C_m (x) = m^{d-2} C_1 (mx)$ for $x \in \R^d\setminus \{ 0\}$ and 
$C_1 \in L^1 (\R^d , dx)$ (cf. \cite{GlJa} p. 126). Therefore

\begin{eqnarray}
\label{7.13eqa}
& & \int_{{(\R^+)}^n} \int_{\R^d} \int_{\R^{dn}} 
    \left\vert \prod^n_{l=1} C_{m_l} (y_l) \vp_l (x-y_l) \right\vert 
    dy_1 \ldots dy_n dx \vert \underline{\rho} \vert
    (d\underline{m}^2) \nonumber\\
& = & \int_{{(\R^+)}^n} \int_{\R^d} \int_{\R^{dn}} 
    \left\vert \prod^n_{l=1} C_1 (y'_l) \vp_l (x-{y'_l \over m_l}) 
    \right\vert dy'_1 \ldots dy'_n dx 
    \bigotimes^n_{l=1} {\vert \rho \vert \over m^2_l}
    (d{m}^2_l) \nonumber\\
& \le & {\left[ {\Vert C_1 \Vert}_{L^1 (\R^d, dx)} \int_{\R^+} 
        {1 \over m^2} \vert \rho \vert (dm^2)
        \right]}^n \prod^{n-1}_{l=1} 
        {\Vert \vp_l \Vert}_{L^\infty (\R^d , dx)} 
        {\Vert \vp_n \Vert}_{L^1 (\R^d , dx)} \nonumber \\
&< & \infty 
\end{eqnarray}
where we have applied

$$\int_{\R^d} \left\vert \prod^n_{l=1} \vp_l (x-z_l) \right\vert
  dx \le \prod^{n-1}_{l=1} {\Vert \vp_l \Vert}_{L^\infty 
  (\R^d , dx)} {\Vert \vp_n \Vert}_{L^1 (\R^d , dx)} \ \  ,
  z_1 \ldots z_n \in \R^d.
  $$

Now (\ref{7.13eqa}) allows us to apply Fubini's theorem to the LHS
of (\ref{7.11eqa}). The LHS and the RHS of (\ref{7.11eqa}) 
are therefore equal for 
$\vp = \vp_1 \otimes \ldots \otimes \vp_n$. At the same time
(\ref{7.13eqa}) shows, by the nuclear theorem, that both sides of
(\ref{7.11eqa}) denote tempered distributions. These give, by the 
above argument,
equal values if evaluated on  test functions
$\vp = \vp_1 \otimes \ldots \otimes \vp_n$. Therefore, by the nuclear 
theorem again, the distributions on both sides of (\ref{7.11eqa}) are equal.
\kasten

\subsection{The Schwinger functions with "sharp masses" as Laplace transforms}
From now on we assume 
$\underline{y} = (y_1 , \ldots , y_n) \in {(\R^d)}^n_<$, 
$\underline{m} = (m_1 , \ldots , m_n) \in {(\R^+)}^n$. Then

$$S^T_{\underline{m},n} (y_1 , \ldots , y_n) = 
  \int_{\R^d} C_{m_1} (x-y_1) \ldots C_{m_n} (x-y_n)\, dx
$$
is well--defined as a function, since the singularities of the functions
$C_{m_l}{(x - y_l )}$ are all separated from each other 
and are therefore all
integrable and, furthermore, for large $x$ the integrand falls off exponentially to
zero (cf. \cite{GlJa} p. 126).

Taking into account that

$$C_{m_l} (x-y_l) = (2 \pi)^{-d} \int_{\R^d}
  e^{-k^0_l \vert x^0 - y^0_l \vert + i \vec k_l (\vec x
  - \vec y_l)} \d^+_{m_l} (k_l) dk_l \ ,
  $$
we get by Fubini's theorem
\begin{eqnarray}
\label{7.14eqa}
    S^T_{\underline{m},n} (y_1 , \ldots , y_n)
& = & (2 \pi )^{d-1-dn} \int_{\R^{d}}
      \left[ \int_\R \prod^n_{l=1} e^{-k^0_l \vert x^0 - y^0_l \vert}
      dx^0 \right] \nonumber\\
& \times  & e^{-i\sum^n_{l=1} \vec k_l \vec y_l} \prod^n_{l=1} \d^+_{m_l} (k_l)
      \d \left( \sum^n_{l=1} \vec k_l \right) \bigotimes^n_{l=1} dk_l
\end{eqnarray}
where we have also applied the distributional identity 
$\FC \left( {1 \over \sqrt{2 \pi}^{d-1}} \right) = \d$ in the
distribution space $\SC'_{\C} (\R^{d-1})$. 

The RHS of (\ref{7.14eqa}) has to be considered as an integral over a
submanifold of $\R^{dn}$ determined by the $\d$--distributions. 
If not stated ~otherwise, all products of distributions which
occur in the following are defined in this way.

The expression in the brackets [...] in (\ref{7.14eqa}) equals

\begin{eqnarray}
\label{7.15eqa}
& & {1 \over \sum^n_{l=1} k^0_l} \prod^n_{l=2} 
    e^{-k^0_l (y^0_l - y^0_1)} + \sum^{n-1}_{j=1} \prod^{j-1}_{l=1}
    e^{-k^0_l (y^0_j -y^0_l)} (y^0_{j+1} - y^0_j) \nonumber\\
& & \times \int^1_0 e^{-[ ( \sum^j_{l=1} k^0_l)s - (\sum^n_{l=j+1}
    k^0_l) (1-s)] (y^0_{j+1} - y^0_j)} ds \nonumber \\
& & \times \prod^n_{l=j+2} e^{-k^0_l (y^0_l - y^0_{j+1})} + 
    {1 \over \sum^n_{l=1} k^0_l} \prod^{n-1}_{l=1} 
    e^{-k^0_l (y^0_n - y^0_l)} \, .
\end{eqnarray}

If we insert (\ref{7.15eqa}) into (\ref{7.14eqa}), then the RHS of
(\ref{7.14eqa}) splits up into $n+1$ summands, say

\begin{equation}
\label{7.16eqa}
S_{\underline{m},n}^T (y_1 , \ldots , y_n) = I_0 (y_1  \ldots y_n)
+ \sum^{n-1}_{j=1} (y^0_{j+1} - y^0_j) I_j (y_1  \ldots  y_n) 
+ I_n (y_1  \ldots  y_n)
\end{equation}

We are going to write each of these summands in the form of
(\ref{7.9eqa}). This is easy for $I_0$ and $I_n$:

\beas
&&    I_0 (y_1 , \ldots , y_n)\\
& = & (2 \pi )^{d-1-dn} \int_{\R^{dn}}
   \prod^n_{l=2} e^{-k^0_l y^0_l - i \vec k_l \vec y_l}
   e^{-(-\sum^n_{l=2}k^0_l) y^0_1 - i \vec k_1 \vec y_1}\\
& \times & {1 \over \sum^n_{l=0} k^0_l} \prod^n_{l=1} \d^+_{m_l} (k_l)
   \d \left( \sum^n_{l=1} \vec k_l \right) \bigotimes^n_{l=1} dk_l\\
& = & (2 \pi )^{d-1-dn}  
   \int_{\R^{d(n-1)}  \times \R^{d-1}}
   \prod^n_{l=2} e^{-k^0_l y^0_l - i \vec k_l \vec y_l}
   e^{-(-\sum^n_{l=1}k^0_l) y^0_1 - i \vec k_1 \vec y_1}\\
& \times  & {1 \over \om_1 + \sum^n_{l=2} k^0_l} \prod^n_{l=2} \d^+_{m_l} (k_l)
   \d \left( \sum^n_{l=1} \vec k_l \right) \bigotimes^n_{l=2} dk_l
   \otimes {d\vec k_1 \over 2 \om_1}\\
& = & (2 \pi )^{d-1-dn} 
   \int_{\R^{dn}} e^{-\sum _{l=1}^n k^0_l y^0_l + i \vec k_l \vec y_l}
   {1 \over 2 \om_1 (\om_1 - k^0_1)} \\
&   & \times \prod^n_{l=2} \d_{m_l}^+(k_l) \d
   \left( \sum^n_{l=1} k_l \right) \bigotimes^n_{l=1} dk_l \, ,
\eeas
where we have introduced $\om_l = \sqrt{\vec k^2_l + m^2_l}$ ,
$l = 1 \ldots n$. Thus, $I_0$ is the Laplace transform of the tempered
distribution

\begin{equation}
\label{7.17eqa}
(2\pi )^{d-1-{dn \over 2}} {1 \over 2 \om_1 (\om_1 - k^0_1)} \prod^n_{l=2}
\d^+_{m_l} (k_l) \d \left( \sum^n_{l=1} k_l \right)
\end{equation}
restricted to ${( E_d)}^n_<$. By an ~analogous calculation we find
that $I_n$ is the Laplace transform of the tempered distribution

\begin{equation}
\label{7.18eqa}
(2\pi )^{d-1-{dn\over 2}} {1 \over 2 \om_n (\om_n + k^0_n)} \prod^{n-1}_{l=1}
\d^-_{m_l} (k_l) \d \left( \sum^n_{l=1} k_l \right)
\end{equation}
restricted to ${( E_d)}^n_<$, where $\d ^-_{m_l}$ is defined as $\theta \d ^+_{m_l}$.

To check the temperedness of the
distributions (\ref{7.17eqa}) and (\ref{7.18eqa}) we observe that for 
$\underline{k} = (k_1 \ldots k_n)$ in the support of these
distributions the denominators are larger than $4 m^2_0$, where 
$m_0 \le m_1,\ldots ,m_n$. The
spectral condition can be directly deduced from these formulas
 (cf. the proof of proposition \ref{7.1prop} below).

Let us now turn to the more complicated calculations for the $I_j$'s
$j=1 \ldots n-1$.
\beas
    I_j (y_1 \ldots y_n)
& = & (2\pi )^{d-1-dn}  \int_{\R^{dn}}
   \prod^{j-1}_{l=1} e^{k^0_l y^0_l - i \vec k_l \vec y_l} \\
&  &\times \int^1_0 e^{[ (\sum^j_{l=1} k^0_l)s + (\sum^n_{l=j+1} k^0_l)
   (1-s) - (\sum^{j-1}_{l=1} k^0_l) ] y^0_j - i \vec k_j \vec y_j}\\
& & \times e^{ -[ (\sum^j_{l=1} k^0_l)s + (\sum^n_{l=j+1} k^0_l)
   (1-s) - (\sum^{n}_{l=j+2} k^0_l) ] y^0_{j+1} - i \vec k_{j+1} 
   \vec y_{j+1}} ds \\
& & \times \prod^n_{l=j+2} e^{-k^0_l y^0_l - i \vec k_l \vec y_l}
   \prod^n_{l=1} \d^+_{m_l} (k_l) \d \left( \sum^n_{l=1} \vec k_l \right)
   \bigotimes^n_{l=1} dk_l \ .
\eeas

We may interchange $ds$ and $\bigotimes^n_{l=1} dk_l$ integrations by
Fubini's theorem. Furthermore, we integrate over
$\d^+_{m_j} (k_j) \d^+_{m_{j+1}} (k_{i+1}) dk^0_j dk^0_{j+1}$ and
change coordinates $k^0_l \mapsto - k^0_l$ for $l = 1 \ldots j-1$,
getting the RHS to be equal to

\beas
& & (2\pi )^{d-1-dn} \int^1_0 \int_{\R^{d(n-2)} \times
   \R^{(d-1)2}} \prod^{j-1}_{l=1} e^{-k^0_l y^0_l - i \vec k_l \vec
   y_l} \\
& &  \times \prod^{j+1}_{l=j} e^{-\tilde k^0_l y^0_l - i \vec k_l \vec
   y_l} \prod^{n}_{l=j+1} e^{- k^0_l y^0_l- i \vec k_l\vec y_l} \\
& &  \times \prod^{j-1}_{l=1} \d^-_{m_l} (k_l) \prod^n_{l=j+2}
   \d^+_{m_l}(k_l) \bigotimes^n_{l=1 , l \ne j,j+1}
   dk_l \bigotimes^{j+1}_{l=j} {d\vec k_l \over 2 \om_l} ds \ ,
\eeas
where we have used the following notations

\beas
\tilde k^0_j &=& \tilde k^0_j (k^0_1 , \ldots , k^0_{j-1} , \vec k_j ,
  \vec k_{j+1} , k^0_{j+1} \ldots k^0_n)\\
& := & - \{ (
  -\sum^{j-1}_{l=1} k^0_l + \om_j ) s + ( \om_{j+1}
  + \sum^n_{l=j+2} k_l^0 ) (1-s) + \sum^{j-1}_{l=1} k^0_l
  \} \  ;  \\ 
\tilde k^0_{j+1} & = & \tilde k^0_{j+1} (k^0_1 , \ldots , k^0_{j-1} , 
  \vec k_j , \vec k_{j+1} , k^0_{j+2} \ldots k^0_n) \\
&:=&   \{ ( -\sum^{j-1}_{l=1} k^0_l + \om_j ) s + 
  ( \om_{j+1} + \sum^n_{l=j+2} k^0_l ) (1-s) - 
  \sum^{n}_{l=j+2} k^0_l  \}.
\eeas

We remark that 
$\sum^{j-1}_{l=1} k^0_l + \tilde k^0_j + \tilde k^0_{j+1} + 
\sum^n_{l=j+2} k^0_l = 0$. Therefore we may introduce new
''integrations'' over new variables $k^0_j$, $k^0_{j+1}$ using the
measure
\beas
& & \d (k^0_j - \tilde k^0_j (k^0_1 , \ldots , k^0_{j-1} , 
  \vec k_j , \vec k_{j+1} , k^0_{j+2} , \ldots , k^0_n)) \d
  \left( \sum^n_{l=1} k^0_l \right) dk^0_j dk^0_{j+1} \\
& = & \d (a (k_j , k_{j+1})s - b (k_{j+1})) \d 
   \left( \sum^n_{l=1} k^0_l \right) dk^0_j dk^0_{j+1} \ ,
\eeas
where
\beas
  a(k_j , k_{j+1}) & = & -k^0_j - \om_j - k^0_{j+1} + \om_{j+1} \\
  b(k_{j+1})       & = & -k_{j+1}^0 + \om_{j+1} \ .  
\eeas
In this way we get
\begin{eqnarray}
\label{7.19eqa}
& {} & I_j(y_1,\ldots ,y_n)  \nonumber \\
& = & (2\pi)^{d-1-dn} \int^1_0 \int _{\R ^{dn}}
     \prod^n_{l=1} e^{-k^0_l y^0_l - i \vec k_l \vec y_l}
     \prod^{j-1}_{l=1} \d^-_{m_l} (k_l) \nonumber\\
& \times & {\d (a(k_j , k_{j+1})s - b(k_{j+1})) \over 4 \om_j \om_{j+1}}
     \prod^n_{l=j+2} \d^+_{m_l} (k_l) \d \left( \sum^n_{l=1} k_l
     \right) \bigotimes^n_{l=1} dk_l ds \ .
\end{eqnarray}

For $n \ge 3$ or $n=2$, $m_1 \ne m_2$ $a(k_j , k_{j+1}) \ne 0$ holds
almost everywhere with respect to the measure
$\prod^{j-1}_{l=1} \d^-_{m_l} (k_l) \prod^n_{l=j+2} \d^+_{m_l} (k_l)
\d \left( \sum^n_{l=1} k_l \right) \bigotimes^n_{l=1} dk_l$. In these
cases we may, by Fubini's theorem again, change the order of the $ds$ and
$\bigotimes^n_{l=1} dk_l$ integrations. This together with

$$\int^1_0 \d (as - b) ds = {1 \over \vert a \vert} \left[
  1_{\{ 0 < b < a \}}(a,b) + 1_{\{ a < b < 0 \}} (a,b)
  \right] \quad \hbox{ for } \ a\ne 0
$$
inserted into (\ref{7.19eqa}), allows us to conclude that $I_j$ is the Laplace--
transform of the distribution
\begin{eqnarray}
\label{7.20eqa}
& &     H_j (k_1 \ldots k_n) \nonumber \\
& = &  (2\pi)^{d-1-{dn\over 2}} {\textstyle
       {\left[ 1_{\{ 0 < b(k_{j+1}) < a(k_j , k_{j+1})\}} (k_j,k_{j+1})
+ 1_{\{ a(k_j , k_{j+1})< b (k_{j+1})< 0 \}}(k_j , k_{j+1})\right] 
\over 4 \om_j \om_{j+1}
       \vert a(k_j , k_{j+1})\vert} }\nonumber\\
& \times &  \prod^{j-1}_{l=1} \d_{m_l}^- (k_l) 
\prod^n_{l=j+2} \d^+_{m_l} (k_l) 
\d \left( \sum^n_{l=1} k_l \right)
\end{eqnarray}
restricted to ${(E_d)}^n_<$. 

The temperedness of $H_j$ can be derived from 
an integral representation like that in (\ref{7.19eqa}), where the 
exponential
functions have to be replaced by a test function $\vp \in \SC (\R ^d)$:
\bea
\label{7.20beqa}
\vert <H_j,\vp >\vert 
& \le &  (2\pi )^{d-1-{dn\over 2}} (\prod_{l=1}^n
m_l )^{-1} \nonumber \\
&\times & \left( \int _{\R^{(d-1)(n-1)}}{1\over (1+\vert \vec{\underline{k}}
\vert ^2)^{dn/2} } \ d \vec{ \underline{k}} \right) \Vert \vp \Vert _{0,dn} \ ,
\eea
where $\Vert \vp \Vert _{0,dn} := \sup _{\underline{k} \in \R ^{dn}} \vert (1+
\vert \underline{k} \vert ^2)^{dn/2} \vp (\underline{k}) \vert$.

Since $I_j(y_1,\ldots ,y_n)$ is the Laplace--transform of $H_j(k_1,\ldots ,k_n)$, 
$(y^0_{j+1} - y^0_j) I_j (y_1, \ldots ,y_n)$ is the Laplace transform of
the tempered distribution $((\p^0_{j+1}\linebreak - \p^0_j) H_j(k_1,\ldots ,k_n))$
where $\p^0_l = {\p \over \p k^0_l}$ $ l=1 \ldots n$. Terms that
depend only on $k^0_j + k^0_{j+1}$ give a zero contribution when
derived with respect to $\p^0_{j+1} - \p^0_j$. This applies to 
$a(k_j , k_{j+1})$ and $\d \left( \sum^n_{l=1} k_l \right)$. Thus,
only the derivatives of the characteristic functions in
(\ref{7.20eqa}) contribute to $(\p^0_{j+1} - \p^0_j) H_j$.

Taking into account
\beas
  1_{\{ 0< b(k_{j+1}) < a (k_j , k_{j+1})\}}
& = & 1_{\{ k^0_j < - \om_j \}} 1_{\{ k^0_{j+1} <  \om_{j+1} \}},\\
   1_{\{ a (k_j , k_{j+1})< b(k_{j+1}) < 0\}}
& = & 1_{\{ k_j^0 > - \om_j \}} 1_{\{ k_{j+1}^0 >  \om_{j+1} \}},
\eeas
${d \over dx} 1_{\{ 0<x\}}(y) = \d(y)$ and also $(2\om_l )^{-1}$
$\d (k^0_l \mp \om_l ) = \d^\pm_{m_l} (k_l)$, we calculate
\begin{eqnarray}
\label{7.21eqa}
&  & {1 \over 4\om_j \om_{j+1}} (\p^0_{j+1} - \p^0_j) 
     1_{\{ 0< b(k_{j+1}) < a (k_j , k_{j+1})\}}(k_j , k_{j+1})
     \nonumber\\
& = &  (-2\om_j )^{-1} 1_{\{ k^0_j < -\om_j \}}(k_j)
      \d^+_{m_{j+1}} (k_{j+1}) \nonumber \\
&   &+ \d^-_{m_j} (k_j)
      (2\om_{j+1})^{-1}  1_{\{ k^0_{j+1} < -\om_{j+1} 
      \}} (k_{j+1})  \nonumber \\
&  & {1 \over 4\om_j \om_{j+1}} (\p^0_{j+1} - \p^0_j) 
     1_{\{ a (k_j , k_{j+1})< b(k_{j+1}) < 0\}}(k_j , k_{j+1})
     \nonumber\\
& = & (2\om_j )^{-1} 1_{\{ k^0_j > -\om_j \}}(k_j)
      \d^{+}_{m_{j+1}} (k_{j+1}) \nonumber \\
&   & + \d^-_{m_j} (k_j)
      (2\om_{j+1})^{-1}  1_{\{ k^0_{j+1} < \om_{j+1}\}}(k_{j+1})  \nonumber \\
&&
\end{eqnarray}

Adding up both sides of (\ref{7.21eqa}) yields 
 
\begin{eqnarray}
\label{7.22eqa}
&   & (\p^0_{j+1} - \p^0_j) H_j(k_1,\ldots,k_n) \nonumber \\
& = & (2\pi )^{d-1-{dn\over 2}}  
     {\textstyle \Biggl\{ 
     { \hbox{sign } (k^0_j + \om_j ) \d^+_{m_{j+1}}
     (k_{j+1}) \over 2 \om_j \vert k^0_j + \om_j + k^0_{j+1} 
     - \om_{j+1} \vert} 
     + {\d^-_{m_j} (k_j) \hbox{sign } (\om_{j+1} - k^0_j) \over
     2 \om_{j+1} \vert k^0_j + \om_j + k^0_{j+1} - \om_{j+1} \vert}
     \Biggr\} } \nonumber \\
&  \times & \prod^{j-1}_{l=1} \d^-_{m_l} (k_l) \prod^n_{l=j+2} 
\d^+_{m_l} (k_l) \d \left( \sum^n_{l=1} k_l \right) \nonumber \\
& = & (2\pi )^{d-1-{dn\over 2}} \prod^{j-1}_{l=1} \d^-_{m_l} (k_l) 
     \Biggl\{ 
     {\d^+_{m_{j+1}} (k_{j+1}) \over 2 \om_j (k^0_j + \om_j)} 
     + {\d^-_{m_j} (k_j)  \over 2 \om_{j+1} ( \om_{j+1} - k^0_{j+1})}
     \Biggr\} \nonumber \\
&\times &     \prod^n_{l=j+2} \d^+_{m_l} (k_l) \d 
     \left( \sum^n_{l=1} k_l \right) 
\end{eqnarray}

A closer analysis shows, that the singularities on the RHS of
(\ref{7.22eqa}) have to be understood in the sense of Cauchy's
principal value. 

Keeping in mind that
$${1 \over 2\om (\om + k^0)} + {1 \over 2 \om (\om - k^0)} = 
  {(-1) \over k^2 - m^2} \ \ , \ \om = \sqrt{\vec k^2 + m^2} \ ,
$$
by adding up (\ref{7.17eqa}), (\ref{7.22eqa}) for 
$j=1 \ldots n-1$ and (\ref{7.18eqa}) (recall also (\ref{7.16eqa}) ) we
get the following
\begin{Proposition}
\label{7.1prop}
For $\underline{m} = (m_1 \ldots m_n) \in (\R _+)^n$, let 
$\hat W^T_{\underline{m},n}$ denote the distribution
\be
\label{7.23aeqa}
(2\pi )^{d-1-{dn\over 2}} \Biggl\{ \sum^n_{j=1} \prod^{j-1}_{l=1} \d^-_{m_l}
     (k_l) {(-1) \over k^2_j - m^2_j} \prod^n_{l=j+2} \d^+_{m_l}
     (k_l) \Biggr\} \d \left( \sum^n_{l=1} k_l \right) 
\ee
for $\ n\ge 3 \hbox{ or } n=2 \ , \ m_1 \ne m_2 $ and
\be
\label{7.23beqa}
 (2\pi )^{-1} \biggl\{ (2\om_1 )^{-2} \d^-_{m_1}
     (k_1) - (2\om_1 )^{-1} (\p_1^0 \d^-_{m_1} ) (k_1) \biggr\}
     \d (  k_1 + k_2)
\ee
for $  n=2 \ , \ m_1 = m_2$.

Then
\begin{enumerate}
\item $\hat W^T_{\underline{m},n}$ is tempered and ~fulfils the
strong spectral condition with the mass gap 
$m_0 \le \min \{ m_l : l = 1 \ldots n \}$
\item $S^T_{\underline{m},n}$ is the restriction of 
$\LC (\hat W^T_{\underline{m},n})$ to the Euclidean region
${(E_d)}^n_<$ in the sense of (\ref{7.6eqa}). This property determines 
$\hat W^T_{\underline{m},n}$ uniquely.
\item $W^T_{\underline{m},n} = \FC^{-1} (\hat W^T_{\underline{m},n})$
is Poincar\'e invariant.
\end{enumerate}
\end{Proposition}

{\bf Proof.}
We only deal with the case $n \ge 3$ or $n=2$ $m_1 \ne m_2$.
Concerning the support properties, let us concentrate on the $j$'th summand
in (\ref{7.23aeqa}). Let $\underline{k} = (k_1, \ldots ,k_n)$ be in the
support of this summand. For $r < j$,
$\sum^r_{l=1} k_l \in V^{*-}_{m_0}$ holds, since each $k_l$, $l = 1\ldots r$,
is in this cone. For $n-1 \ge r \ge j$ we get
$\sum^r_{l=1} k_l = - \sum^n_{l=r+1} k_l \in V^{*-}_{m_0}$, since each 
$k_l$, $l = r+1\ldots n$, is in $V^{*+}_0$ and thus 
$-k_l \in V^{*-}_0$.

The facts that $\hat W^T_{\underline{m},n}$ is tempered, and that 
$S^T_{\underline{m},n}$ is the restriction of 
$\LC ( W^T_{\underline{m},n})$ to ${(E^n_d)}^n_<$ summarize the above
discussion. We have worked in the notations of (\ref{7.9eqa}) rather
than in that of (\ref{7.6eqa}), but, as already remarked, these two
relations are equivalent. The other statements follow from Theorem
\ref{7.2theo}.
\kasten

The derivation of (\ref{7.23beqa}) is relatively easy. In the
following we will not use this formula and therefore leave the
calculation as an exercise.

Nevertheless, the formula (\ref{7.23beqa}) has some consequences: If we
take a noise $F$ with mean zero and 
$X=(-\D + m^2_1)^{-1} F = (2\pi )C_{m_1} * F$, then it is easy to see that 
(\ref{7.23beqa}) gives the Fourier transform of the 2--point Wightman
function of the model. Since the distribution in (\ref{7.23beqa}) does
not admit a K\"allen--Lehmann representation, not all of the 
one-particle and free ''states'' of this model have positive norm. Thus, a
good physical interpretation of such models is impossible, even if $F$ has
zero Poisson part.  

Two more details may be of interest:

\begin{Remark}
\label{7.3rem}
\begin{enumerate}
\item Let $L^{\uparrow}_+(\R^d)$ denote the proper orthochronous Lorentz group.
For $\L \in L^{\uparrow}_+(\R^d)$ and $a \in \R^d$, we recall that 
the Poincar\'e
group acts on functions $\vp (k_1, \ldots, k_n)$ defined on the momentum
space $(\R^d)^n$ as follows

$$(T_{\{ \L , a\}} \vp ) (k_1, \ldots ,k_n) = \vp ((\L ^*)^{-1} k_1 ,
  \ldots , (\L ^*)^{-1} k_n) e^{i< \sum^n_{l=1} k_l , a>_M}.
  $$
where the adjoint $\L^*$ is w.r.t. $< \ , \ >_M$,
 the Minkowski inner product. From
(\ref{7.23aeqa}) we know that the  $\hat W^T_{\underline{m},n}$ are
''manifestly'' Poincar\'e invariant.
\item Only if $m_1 = m_2 = \ldots = m_n$ we can expect 
$ W^T_{\underline{m},n}$ to be a ''local'' 
distribution (cf. Theorem \ref{7.2theo}).
\end{enumerate}
\end{Remark}

\subsection{The analytic continuation of the truncated Schwinger functions}
Proposition \ref{7.1prop} immediately implies

\begin{Theorem}
\label{7.3theo}
Suppose that the distribution 
$\hat W^T_n := c_n \int_{(\R^+)^n} \hat W^T_{\underline{m},n}
\underline{\rho} (d\underline{m}^2)$, i.e.

\begin{equation}
\label{7.24eqa}
< \hat W^T_n , \vp> = c_n \int_{(\R^+)^n} <\hat W^T_{\underline{m},n}
 , \vp > \underline{\rho} (d\underline{m}^2) \ \ 
\vp \in \SC (\R^{dn})
\end{equation}
 is well--defined for all $\vp \in \SC (\R^{dn})$ and furthermore
$\hat W^T_n \in \SC' (\R^{dn})$. Then

\begin{enumerate}
\item $\hat W^T_n$ ~fulfils the strong spectral condition with
respect to the mass gap $m_0$, where $m_0$ is as in Condition
\ref{7.1con}.
\item $S^T_n$ is the restriction of the Laplace transform 
$\LC (\hat W^T_n)$ to the Euclidean region 
${(E_d)}^n_<$. This determines $\hat W^T_n$ uniquely. Furthermore 
$W^T_n =\FC^{-1} (\hat W^T_n)$ is the boundary-value of the analytic
function $\LC (\hat W^T_n) (\underline{z}) , \underline{z} \in T^n$, for
$\Im \underline{z} \to 0$ as described in Theorem \ref{7.1theo}. In
this sense, we call the truncated $n$--point Wightman distribution $W^T_n$
the analytic continuation of $S^T_n$ to the Minkowski space--time.
\item $W^T_n$ is a Poincar\'e invariant, local, hermitian distribution,
which ~fulfils, in addition, the cluster--property of the truncated
Wightman distributions, i.e. for
$\vp_1 \ldots \vp_{n+m} \in \SC (\R^{dn})$ and a spacelike $a\in \R ^d$ 
($a^2<0$) we have
\begin{equation}
\label{7.25eqa}
W^T_{m+n} {(\vp_1 \otimes \ldots \vp_m \otimes T_{\l a}
(\vp_{m+1} \otimes \ldots \otimes \vp_{m+n}))}
\to 0 \ \ , \mbox{if } \l \to \infty ,
\end{equation}
where $T_{\l a}$ is the translation by $\l a$.
\end{enumerate}
\end{Theorem}

{\bf Proof.}
\begin{enumerate}
\item By Proposition \ref{7.1prop} all the 
$\hat W^T_{\underline{m},n}$, $\underline{m} = (m_1 \ldots m_n)$,
$m_l \in $ supp$\rho$ $l = 1 \ldots n$, ~fulfil the strong spectral
condition with the mass gap $m_0>0$ given by Condition \ref{7.1con}. Since $\hat W^T_n$ is a superposition of such
$\hat W^T_{\underline{m},n}$, the same applies to $\hat W^T_n$.
\item As remarked before, $\tilde e (\underline{k},\underline{y})$
$:= (2\pi)^{-{dn\over 2}}e^{-\sum^n_{l=1} k^0_l y^0_l + 
i \vec k_l \vec y_l}$ on the support
of $\hat W^T_n$ behaves like a fast falling function in
$\underline{k} \in \R^{dn}$, whenever  
$\underline{y} = (y_1,\ldots ,y_n)\in (\R^d)^n_<$. Therefore, 
the following equations hold:
\beas
    <\hat W^T_n , \tilde e ( \cdot , \underline{y})>
& = & c_n \int_{(\R^+)^m} <\hat W^T_{\underline{m},n} , \tilde e
   ( \cdot , \underline{y}) > \underline{\rho} (d\underline{m}^2) \\
& = & c_n \int_{(\R^+)^m}  S^T_{\underline{m},n} (y_1, \ldots ,y_n)
    \underline{\rho} (d\underline{m}^2) \\
& = & S^T_n (y_1, \ldots , y_n ).
\eeas
The second equality is valid by Proposition \ref{7.1prop}, where the
RHS makes sense $d\underline{y}$ a.e. by Lemma \ref{7.1lem}, which also 
implies the
third equality. Theorem \ref{7.1theo} and \ref{7.2theo} now imply $2$.
\item Except for the cluster--property, everything follows from $2$
and Theorem \ref{7.2theo}. For the cluster--property, we refer to
\cite{OS1}, Theorem \ref{vier.1theo} and Corollary \ref{vier.1cor}
(see alternatively \cite{RS} Vol. III p. 324).
\end{enumerate}
\kasten

\begin{Corollary}
\label{7.1cor}
Let ${\{ W_n \}}_{n \in \N_0}$ be the Wightman distributions
determined by the truncated sequence ${\{ W^T_n \}}_{n \in \N}$ and
$W_0 = 1$. Then $\hat W_n$ ~fulfils the spectral condition, for $n\in \N$. 
The
statements $1$, $2$ and $3$ of Theorem \ref{7.3theo} hold, if the $W^T_n$,
$S^T_n$ are replaced by $W_n$, $S_n$, respectively and the cluster--
property of the truncated Wightman function is replaced by that for
the Wightman functions (see \cite{StWi} or \cite{OS1}).
\end{Corollary}

For a proof of ~Corollary \ref{7.1cor} we refer to similar discussions
in Section 4 and to \cite{Ar1}. 

Let us now turn to the question of the temperedness of the formal
expressions (\ref{7.24eqa}). It is e.g. not difficult to see that for
$\rho$'s that have compact support in $\R_+$, temperedness
follows (c.f. (\ref{7.20beqa}) ). Nevertheless, in these cases we cannot 
expect the two--point
Wightman function to admit a K\"allen--Lehmann representation. The
reader is asked to convince herself/himself that there is no such
representation e.g. for the case $\rho (dm^2)=f(m^2)dm^2$, $f>0$, where
$f\in \SC (\R)$ has compact support in $\R^+$. In this case again, we 
cannot give a good physical interpretation, even not for 
one--particle or free states.

Therefore we restrict ourselves to the $\rho$'s obtained in Section 6,
i.e.

\begin{equation}
\label{7.26eqa}
\rho_\a (dm^2) = 2 \sin \pi \a 1_{\{ m^2> m^2_0\}}(m^2)
{dm^2 \over (m^2 - m^2_0)^\a} \ \ \a \in (0,1)
\end{equation}
Again it is possible to show the temperedness of the distributions $\hat
W_{n,\a }^T$, defined by (\ref{7.24eqa} ) with $\rho = \rho _\a $ if $\a
\in (1/2,1)$ by a direct estimate, using (\ref{7.20beqa}). 

Let us therefore turn to the case $\a \in (0,\half ]$.
We introduce the notations

\beas
   \mu^+_\a (k) 
& = & (2\pi)^{-d/2}\sin \pi\a 1_{\{ k^2 > m^2_0 , k^0 >0\}}(k)
   {1 \over (k^2 - m^2_0)^\a} \\
   \mu^-_\a (k)
& = &  (2\pi)^{-d/2} \sin \pi\a 1_{\{ k^2 > m^2_0 , k^0 <0\}}(k)
   {1 \over (k^2 - m^2_0)^\a} \\
   \mu_\a (k) 
& = & (2\pi)^{-d/2}\left( \cos \pi \a 1_{\{ k^2 > m^2_0 \}} (k)+ 
   1_{\{ k^2 < m^2_0 \}} (k) \right) 
   {1 \over \vert k^2 - m^2_0\vert^\a}
\eeas
and we get:

\begin{Proposition}
\label{7.2prop}
Let $\hat W_{n,\a}^T$ be defined as in (\ref{7.24eqa}) with 
$\underline \rho = \rho _\a$, $\a \in (0,\half )$. Then 
$\hat W_{n,\a}^T$ is tempered and equal to
\begin{equation}
\label{7.27eqa}
c_n (2\pi)^d 2^{n-1} 
\left\{ \sum^n_{j=1} \prod^{j-1}_{l=1}
\mu^-_\a (k_l) \mu_\a (k_j) \prod^n_{l=j+1} \mu^+_\a
(k_j) \right\} \d \left( \sum^n_{l=1} k_l \right) \ n \ge 2
\end{equation}
\end{Proposition}

{\bf Proof.}
Despite the fact, that one cannot apply Fubini's theorem because of the
presence of the Cauchy principal values in (\ref{7.23aeqa}), it can be
shown by a regularization (of the $\rho_\a$ and the $\hat
W_{\underline{m},n}^T)$ and a passage to the limit, that the integration
w.r.t. $\underline{\rho}  _\a(d\underline{m})$ and the evaluation with a
test function $\vp \in \SC (\R ^{dn})$ can be interchanged in 
(\ref{7.24eqa}). Therefore, we get for $\hat W_{n,\a}^T$:
\beas
&  & c_n (2\pi)^{d-1-{dn\over 2}} 2^{n-1} \Biggl\{ \sum^n_{j=1} 
\prod^n_{l=1}\left[ \sin \pi\a
 \int_{m_0^2}^\infty { \d^-_{m_l} (k_l) \ dm^2_l\over (m_l^2-m_0^2)^\a}
 \right]  \Biggr. \\
&\times &  \Biggl. \left[ \int_{\R^+}
  {\rho _\a (dm_j^2) \over m^2_j - k^2_j} \right]  \prod^n_{l=j+1}\left[
 \sin \pi \a  \int_{m_0^2}^\infty  {\d^+_{m_l} (k_l) \
 dm^2_l\over (m_l^2-m_0^2)^\a} \right] \Biggr\} \d \left( \sum^n_{l=1}
  k_l \right) \\
& = & c_n (2\pi)^{d-1-{d\over 2}} 2^{n-1} \Biggl\{ \sum^n_{j=1} \prod^{j-1}_{l=1}
   \mu^-_\a (k_l)  \Biggr. \Biggl. \left[ \int_{\R^+}
  {\rho _\a (dm_j^2) \over m_j^2 - k^2_j}  \right] \\
& \times &\prod^n_{l=j+1}
  \mu^+_\a (k_l)  \Biggr\} \d \left( \sum^n_{l=1}
  k_l \right) \ .
\eeas
But (cf. \cite{Bro} p.70)
\beas
& &   \int^\infty_{m^2 > m^2_0} {1 \over (m^2 - k^2)} 
   {1 \over (m^2 - m^2_0)^\a} dm^2 \\
& = & \int^\infty_0 {1 \over (x-(k^2 - m^2_0)) x^\a} dx \\
& = & {1 \over \vert k^2 - m^2_0\vert^\a} \int^\infty_0
   {1 \over y \mp 1} dy \\
& = & \cases{\pi \cot \pi\a {1 \over \vert k^2 - m^2_0 \vert^\a}
          &if $k^2-m^2_0 > 0$ \ ; \cr
          \pi (\sin \pi\a)^{-1} {1 \over \vert k^2-m^2_0 \vert^\a}
          &if $k^2 - m^2_0 < 0 $ \ . \cr}
\eeas
Thus, we get (\ref{7.27eqa}). It is not difficult to show the
temperedness of (\ref{7.27eqa}) by application of a Cauchy--Schwarz
inequality, making also use of the fact that 
${(\mu^+_\a)}^2 , {(\mu^-_\a)}^2 , {(\mu_\a)}^2$ are locally
integrable functions.
\kasten

Poincar\'e-- invariance in (\ref{7.27eqa}) follows from 
Theorem \ref{7.3theo}, but
is also ''manifest'' (cf. Remark \ref{7.3rem} $1$.). Let us make sure
that (\ref{7.27eqa}) in the case $n=2$ yields the same result as
Section 6. We have
\begin{eqnarray}
\label{7.28eqa}
&   & c_2 (2\pi)^d 2 \left\{ \mu_\a (k_1) \mu^+_\a (k_2) + 
      \mu^-_\a (k_1) \mu_\a (k_2) \right\} \d (k_1+k_2) \nonumber\\
& = & c_2 4 \sin \pi \a \cos \pi \a \left[
      1_{\{ k^2_1 > m^2_0 , k^0_1 < 0\}} (k_1)
      {1 \over (k^2_1 - m^2_0)^\a} \right]^2 \d (k_1 + k_2)
      \nonumber\\
& = & c_2 2 \sin 2\pi \a 
      1_{\{ k^2_1 > m^2_0 , k^0_1 < 0\}} (k_1)
      {1 \over (k^2_1 - m^2_0)^{2\a}}  \d (k_1 + k_2) 
\end{eqnarray}
for $\a \in (0,\half )$,
where we have applied $\sin \pi\a \cos \pi\a = \half \sin (2\pi \a)$.
(\ref{7.28eqa}) differs only by a time reflection $\theta$ from the 
distribution defined in Remark \ref{sechs.4rem}, which arises from different conventions in the
definition of the Laplace--transform in Section 6 and Section 7. Thus,
we have the same result as in Section 6. (Note that the additional factor 
$(2\pi)^{-{d\over 2}}$ in Remark 6.7 arises from the fact, that the
normalization factor of the Fourier transform in one argument is
$(2\pi)^{-{d\over 2}}$ while for the Fourier transform in 2 arguments it
is $(2\pi)^{-d}$).

\begin{Corollary}
\label{7.2cor}
For $n \ge 3$, Proposition \ref{7.2prop} also applies to
$\hat W^T_{n,\half}$. A technical calculation shows, that
(\ref{7.27eqa}) is also tempered for $\a = \half$. Furthermore, the 2-
point-function $ W_{2,\half}^T=(2\pi )c_2\FC ^{-1}[\d _{m_0}^-(k_1)\d (k_1+k_2)]$
is the well-known 2-point function of the
relativistic free field. Therefore, also in the case $\a = \half $
Theorem \ref{7.1theo} applies.
\end{Corollary}

\begin{Remark}
\label{7.4rem}
For $0 < \a < \half$, $c_1 = 0$, $\hat W_{2,\a} = \hat W^T_{2,\a}$ admits
a K\"allen--Lehmann representation. Therefore the corresponding
Gaussian Euclidean field with covariance--function $S_{2,\a}$ is
reflection positive (but not Markov, the latter being seen from a general
theorem of Pitt \cite{Pi}).

For $\a = \half$ the
corresponding Gaussian Euclidean field is the Markov free field of
mass $m_0$ (\cite{N2}). This can be also taken from the Equation
(\ref{7.28eqa}) by the following considerations: For $\a
\uparrow \half$, on one hand we have that the coefficient 
$\sin(2\pi\a)\downarrow 0$. This implies that the Fourier transformed
truncated 2-point functions $\hat \W^T_{2,\a}$ vanish on open sets in
$\R^{d2}$ which do not intersect the mass-shell $\{ k_1^2=m_0^2,
k_1^0 <0, k_1+k_2=0 \} $. On the other hand, the singularity of the 
$\hat \W^T_{2,\a}$ on this mass-shell causes non-integrability for $\a\uparrow \half$. 
Combining this two aspects in a quantitive calculation \cite{Go}, one can show
that
$$\lim _{\a\uparrow\half}  2 \sin 2\pi \a 
      1_{\{ k^2_1 > m^2_0 , k^0_1 < 0\}} (k_1)
      {1 \over (k^2_1 - m^2_0)^{2\a}} = 2\pi \d _{m_0}^-(k_1)\ ,$$
where the limit is the weak limit in $\SC '(\R^d)$.

Let us now have a look on (\ref{7.28eqa}) for $1> \a > \half$. First of
all we note, that (\ref{7.28eqa}) is no more tempered, since the
exponent $-2\a$ is smaller than -1 and (\ref{7.28eqa}) is not locally
integrable. Therefore, formula(\ref{7.28eqa}) in this case can not represent
$\hat W_{2,\a}^T$. Nevertheless, one may speculate that (\ref{7.28eqa}) 
still holds if $k_1$ stays away from the mass-shell $\{ k_1^2=m_0^2,
k_1^0 <0, k_1+k_2=0 \} $. If this were true, an interesting observation can be made:
Since the function $\sin 2\pi\a $ at $\a = \half $ changes it´s sign from
$+$ to $-$, the corresponding truncated 2-point Schwinger  function
$S_{2,\a }^T$ for $\a\in (\half,1)$ would be "reflection
negative", rather than "reflection positive".

Thus, it seems, as if the Markov--property in the case of
Gaussian Euclidean random fields would appear at the ''boundary'' of
reflection--positivity. Nevertheless, a proper ~treatment of this problem has to be
left to future work. 
\end{Remark}

\subsection{Positivity in the scattering region}
Let us concentrate on $\a = \half$ and $d = 4$, $c_1 = 0$. 
We thus consider the truncated Wightman functions of the convoluted
 generalized white noise 
$X=(-\D + m^2_0)^{-\half} F$. If $F$ is Gaussian, $X$ is the free
Markov field of mass $m_0$ and the analytic continuation of its only
nonzero Schwinger function $S^T_2$ yields that $S^T_2$ on 
${(E_d)}^2_<$ is the Laplace ~transform of

$$\hat W^T_2 (k_1 , k_2) = (2\pi)^{d+1} c_2
  \d^-_{m_0} (k_1) \d (k_1 + k_2) \ .
  $$
This is the Fourier transform of the 2--point function of the
relativistic free field of mass $m_0$ \cite{N2}. Let 
${\{ W^g_n\}}_{n \in \N}$ be the sequence of Wightman functions
composed from the truncated sequence $\{ W^T_n\}_{n\in\N }$, $W^T_n = 0 $
$n \ne 2$, $W^T_2 = \FC^{-1} (\hat W^T_2)$. $W^g_n$ is composed from the
$W^T_2$ in the way of Corollary \ref{drei.2cor}. For 
$\vp \in \SC_{\C} (\R^{dn})$ define $\vp^*$ by $\vp^* (x_1 \ldots x_n)$
$= \bar f (x_n \ldots x_1)$. It is well known, that in this case
positivity holds for the $\{ W_n^g \} _{n\in \N _0}$, i.e. for $\vp^0 \in \C$,
$\vp_l \in \SC_{\C} (\R^{dl})$, $l = 1 \ldots n$, we can define a
seminorm for the vector 
$\Psi = (\vp^0 , \vp_1 \ldots \vp_n , 0 \ldots)$ in the Bochner algebra
$\bigoplus^\infty_{n=0} \SC_{\C} (\R^{dn}) =: \underline{\SC }$ as
\begin{equation}
\label{7.29eqa}
\Vert \Psi \Vert^2_g := \sum^n_{l,m =0} W^g_{l+m} (\vp^*_l
\otimes \vp_m) \ge 0
\end{equation}
This allows us to look at equivalence classes of vectors with norm larger 
than zero as the
''physical states'' of the theory. Let now $F$ be a generalized white
noise. We define the general non--definite (cf. Section 5) squared
pseudo--norm $\Vert \ \cdot \ \Vert^2$ on $\underline{\SC }$ in analogy to
(\ref{7.28eqa}), where the $W^g_{l+m}$'s are replaced by $W_{l+m}$'s
obtained from the truncated Wightman distributions
$W^T_{m,\half} = \FC^{-1} (\hat W^T_{m, \half})$ defined in Corollary
\ref{7.2cor}. Let us now fix $\ve > 0$ and define a region $U$ in the
Minkowski space time $M_d$ as
$U:= \{ k \in M_d : k^2 \in [m^2_0 - \ve , m^2_0 + \ve ] , k^0 > 0
\}$. Let $\SC (U)$ denote the Schwartz functions on $\R^d$ with
support in $U$. For $\vp \in \SC (\R^d)$ such that 
$\hat \vp \in \SC (U)$ define $\vp (x,t) := \FC^{-1} (\tilde \vp_t)$
$(x)$ where $\tilde \vp_t (k) = \hat \vp (k) e^{i(k^0  - \om)t}$. Let
us concentrate on $\Psi (t) \in \underline{\SC}$ such that 
$\Psi (t) = (\vp^0 , \vp_1 (t) ,\ldots ,\vp_n (t) 0 \ldots 0 \ldots )$ 
where $\vp_r (t) = \bigotimes^r_{s=1} \vp^s_r (t)$, with 
$\vp^s_r (t)$ defined as above. It is well known \cite{He} that the
"wave packet" $\vp^s_r (t)(x)$ is concentrated near the plane $x^0 =t$. In
this sense, we say $\Psi (t)$ approaches the asymptotic region 
$x^0 \to \pm \infty$ if $t$ goes to that limit.

Let us quote \cite{RS} Vol III p. 324 ff. and \cite{He} for the
following basic results of Haag--Ruelle--Theory, based only on
locality, strong spectral condition and invariance of the $W^T_n$ $n
\ge 3$ and the special form of $W^T_2$:
\begin{equation}
\label{7.29aeqa}
{d \over dt} W^T_2 \left( \vp^{s_1 (*)}_{r_1} (t)\otimes  
\vp^{{s_2}(*)}_{r_2}(t) \right)= 0
\end{equation}
and
\begin{equation}
\label{7.29beqa}
W^T_n \left( \vp^{{s_1}(*)}_{r_1} (t)\otimes \ldots \otimes
\vp^{{s_n}(*)}_{r_n}(t) \right) \to 0 \ \mbox{for} \ n \ge 3 \mbox { as }
t\to \pm \infty .
\end{equation}
The limit in (\ref{7.29beqa}) is approached as
$(1+ \vert t \vert)^{{3 \over 2}(n-2)}$ falls to zero in the general case, 
and faster than $(1+ \vert t \vert)^{-N}$ $N \in \N$ falls to zero, 
if the $\vp^{s_i}_{r_i}(0)$'s are non--overlapping in velocity--space, i.e.
$\om^{-1}_i k^{+1}_i \ne \om^{-1}_j k^{-1}_j$ for $k_i \in 
$supp$\vp^{s_i}_{r_i}$ $k_j \in 
$supp$\vp^{s_j}_{r_j}$ $j,i = 1 \ldots n$ $i \ne j$. 

Since we can
expand $\Vert \Psi (t) \Vert^2$ into products of truncated Wightman
functions of the type 
$W^T_2 (\vp^{s_1 (*)}_{r_1} (t) \otimes \vp^{s_2 (*)}_{r_2} (t))$ and those of the
type of the LHS of (\ref{7.29beqa}), we get

\begin{Proposition}
\label{7.3prop}
Let $\Psi (t) \in \underline{\SC}, \Vert \ . \ \Vert^2$ $\Vert \ . \ \Vert_g^2$ as
above. Then

\begin{equation}
\label{7.30eqa}
\Vert \Psi (t) \Vert^2 \to \Vert \Psi (0) \Vert^2_g \  \mbox{ as } t\to
\pm \infty
\end{equation}
The limit here is approached as $(1 + \vert t \vert)^{- {3 \over 2}}\to 0$
in general and faster as $(1 + \vert t \vert)^{-N}\to 0$ for any $N\in
\N$ for a ''non--overlapping'' $\Psi (0)$.
\end{Proposition}

\begin{Remark}
\label{7.5rem}
\begin{enumerate}
\item Let us point out, that the existence of stable one--particle
respectively of free states, i.e. the holding of (\ref{7.29aeqa}),
 is special for $\a = \half$. We do not have an analogue 
of these statements e.g. in the case $\a \in
(0,\half )$.
\item We remark that the result of Proposition \ref{7.3prop} in the
non--overlapping case holds for all dimensions 
$d \ge 2$ \cite{He}. Proposition \ref{7.3prop} can also be generalized
to the dimensions $d \ge 4$, but the classical literature only deals
with the physical--space--time $d=4$.
\item If there is at least one $\vp_s$, such that all 
$\hat \vp^r_s$ $r = 1 \ldots s$ take nonzero values on the mass-shell of
mass $m_0$, then we have $\Vert \Psi (0) \Vert^2_g > 0$ and thus also
$\Vert \Psi (t) \Vert^2$ becomes positive for large $t$. 
\item $\Vert \ . \ \Vert_g$ may also be called the norm of a free or 
noninteracting  state. In the very vague sense of (\ref{7.30eqa}) we may
therefore say, that $\Psi (t)$ approaches a free state. Nevertheless,
we cannot define asymptotic free states as in Haag--Ruelle theory,
since at the moment we have no Hilbert topology in which the $\Psi (t)$'s could
converge. 
\item For further investigations it seems therefore to be
necessary to introduce a suitable auxiliary topology. A promising
candidate is the Krein topology: In \cite{AGW1} we proved that 
the Wightman distributions of our model fulfill
the Hilbert-structure condition of \cite{MS}. The model developed here 
thus fulfills the modified Wightman axioms for "fields in an 
indefinite metric" (see e.g. \cite{Bo} for a definition of such fields).
By the general results on
spaces with indefinite inner product 
( see again\cite{MS} and references therein) we get that
there exists a Hilbert space $(\HC ,(.,.))$ such that 
$\underline{\SC }\subset
\HC $ is dense. If $<.,.>$ denotes the inner product on
$\underline{\SC}$ induced by the
sequence of Wightman distributions $\{ W_n\} _{n\in \N_0}$, then there
is a continuous self-adjoined operator $\eta $ on 
$\HC $ s.t. $\eta ^2 =1$ and
$(.,\eta .)=<.,.>$ holds on $\underline{\SC}$. 
\end{enumerate}
\end{Remark}

Finally, we would like to summarize this section as follows:
In Proposition \ref{7.2prop} and Corollary \ref{7.2cor} we give
explicit formulae for the Fourier--transformed (truncated) Wightman
distributions that belong to the random field 
$X=(-\D+m^2_0)^{-\a} F$, $\a \in (0 , \half]$. The sequence of Wightman
distributions constructed from the former distributions ~fulfils all
Wightman axioms (cf. \cite{Si}, \cite{StWi}) of relativistic QFT,
except for the positivity of the square norm in the state--space, which 
in some
cases does not hold and in others is uncertain (it is certain only for
the case where $F$ is Gaussian). Nevertheless, in the case $\a = \half$ there
exists stable one resp. free states, and therefore Haag--Ruelle theory
allows us to derive a positivity condition for states $\Psi (t)$
approaching the asymptotical respectively scattering regions as
$t \to \pm \infty$.

\

\noindent {\bf Acknowledgements} We thank D. Applebaum, C. Becker, 
P. Blanchard, R. Gielerak, Z. Haba, J. Sch\"afer and Yu. M. Zinoviev 
for stimulating discussions. The financial support of D.F.G. is 
gratefully acknowledged.

\def\HK{ H\o egh--Krohn}

\end{document}